\documentclass[sn-mathphys-num]{sn-jnl}

\usepackage{subfigure}

\usepackage{lmodern}        
\usepackage{amsmath}        
\usepackage{amssymb}

\usepackage{amssymb}

\usepackage{graphicx}%
\usepackage{multirow}%
\usepackage{amsmath,amssymb,amsfonts}%
\usepackage{amsthm}%
\usepackage{mathrsfs}%
\usepackage[title]{appendix}%
\usepackage{xcolor}%
\usepackage{textcomp}%
\usepackage{manyfoot}%
\usepackage{booktabs}%
\usepackage{algorithm}%
\usepackage{algorithmicx}%
\usepackage{algpseudocode}%
\usepackage{listings}%

\usepackage{enumerate}
\usepackage{geometry} 
\usepackage{epsfig}
\usepackage{color}
\usepackage{appendix}

\usepackage[utf8]{inputenc}

\usepackage{bm}
\usepackage{array}
\usepackage{xspace}
\usepackage{verbatim}
\usepackage{hyperref}

\usepackage{etoolbox}
\usepackage[title]{appendix}

\newtheorem{theorem}{Theorem}[section]
\newtheorem{lemma}[theorem]{Lemma}
\newtheorem{proposition}[theorem]{Proposition}
\newtheorem{corollary}[theorem]{Corollary}
\theoremstyle{definition}

\theoremstyle{remark}
\newtheorem{remark}[theorem]{Remark}
\numberwithin{equation}{section}
\newcommand {\beq} {\begin{equation}}
\newcommand {\eeq} {\end{equation}}

\newcommand{\Rmnum}[1]

\apptocmd{\appendix}
  {\renewcommand{\theequation}{\thesection.\arabic{equation}}%
   \setcounter{equation}{0}}
  {}{}

\begin{document}

\title{Analytical estimations of edge states and extended states in large finite-size lattices}


\author[1]{\fnm{Huajie} \sur{Song}}

\author*[1]{\fnm{Haitao} \sur{Xu}}\email{hxumath@hust.edu.cn}

\affil*[1]{\orgdiv{Center for Mathematical Sciences}, \orgname{Huazhong University of Science and Technology}, \orgaddress{\street{Luoyu Road 1037}, \city{Wuhan}, \postcode{430074}, \state{Hubei}, \country{China}}}


\abstract{The bulk–boundary correspondence, one of the most significant features of topological matter, theoretically connects the existence of edge modes at the boundary with topological invariants of the bulk spectral bands. However, it remains unspecified in realistic examples how large the size of a lattice should be for the correspondence to take effect. In this work, we employ the diatomic chain model to introduce an analytical framework to characterize the dependence of edge states on the lattice size and boundary conditions. In particular, we apply asymptotic estimates to examine the bulk–boundary correspondence in long diatomic chains as well as precisely quantify the deviations from the bulk-boundary correspondence in finite lattices due to symmetry breaking and finite-size effects. Moreover, under our framework the eigenfrequencies near the band edges can be well approximated where two special patterns are detected. These estimates on edge states and eigenfrequencies in linear diatomic chains can be further extended to nonlinear chains to investigate the emergence of new nonlinear edge states and other nonlinear localized states. In addition to one-dimensional diatomic chains, examples of more complicated and higher-dimensional lattices are provided to show the universality of our analytical framework.}




\maketitle

\section{Introduction}
\label{sec:intro}

The study of localized solutions in lattices has been a classical topic for decades. In the setting of nonlinear lattices, localized states such as solitary traveling waves and breathers have been extensively studied almost since the start of the modern nonlinear science~\cite{FPU, Toda, Ovchinnikov, Kosevich, Flach, Yaroslav, Malomed, Fibich, IoKi, MacKay, Aubry2, Livi, Jame2, Frie, Io, Lev, Vainchtein} and emerged in a wide range of examples such as optical waveguides~\cite{CDN, Lederer}, Bose-Einstein Condensates~\cite{Trombettoni},  Josephson-junction arrays~\cite{Trias}, granular crystals~\cite{Sen} and biological molecules~\cite{Kuwayama}. 

More recently, topological materials and their equipped edge states have attracted tremendous attention to localized states in linear lattices~\cite{{Bernevig, Kane1, Kane2, Bernevig2, Zhang2020, Klitzing, Laughlin, Halperin, Wen1, Wen2}} (although the interest has later been extended to nonlinear edge states as well). It is known as ``bulk-boundary correspondence" that the existence of edge states in lattices is connected to the bulk topology, which can usually be characterized by topological invariants such as Chern number~\cite{Thouless, Berry, Kohmoto, Hatsugai} or Zak phase~\cite{Zak}. As a result, edge states are known to persist under perturbations as long as the bulk topology is protected. With that being said, the topologically protected states and their robustness have spurred substantial applications in optical, acoustic, mechanical and electrical systems~\cite{Prodan2009, Wang09, Khanikaev13, Rechtsman13, Huber15, Khanikaev16, He16, Shi2017, Rechtsman18, Ozawa19, Ma19}.

Although the lattices in numerical experiments or realistic applications have to be of finite size, existing theoretical results on finite-size lattices are still relatively limited~\cite{Zhou, PhysicaE, Hou23, Liu24, Wang25}. To be specific, some studies assume the lattices to be infinite or cyclic (with periodic boundary conditions), which unfortunately are fundamentally different from non-cyclic finite lattices. In other discussions, finite-size lattices are usually assumed to be large enough and equipped with fixed special boundary conditions (such as free ends). With that being said, although nontrivial bulk topology is expected to imply the existence of edge states in large enough lattices (by bulk-boundary correspondence), it remains theoretically unspecified how the competition between the lattice size and the boundary-induced localization length determines the existence of edge modes. In this work, we introduce an analytical framework to characterize the interdependence of edge states on the lattice size $n$ and boundary stiffness. This allows us to determine not just the critical boundary conditions for a fixed size, but conversely, the minimum lattice size required to support robust edge states against boundary interference.


In addition to edge states with eigenfrequencies (or energy levels in the context of quantum mechanics) outside the spectral bands, we are also able to estimate eigenfrequencies inside the spectral bands and near the band edges for large lattices. Interestingly, these near-band-edge eigenfrequencies are found to follow special patterns that depend on boundary conditions. Moreover, knowledge of these eigenfrequencies and their corresponding eigenstates in large linear lattices also enables theoretically studying the emergence of nonlinear localized states (including nonlinear edge states and middle-localized states) with frequencies exiting from the spectral bands. It should be noticed that the localized states emerging this way are exclusive in nonlinear lattices, and they are different from those frequently mentioned nonlinear edge states continued from linear edge states ~\cite{Leykam16, Zangeneh2019, Smirnova20, SSH21,Rajesh2022, Magnus}. Moreover, attention has been devoted to self-induced nonlinear edge states through envelope approximation or effective parameter mapping~\cite{Hadad2016, Hadad2018, Chaunsali2019, Chaunsali2023}. Compared with existing results, our work employs a fully discrete asymptotic analysis, which not only potentially helps the development of new nonlinear topological materials, but also significantly benefits the theoretical understanding of the emergence of localized states in finite nonlinear lattices where eigenfrequency exiting spectral bands is a typical scenario lacking careful explanations~\cite{Aubry1, Chaunsali20}. 

In this work, we start with an acoustic diatomic chain as a toy model for demonstration, which is simple enough to be theoretically analyzed and accessible enough to be experimentally implemented. However, our results on linear edge states, spectrally embedded eigenfrequencies and nonlinear localized states are universal such that they can be extended to a wide range of models with various setups. Particularly, we provide a detailed discussion for the one-dimensional case and showcase the existence of similar results in higher dimensions partially through numerical computations. 

The structure of this article is as follows. In Sec. \ref{linear model}, we introduce the basic infinite model of a one-dimensional linear diatomic chain and explicitly calculate the eigenfrequencies and eigenstates. In Sec. \ref{sec:semi_infinite} we consider semi-infinite chains where the effects of boundary conditions and topological invariants on edge states are discussed. In Sec. \ref{sec:finite_edgestates} we move to finite chains and introduce the long-chain limit (or semi-infinite limit) to show the existence and estimates of edge states under different boundary setups. In Sec. \ref{sec:extended_states}, estimates on the eigenfrequencies near the band edges and the corresponding eigenstates in long linear chains are discussed. In Sec. \ref{sec:extended topics}, we demonstrate the extensibility of our framework by applying it to nonlinear chains as well as multi-layer chains and two-dimensional lattices, showing how the rigorous 1D linear results serve as a foundation for understanding these complex systems.
In Sec. \ref{sec:conclusion}, the main results are summarized and future challenges are discussed. In order to emphasize our main findings among involved calculations, most of the proofs and results of a more technical nature are presented in the Appendices.

\section{The model of a linear diatomic chain}
\label{linear model}

We start with a simple infinite linear diatomic chain with identical masses $M$ and alternating nearest-neighbour interactions, which can be viewed as an acoustic version of Su-Schrieffer-Heeger (SSH) model~\cite{SSH79}. This is effectively a spring-mass system, whose equations of motion are as follows:
%
\begin{equation}
\begin{split}
  M\frac{d^2}{dt'^2}{x}_{2m} = & K_{1}(x_{2m-1}-x_{2m})+K_{2}(x_{2m+1}-x_{2m}), \\
  M\frac{d^2}{dt'^2}{x}_{2m+1} = & K_{2}(x_{2m}-x_{2m+1})+K_{1}(x_{2m+2}-x_{2m+1})
\end{split}
\label{eq:motion0}
\end{equation}
where $x_j$ denotes the displacement of 
$j$-th mass, $K_1$ and $K_2$ are the two alternating spring stiffness constants, respectively. Unless later specifically explained, it will be assumed that the stiffness constants are non-negative and $K_{2}>K_{1}>0$.

With the scaling $t=\sqrt{\frac{K_1}{M}}t'$, $q_i(t)=x_i(t')$ and $k_j=\frac{K_j}{K_1}$ for $i\in\mathbb{Z}$ and $j=1,2$, the above equations can be rewritten as
\begin{equation}
\begin{split}
  \ddot{q}_{2m} = & k_{1}(q_{2m-1}-q_{2m})+k_{2}(q_{2m+1}-q_{2m}), \\
  \ddot{q}_{2m+1} = & k_{2}(q_{2m}-q_{2m+1})+k_{1}(q_{2m+2}-q_{2m+1}).
\end{split}
\label{eq:infinite_chain}
\end{equation}
In the following discussion $k_1=1$ can be viewed as just a constant but our results indeed also hold for generic $k_1$.
Focusing on the time-periodic solutions of~\eqref{eq:infinite_chain}, we adopt the ansatz ${q}(t)={u}e^{i\omega t}$ where $\omega$ represents the eigenfrequency. Due to the periodic structure of the chain, two adjacent masses form a unit cell and two adjacent unit cells satisfy the following iteration relation
\begin{equation}
\begin{split}
  \left(
  \begin{array}{c}
    u_{2m+1} \\
    u_{2m+2}
  \end{array}
  \right)
  =&
  \frac{1}{k_{1}k_{2}}
  \left(
  \begin{array}{cc}
    -k_{1}^{2} & -k_{1}(\omega^{2}-k_{1}-k_{2}) \\
    k_{1}(\omega^{2}-k_{1}-k_{2}) & (\omega^{2}-k_{1}-k_{2})^{2}-k_{2}^{2}
  \end{array}
  \right)
  \left(
  \begin{array}{c}
    u_{2m-1} \\
    u_{2m}
  \end{array}
  \right)\\
  =&
  \mathcal{T}(\omega)
  \left(
  \begin{array}{c}
    u_{2m-1} \\
    u_{2m}
  \end{array}
  \right).
  \end{split}
  \label{eq:iteration}
\end{equation}
If the eigenvalues of $\mathcal{T}$ are denoted by $\lambda$, then they can be calculated from
\begin{equation}
\label{iterative eigenvalue}
  \lambda^{2}+\lambda(\frac{-\omega^{4}
  +2\omega^{2}(k_{1}+k_{2})-2k_{1}k_{2}}
  {k_{1}k_{2}})+1=0
\end{equation}
where two roots of the above equation satisfy $\lambda_{1}\lambda_{2}=1$ and $\lambda_1+\lambda_2\in\mathbb{R}$. This implies that only the following two situations (A1) and (A2) can take place.
\begin{itemize}
\item (A1): $|\lambda|=1$ ($\omega^2\in [0,2k_{1}]\cup[2k_{2},2k_{1}+2k_{2}]$, these two bands for frequencies are usually called ``acoustic band" and ``optical band" respectively);
\item (A2): $\lambda\in\mathbb{R}$ and $\lambda\neq \pm 1$ ($\omega^2\in (2k_{1},2k_2)\cup(2k_{1}+2k_2,\infty)$). 
\end{itemize}
Without loss of generality, we assume that $|\lambda_1|\leq|\lambda_2|$ and use $a$ to denote the eigenvalue $\lambda_1$, then the eigenfrequency $\omega$ can be expressed as
\begin{equation}
  \omega^{2}=k_{1}+k_{2}
  +\sigma\sqrt{(k_{1}+k_{2}a)
  (k_{1}+k_{2}/a)}, \quad \sigma=\pm 1.
  \label{eq:omega}
\end{equation}
As a special case, the eigenvalue of $\mathcal{T}$ is $1$ or $-1$ (with algebraic multiplicity two and geometric multiplicity one) when the frequency sits exactly at some edge of the two bands, namely $\omega^2\in\{0, 2k_1, 2k_2, 2k_1+2k_2\}$. In the generic situation where $a\neq\pm1$, the spatial profile $u$ can be described through the iterative relation between unit cells as 
\beq
\label{eq:u_cells}
(u_{2m+1}, u_{2m+2})^T=c_1 a^m v_1+c_2 a^{-m} v_2
\eeq
where 
\beq
\label{eq:v1v2_0}
v_1= b_1\left(\begin{array}{c}
  -\sigma\sqrt{k_1^2+k_2^2+k_1k_2(a+1/a)} \\
  {k_1+k_2 a}
 \end{array}\right), \quad 
v_2= b_2\left(\begin{array}{c}
  -\sigma\sqrt{k_1^2+k_2^2+k_1k_2(a+1/a)} \\
  {k_1+k_2 /a}
 \end{array}\right)
\eeq 
are eigenvectors of $\mathcal{T}(\omega)$ for eigenvalues $a$ and $\frac{1}{a}$, respectively. In order to obtain an eigenstate localized at the left edge, an obvious choice is setting $c_2=0$ and $|a|<1$ as well as adding a left end to the infinite chain. In the next section, edge states in such semi-infinite chains and their dependence on the boundary condition will be discussed.

\section{``Edge states'' in semi-infinite linear diatomic chains}
\label{sec:semi_infinite}

Now we consider a semi-infinite diatomic chain with a left end as follows
\begin{equation}
\begin{split}
  \ddot{q}_1= & k_1(q_2-q_1)+k_{3,1}(q_0-q_1), \\
  \ddot{q}_{2m} = & k_{1}(q_{2m-1}-q_{2m})+k_{2}(q_{2m+1}-q_{2m}), \\
  \ddot{q}_{2m+1} = & k_{2}(q_{2m}-q_{2m+1})+k_{1}(q_{2m+2}-q_{2m+1}), \quad m\geq 1
\end{split}
\label{eq:semi_infinite}
\end{equation}
where $q_0=0$ and $k_{3,1}$ represents the stiffness constant connecting the left end. We will show that the existence of left edge states is related to the value of $k_{3,1}$ but not in a sensitive way. Although this is just an ideal semi-infinite model, it demonstrates representative results that can benefit the later discussions on finite-size chains.

When $c_2=0$ and $|a|<1$, the eigenstate with relation \eqref{eq:u_cells} is left localized and $|u_j|$ decays exponentially such that $\sum_{j=1}^{+\infty}|u_j|^2<+\infty$. 
Substituting $(u_1,u_2)^T=c_1 v_1$ and $q_j=e^{i\omega t}u_j$ into the equation above and denoting the corresponding special $a$ for $c_2=0$ as $\tilde{a}$ yield
\begin{equation}
 \frac{u_{1}}{u_{2}}
  =\frac{ -\sigma\sqrt{k_1^2+k_2^2+k_1k_2(\tilde{a}+1/\tilde{a})} } 
  {k_1+k_2 \tilde{a}}
=\frac{ -\sigma {\rm sgn}(k_1+k_2 \tilde{a}) \sqrt{k_1+k_2/\tilde{a}} }{\sqrt{k_1+k_2 \tilde{a}}}
=\frac{k_1}{k_1+k_{3,1}-\omega^2}
\end{equation}
which leads to the equation of $\tilde{a}$
\begin{equation}
  k_{3,1}-k_{2}=\sigma{\rm sgn}(k_1+k_2 \tilde{a})\frac{ k_2\sqrt{k_1+k_2\tilde{a}}}{{\tilde{a}}\sqrt{k_1+k_2/\tilde{a}}}
\label{eq:5a}  
\end{equation}
or
\begin{equation}
\label{eq:5}  
 (k_{3,1}-k_{2})^{2}k_{1} \tilde{a}^{2}
  +[k_{2}(k_{3,1}-k_{2})^{2}
  -k_{2}^{3}] \tilde{a}-k_{1}k_{2}^{2}=0.
\end{equation}

\begin{remark}
\label{remark:root_a}
When $k_1<k_2$, the roots of equation~\eqref{eq:5} can be obtained as follows:
\begin{itemize}
\item If $|k_{3,1}-k_2|>k_2$, then equation~\eqref{eq:5} has roots $\tilde{a}_1$ and $\tilde{a}_2$ such that $0<\tilde{a}_1<1$ and $\tilde{a}_2<-1$;
\item If $0<|k_{3,1}-k_2|<k_2$, then the roots of equation~\eqref{eq:5} satisfy $0>\tilde{a}_1>-1$ and $\tilde{a}_2>1$;
\item If $k_{3,1}=k_2$, then the equaiton~\eqref{eq:5} becomes degenerate and the only root is $\tilde{a}=-\frac{k_1}{k_2}>-1$;
\item If $|k_{3,1}-k_2|=k_2$, then the roots of equation~\eqref{eq:5} are $\tilde{a}=\pm 1$. 
\end{itemize}
When $k_{3,1}\neq 2k_2$ and $k_{3,1}\neq 0$, there is always one root $|\tilde{a}|<1$ for~\eqref{eq:5} hence the semi-infinite chain~\eqref{eq:semi_infinite} bears one left edge state. To be specific, the eigenfrequency of the left edge state is within the bandgap ($\omega^2 \in (2k_1, 2k_2)$) when $0<k_{3,1} <2k_2$ and above the optical band ($\omega^2\in (2k_1+2k_2, +\infty)$) when $k_{3,1}>2k_2$.  
In particular, when $k_{3,1}=k_2$, there is a left edge state with eigenfrequency exactly at the middle of the bandgap ($\omega^2=k_1+k_2$), which corresponds to the zero-energy edge state in SSH model.
After solving $\tilde{a}$ from equation~\eqref{eq:5}, one can find the corresponding $\sigma$ ($\sigma=\pm 1$) from equation~\eqref{eq:5a} to obtain the eigenfrequency for the edge state.
\end{remark}
According to the bulk-boundary correspondence, the existence of left edge states in the chain~\eqref{eq:semi_infinite} depends on the bulk topology. To be specific, if we assume $k_1>k_2$ instead of $k_1<k_2$, then roots of equation~\eqref{eq:5} can be obtained as follows and the existence of left edge states can be identified.
\begin{remark}
\label{remark:root_a_2}
When $k_1>k_2$, the roots of equation~\eqref{eq:5} can be obtained as follows:
\begin{itemize}
\item If $|k_{3,1}-k_2|>k_2$, equation~\eqref{eq:5} has roots $-1<\tilde{a}_1<0$ and $0<\tilde{a}_2<1$;
\item If $0<|k_{3,1}-k_2|<k_2$, both roots of equation~\eqref{eq:5} satisfy $|\tilde{a}_j|>1$ ($j=1,2$);
\item If $k_{3,1}=k_2$, then the equaiton~\eqref{eq:5} becomes degenerate and the only root is $\tilde{a}=-\frac{k_1}{k_2}<-1$;
\item If $|k_{3,1}-k_2|=k_2$, then the roots of equation~\eqref{eq:5} are $\tilde{a}=\pm 1$. 
\end{itemize}
When $k_{3,1}> 2k_2$, equation~\eqref{eq:5} has roots $-1<\tilde{a}_1<0$ and $0<\tilde{a}_2<1$ hence there exist two left edge states. However, although the eigenfrequency of one left edge state is within the bandgap, it can never be exactly at the middle of the bandgap ($\tilde{a}_1 \in (-1, -\frac{k_2}{k_1}) $ and $\tilde{a}_1\neq -\frac{k_2}{k_1}$). 
\end{remark}

\begin{figure}[!htp]
\centering
\subfigure[]{
\includegraphics[height = 5.5cm, width = 6.5cm]{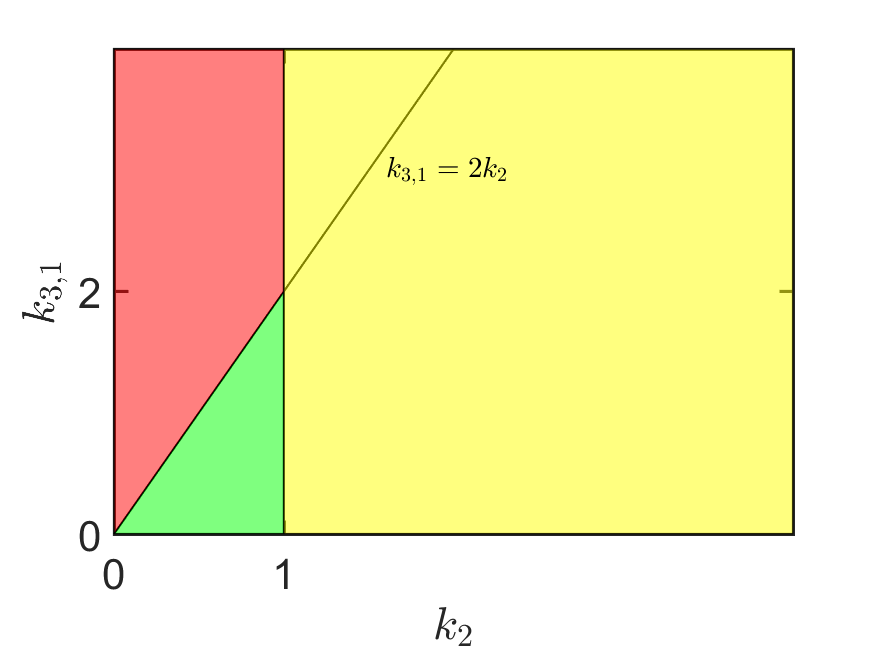}
}
\hspace{2mm}
\subfigure[]{
\includegraphics[height = 5.5cm, width = 6.5cm]{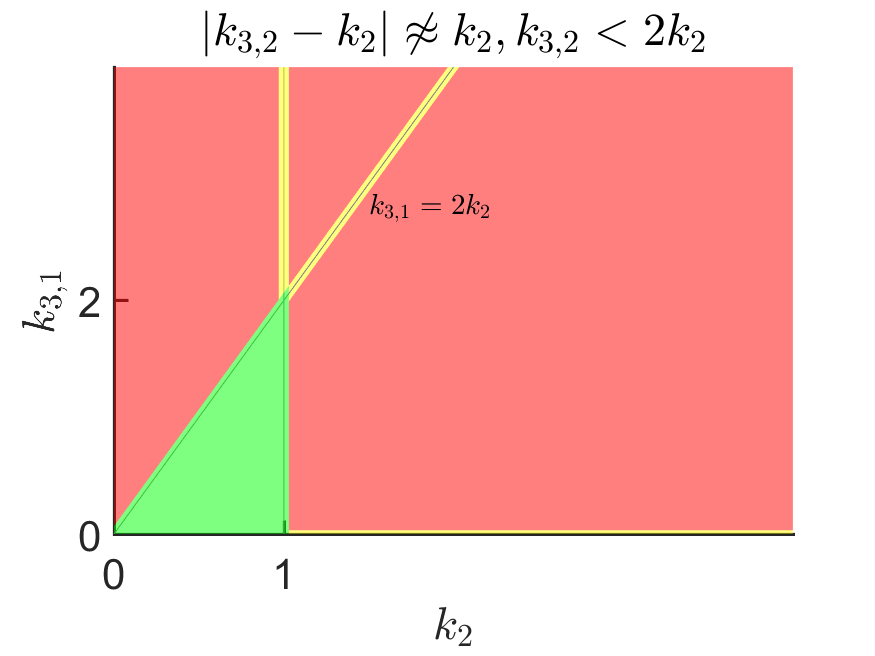}
}

\subfigure[]{
\includegraphics[height = 5.5cm, width = 6.5cm]{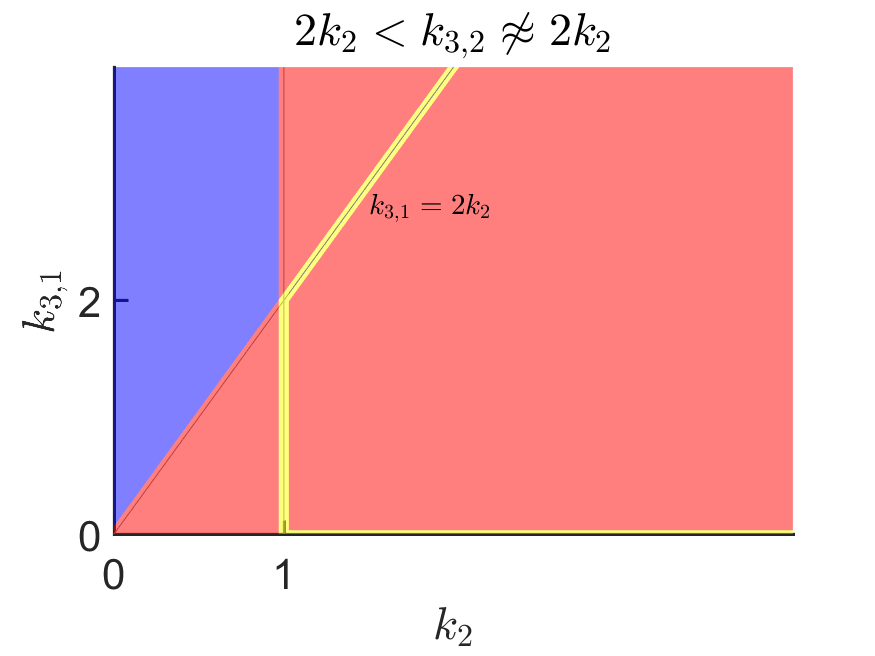}
}
\hspace{2mm}
\subfigure[]{
\includegraphics[height = 5.5cm, width = 6.5cm]{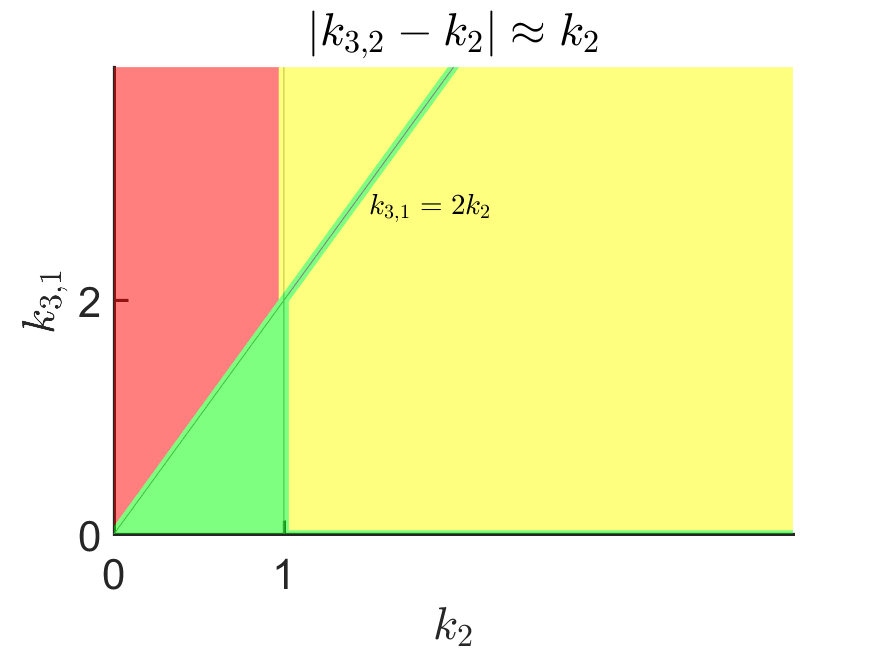}
}
\caption{Here we illustrate the dependence of the number of edge states on the bulk topology and boundary conditions in diatomic chains with $k_{1}=1$, where different colors represent different numbers of edge states. Specifically, green represents $0$, red represents $2$, yellow represents $1$ and blue represents $4$. The panel (a) represents the number of edge states in the semi-infinite chain \eqref{eq:semi_infinite}, while panels (b) to (d) correspond to finite chain \eqref{eq:finite}.
In panel (a), we show the number of edge states with varying $k_{3,1}$ and $k_{2}$ where the whole area is divided into three parts with two lines, namely $k_{3,1}=2k_{2}$ and $k_{2}=1$. The panels (b) to (d) are similarly plotted for varying $k_{3,1}$ and $k_2$ but with different assumptions on $k_{3,2}$.
}
\label{fig:num_edge_states}
\end{figure}

In order to understand the existence of edge states from a topological perspective, we can calculate the Zak phase which is a Berry phase for Bloch states in our diatomic chain. Since the Bloch states satisfy $(u_{j+2l}, u_{j+1+2l})^T=e^{i l\theta}(u_{j}, u_{j+1})^T$ where $\theta\in [-{\pi}, {\pi}]$, we define $\hat{v}_1(\theta)=\frac{v_1(a)}{|v_1(a)|}$ with $a=e^{i\theta}$ and compute the Zak phase as 
\begin{equation}
\label{Zak}
\begin{split}
\gamma&=i\int_{-{\pi}}^{{\pi}} (\hat{v}_1, \frac{\partial \hat{v}_1}{\partial \theta}) d\theta \\
&=-\int_{-{\pi}}^{{\pi}}  \frac{k_2(k_2+k_1\cos\theta)}{2(k_1+k_2+2k_1 k_2 \cos\theta)} d\theta \\
&=i\oint_{|a|=1} k_2\frac{ 2k_2 a+k_1 +k_1 a^2 }{ 4(k_1 a+k_2) (k_1+k_2 a) a} da \\
&=\begin{cases}
\pi~({\rm mod}~2\pi), & {\rm if}~k_1<k_2, \\
0~({\rm mod}~2\pi), & {\rm if}~k_1>k_2.
\end{cases}
\end{split}
\end{equation}
Similar to the situation in SSH model, the Zak phase of our diatomic chain is nontrivial when $k_1<k_2$. According to Remark~\ref{remark:root_a} and Remark~\ref{remark:root_a_2}, the existence of left edge states depends on both ${\rm sgn}(k_1-k_2)$ and ${\rm sgn}(k_{3,1}-2k_2)$, which can be illustrated in the panel (a) of Fig.~\ref{fig:num_edge_states}. To be more explicit, the number of left edge states is even (zero or two) only when $k_1>k_2$ and particularly zero when ${\rm sgn}(k_1-k_2)=1={\rm sgn}(2k_2-k_{3,1})$. That is to say, the bulk-boundary correspondence in ideal semi-infinite chains is similar to that in the SSH model. 
In the next section, we will advance to realistic finite chains to discuss edge states and their bulk-boundary correspondence.

\section{``Edge states'' in long linear diatomic chains}
\label{sec:finite_edgestates}

\subsection{Definition of edge states in finite chains}
\label{subsec:def_edgestates}

Suppose now we consider a diatomic chain of length $2n$ with equations of motion
\begin{equation}
\begin{split}
  \ddot{q}_1= & k_1(q_2-q_1)+k_{3,1}(q_0-q_1), \\
  \ddot{q}_{2m} = & k_{1}(q_{2m-1}-q_{2m})+k_{2}(q_{2m+1}-q_{2m}), \\
  \ddot{q}_{2m+1} = & k_{2}(q_{2m}-q_{2m+1})+k_{1}(q_{2m+2}-q_{2m+1}), \quad 1\leq m\leq n-1 \\
  \ddot{q}_{2n} = & k_{1}(q_{2n-1}-q_{2n})+k_{3,2}(q_{2n+1}-q_{2n})
\end{split}
\label{eq:finite}
\end{equation}
where $q_0=q_{2n+1}=0$ represent two fixed ends and $\{ k_{3,1}, k_{3,2} \}$ are the stiffness constants for the ``springs'' connecting the ends. Here we consider finite chains with Dirichlet boundary conditions for simplicity, while our following analysis for finite chains can also be applied to other types of boundary conditions. Again focusing on time-periodic solutions of equations~\eqref{eq:finite}, we assume $q_j=e^{i\omega t} u_j$ and obtain the matrix form of the system as
\begin{equation}
  -\omega^2 u=\mathcal{L}{u}
  \label{eq:motion1}
\end{equation}
where
$\mathcal{L}=\left(\begin{array}{cccccc}
  -k_1-k_{3,1} & k_1 & &   & & \\
  k_{1} & -k_1-k_2 & k_2  & &  &\\
  & k_2 & -k_1-k_2 &  k_1  & & \\
  & &  \cdots & \cdots & \cdots  &\\
  & & & k_2 & -k_1-k_2 & k_1 \\
  & & & & k_1 & -k_1-k_{3,2}
 \end{array}\right)$ and ${u}=(u_{1},u_{2},
\cdots,u_{2n})^{\top}$. 
\begin{lemma}
\label{lm:L}
$\mathcal{L}$ is a diagonally-dominant and symmetric tridiagonal matrix. It has $2n$ distinct nonpositive (negative if $k_{3,1}^2+k_{3,2}^2\neq 0$) eigenvalues $-(\omega^{(1)})^2, -(\omega^{(2)})^2, \cdots, -(\omega^{(2n)})^2$.
\end{lemma}
\begin{corollary}
    In the chain \eqref{eq:finite}, if two eigenfrequencies satisfy $\omega^{(j)}<\omega^{(k)}$ for some $(k_{3,1}, k_{3,2})$, then $\omega^{(j)}<\omega^{(k)}$ for all $(k_{3,1}, k_{3,2})$.   
\end{corollary}
\begin{lemma}
\label{lm:increasing}
(Proof in Appendix \ref{proof:lemma_increasing_freq})
Every eigenvalue of $\mathcal{L}$ decreases as $k_{3,1}$ (or $k_{3,2}$) increases. In other words, every eigenfrequency $\omega$ satisfies $\frac{\partial \omega}{\partial k_{3,1}}> 0 < \frac{\partial \omega}{\partial k_{3,2}}$.
\end{lemma}
\begin{lemma}
\label{lm:symmetry}
If $k_{3,1}=k_{3,2}$, then the chain \eqref{eq:finite} is symmetric about the middle. As a result, if $\vec{u}=(u_1,u_2,\cdots,u_{2n})^T$ is an eigenvector of $\mathcal{L}$ for eigenvalue $-\omega^2$, then $\vec{u'}=(u_{2n},u_{2n-1},\cdots,u_{1})^T$ also satisfies $-\omega^2 \vec{u'}=\mathcal{L}\vec{u'}$. Since the eigenvalues of $\mathcal{L}$ are distinct, there exists some $c\neq 0$ such that $\vec{u}=c\vec{u'}$.
\end{lemma}
It can be immediately observed that the iteration relation \eqref{eq:iteration} between adjacent unit cells holds regardless of the finiteness of the lattice. Following the same notations in \eqref{eq:u_cells}, \eqref{eq:v1v2_0}, we suppose $(u_1, u_2)^T=c_1 v_1+c_2 v_2$ and $a\neq\pm 1$, then $(u_{2m+1}, u_{2m+2})^T=c_1 a^m v_1+c_2 a^{-m} v_2$ and system \eqref{eq:finite} can be rewritten as 
\begin{eqnarray}
\label{eq:bc_1}
  \frac{u_{1}}{u_{2}}&
  =&\frac{c_{1}v_{1,1}+c_{2}v_{2,1}}
  { c_{1}v_{1,2}+ c_{2}v_{2,2}}=\frac{k_{1}}
  {k_{1}+k_{3,1}-\omega^{2}},
  \\
  \label{eq:bc_2}
  \frac{u_{2n-1}}{u_{2n}}&
  =&\frac{c_{1}v_{1,1}a^{n-1}+c_{2}v_{2,1}a^{1-n}}
  { c_{1}v_{1,2}a^{n-1}
  + c_{2}v_{2,2}a^{1-n}}
  =\frac{k_{1}+k_{3,2}-\omega^{2}}{k_{1}}
\end{eqnarray}
where $v_{j,k}$ denotes the $k$-th component of $v_j$ in \eqref{eq:v1v2_0}. 
\begin{remark}
\label{rmk:left_edge_state} ({\bf Genuine edge states})
We notice that eigenstates satisfying $|a|<1$ and $c_2=0$ are left-localized and they will exponentially decay to zero if the chain can continue infinitely in the right direction. In what follows, we will denote such edge states ($c_1=0$ or $c_2=0$) in finite chains as ``genuine edge states" since they are also edge states in semi-infinite chains. 
\end{remark}

\begin{figure}[!htp]
\centering
\subfigure[]{
\includegraphics[width=6.5cm, height=5.5cm]{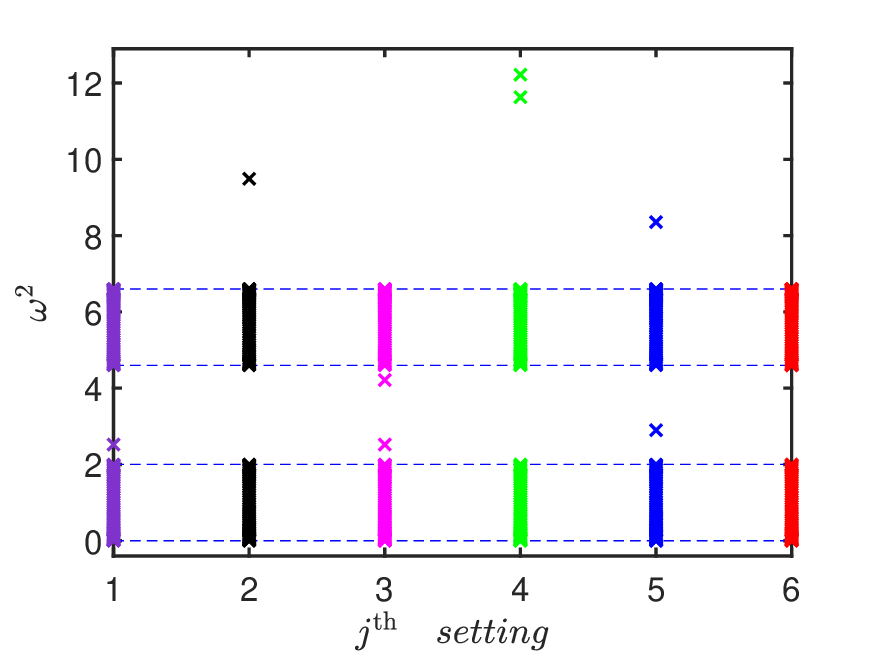}
}
\hspace{2mm}
\subfigure[]{
\includegraphics[height = 5.5cm, width = 6.5cm]{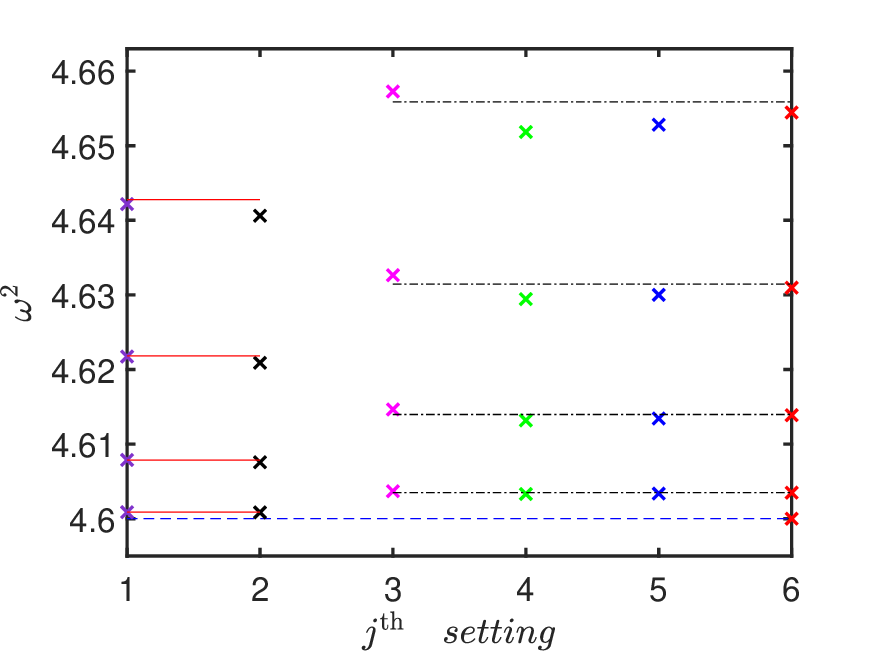}
}
\caption{
 Here panel (a) shows the eigenfrequencies of a linear diatomic chain \eqref{eq:finite} with $k_1=1$, $k_2=2.3$ and $2n=100$ under different boundary stiffness settings. The crosses in each column of panel (a) correspond to the eigenfrequencies under a setting with a different pair of $(k_{3,1}, k_{3,2})$, ranging among $\{ (1.3, 4.6), (8.3, 4.6), (1.3, 3.5), (11.1, 10.5), (7.1, 1.8), (4.6, 4.6) \}$, from left to right. It can be observed that no more than two eigenfrequencies are outside the bands where blue dashed lines represent the band edges. Panel (b) just zooms in panel (a) on the region near the lower edge of the optical band where red solid and black dot-dash auxiliary lines are added to better demonstrate the two patterns of the eigenfrequencies. We find that the near-band-edge eigenfrequencies in the first two columns are almost the same while those in the last four columns are very close. 
}
\label{fig:eigenfrequencies}
\end{figure}

According to our discussion from the previous section, the value of $a$ for left edge states with $c_2=0$ (namely $\tilde{a}$) can be solved from \eqref{eq:5a} as a function of $k_{3,1}$, which implies that admissible $k_{3,2}$ in \eqref{eq:bc_2} for ``genuine left edge states'' (this $k_{3,2}$ is denoted by $\tilde{k}_{3,2}$) can also be determined as a function of $k_{3,1}$:
\beq
\label{eq:tilde_k32}
\tilde{k}_{3,2}=\omega^2(\tilde{a})-k_1+k_1\frac{v_{1,1}(\tilde{a})}{v_{1,2}(\tilde{a})}=\tilde{k}_{3,2}(\tilde{a})=\tilde{k}_{3,2}(k_{3,1}).
\eeq
With that being said, it seems that finite diatomic chains \eqref{eq:finite} with generic choices of $\{k_{3,1}, k_{3,2}\}$ usually don't bear ``genuine edge states''.  However, in a long ($n$ being large) linear diatomic chain (with $k_1<k_2$), numerical experiments suggest the following typical scenario:
\begin{itemize}
\item (L1)
Most of the eigenfrequencies $\omega^{(j)}$ belong to the two bands $[0,\sqrt{2k_1}]\cup[\sqrt{2 k_2}, \sqrt{2k_1+2k_2}]$ (usually known as ``acoustic band'' and ``optical band'') and their corresponding eigenstates $u^{(j)}$ are non-localized;
\item (L2)
No more than two eigenfrequencies $\omega^{(j)}$ fall outside the bands and their eigenstates $u^{(j)}$ are usually localized; 
\item (L3)
Eigenfrequencies inside the bands and near the band edges follow special patterns.
\end{itemize}
Examples of eigenfrequencies in different finite chains are plotted in panel (a) of Fig.~\ref{fig:eigenfrequencies} which demonstrate (L1) and (L2) (theoretical justifications will be provided later in Theorem~\ref{theorem:edgestates} and Corollary~\ref{corollary:number_of_edgestates}). Besides, as an illustration of (L3), panel (b) of Fig.~\ref{fig:eigenfrequencies} suggests that the eigenfrequencies inside the optical band and near its lower edge ($2k_2>\omega^2\approx 2k_2$) mainly have two patterns, which will be later explained in Lemma~\ref{lm:op_eigenfrequencies} and Lemma~\ref{lm:ac_eigenfrequencies}.
For frequencies outside the bands, although $c_1$ and $c_2$ in the corresponding eigenstates generically do not vanish, we will show that they commonly lead to `` edge states'' for naked eyes with a quite wide range of choices for boundary stiffness constants $k_{3,1}$ and $k_{3,2}$.

\begin{figure}[!htp]
\centering
\subfigure[]{
\includegraphics[height = 5.5cm, width = 6.5cm]{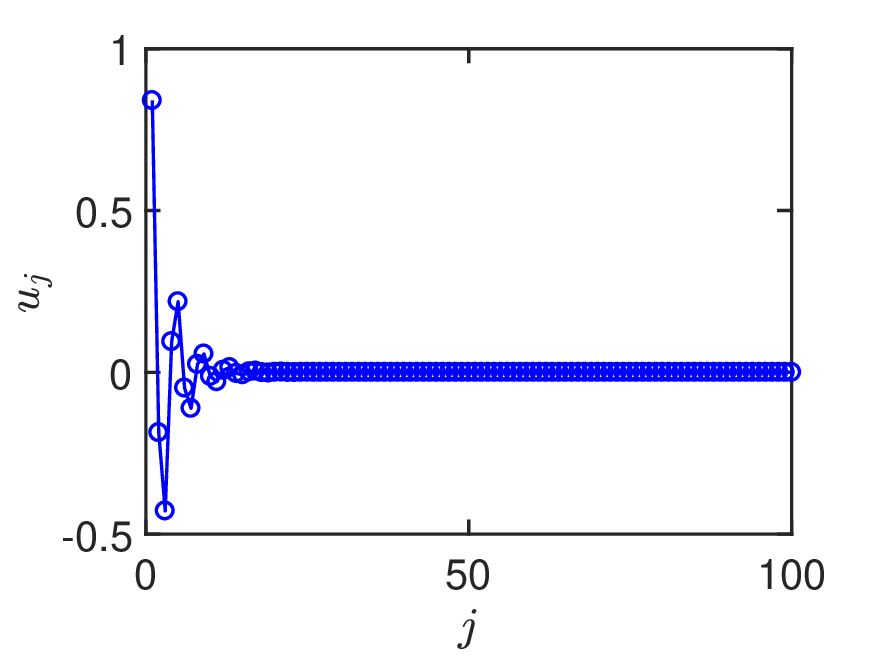}
}
\hspace{2mm}
\subfigure[]{
\includegraphics[height = 5.5cm, width = 6.5cm]{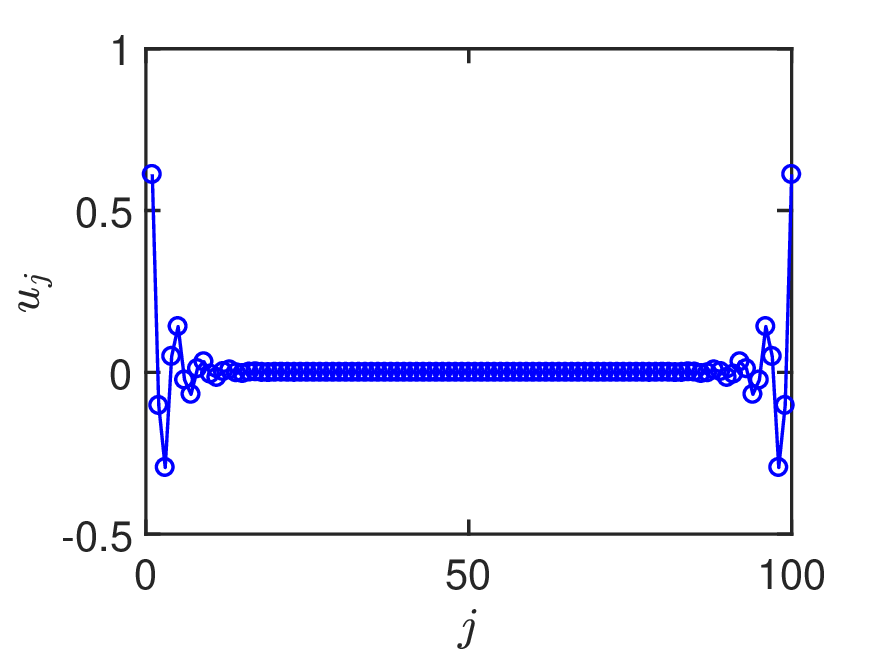}
}

\subfigure[]{
\includegraphics[height = 5.5cm, width = 6.5cm]{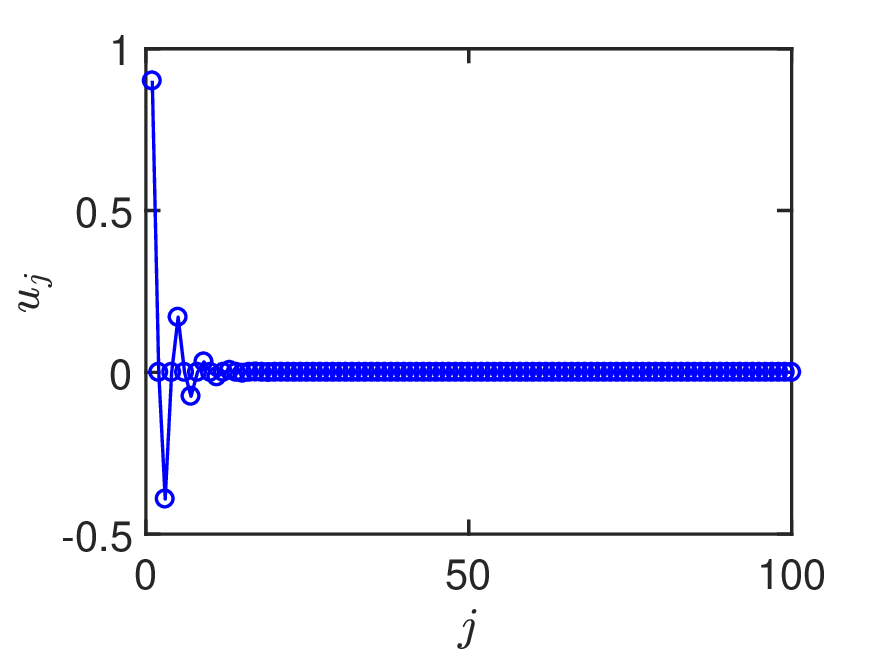}
}
\hspace{2mm}
\subfigure[]{
\includegraphics[height = 5.5cm, width = 6.5cm]{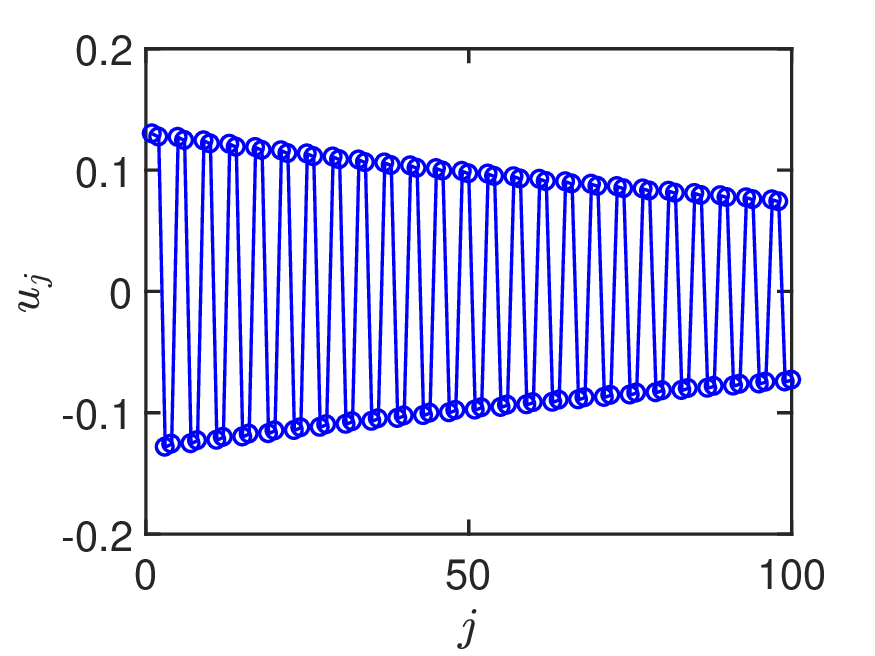}
}

\subfigure[]{
\includegraphics[height = 5.5cm, width = 6.5cm]{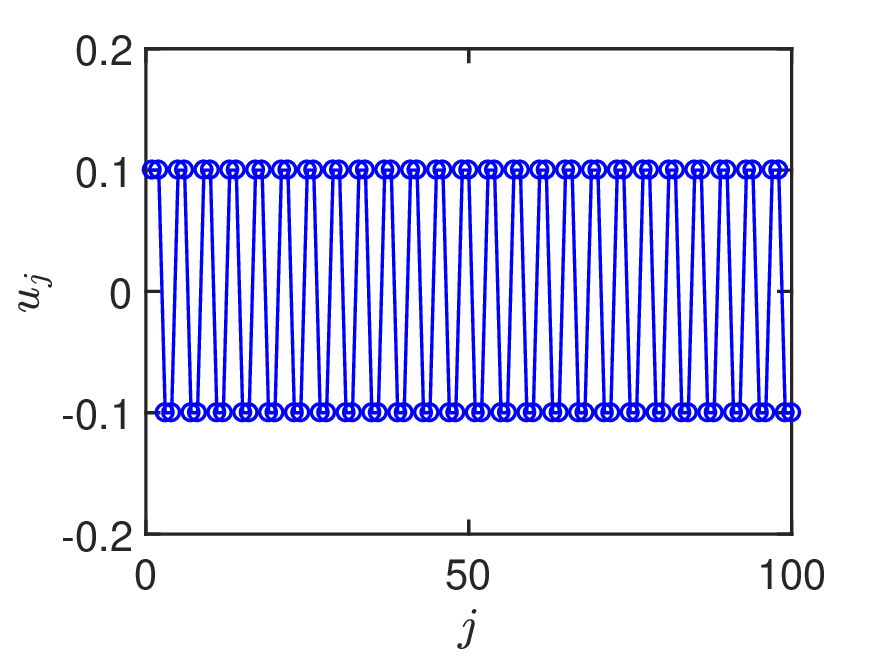}
}
\hspace{2mm}
\subfigure[]{
\includegraphics[height = 5.5cm, width = 6.5cm]{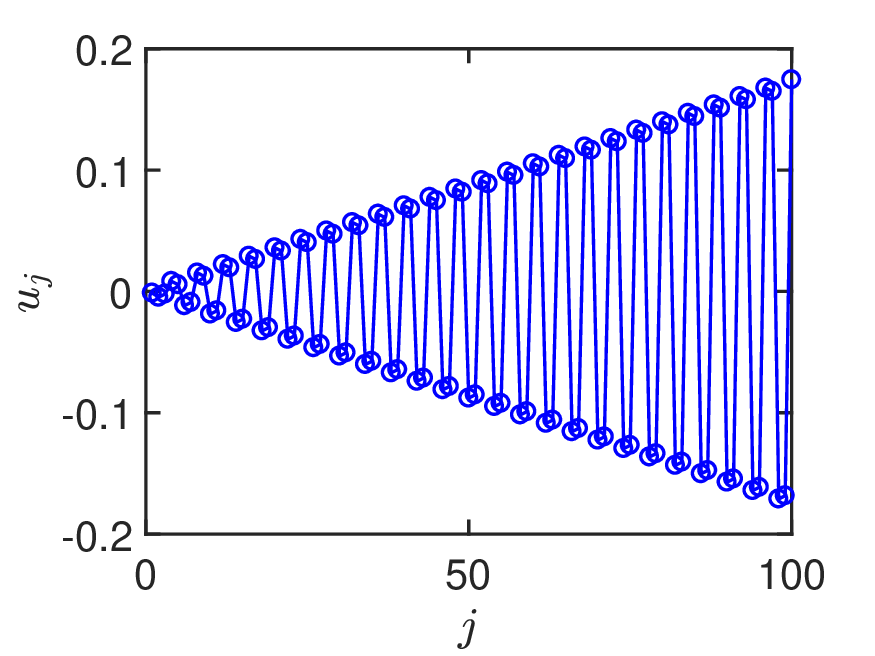}
}
\caption{Here we plot some representative eigenstates in a linear diatomic chain~\eqref{eq:finite} with $2n=100$, $k_{1}=1$ and $k_{2}=2.3$.  The panel (a) shows a left edge state for the
generic case with $k_{3,1}=1.3$ and $k_{3,2}=3.5$ while the panel (b) shows an eigenstate localized at both ends when $k_{3,1}=k_{3,2}=1.5$. In the panel (c), we plot a left edge state with $k_{3,1}=k_{2}$ and $k_{3,2}=1.2$. The panel (d) illustrates an eigenstate decays ``slowly'' with $a\approx-0.9910345\approx-1$ when $4.58=k_{3,1}\approx 2k_{2}\approx k_{3,2}=4.62$. The panel (e) in the lower left (panel (f) in the lower right) shows an eigenstate with $\omega^{2}=2k_{2}$ and $k_{3,1}=2k_{2}=k_{3,2}$ ($\omega^{2}=2k_{1}$,  $k_{3,1}=2k_{2}$ and $k_{3,2}=\frac{2k_{1}k_{2}}{(2n-1)(k_{2}-k_{1})+k_{2}}$).  }
\label{fig:edge state}
\end{figure}
\begin{remark}
\label{rmk:order} ({\bf Asymptotic notations})
In a chain of length $2n$, the possible eigenfrequencies and eigenstates (as well as quantities $a$, $c_1$, $c_2$ and so on) are functions of $n$. When $n$ is a large number, we can estimate the order of these relevant quantities in terms of $n$. Some notations  in this work regarding the orders are as follows:
\begin{itemize}
\item $f(n)\approx g(n)$ means $f(n)$ and $g(n)$ are the same at the leading order, namely $f(n)=g(n)=0$ or $\lim\limits_{n\to+\infty}\frac{f(n)-g(n)}{f(n)}=0$.
\item $f(n)=\Theta(g(n))$ or $f(n)\sim g(n)$ means $|f(n)|$ and $|g(n)|$ are at the same order, namely there exist positive constants $c_1$ and $c_2$ such that $c_1 |g(n)| \leq |f(n)|\leq c_2 |g(n)|$ for $n$ large enough.
\item $|f(n)| \ll |g(n)|$ means the order of $|f(n)|$ is much smaller than that of $|g(n)|$, namely $\lim\limits_{n\to+\infty}\frac{f(n)}{g(n)}=0$.
\item $|f(n)| \leq \Theta(|g(n)|)$ or $|f(n)|=\mathcal{O}(|g(n)|)$ means the order of $|f(n)|$ is no more than that of $|g(n)|$, namely there exists some positive $c$ and such that $|f(n)|\leq c |g(n)|$ for ${n}$ large enough.
\item $|f(n)| \geq \Theta(|g(n)|)$ or $|f(n)|=\Omega(|g(n)|)$ means the order of $|f(n)|$ is no less than that of $|g(n)|$.
\end{itemize}
\end{remark}
When the left end of an eigenstate $u$ follows $(u_1, u_2)^T=c_1 v_1 +c_2 v_2$, its right end (when $a\neq \pm 1$) can be described as 
\begin{equation}
\label{eq:right_end_eigenstate}
(u_{2n-1},u_{2n})^{T}=c_{1}a^{n-1}\vec{v}_{1}+c_{2}a^{1-n}\vec{v}_{2}=c_1 a^{n-1}(
 \vec{v}_1+\frac{c_2}{c_1} a^{2-2n}\vec{v}_2).
\end{equation}
\begin{remark}
\label{rmk:edge states} ({\bf Left edge states in long chains})
If the eigenstates with \eqref{eq:u_cells} satisfy 
\beq
\label{eq:left_edge_state_condition}
|a|\not\approx 1, \quad |\frac{c_2}{c_1}a^{2-2n}|=\mathcal{O}(1),
\eeq
then we denote them by ``{\bf left edge states}'' in finite chains~\eqref{eq:finite}. The definition of ``{\bf right edge states}'' in long finite chains can be similarly obtained by symmetry. 
\begin{itemize}
\item The first condition $|a|\not\approx 1$ guarantees the edge states decay fast enough spatially. It should be noticed that states with $c_2=0$ and $|a|<1$ in semi-infinite chains must be left edge states but $1\approx |a|<1$ in finite chains could lead to states that are not left-localized enough (see panel (d) of Fig.~\ref{fig:edge state} for an example). 
\item The second condition $|\frac{c_2}{c_1}a^{2-2n}|=\mathcal{O}(1)$ means that $c_1 a^{m-1} \vec{v}_1$ plays a dominant role in $(u_{2m-1}, u_{2m})^T$ for forming the shape of $u$, which restricts us to left localized states and rules out the possibility of two-sided localized states. Since $c_1$ can be chosen as $1$ without loss of generality and $|a^{2-2n}|\sim |a^{-2n}|$, this condition can usually be simplified as $|c_2 a^{-2n}|=\mathcal{O}(1)$. At the same time, we will see that this condition coincides with the expected radius of convergence in later perturbative analysis (Remark~\ref{rmk:IFT_generic}).

\end{itemize}
%
\end{remark}
\begin{remark}
\label{rmk:def two-sided edge states}
    In this work, the eigenstates in long chains~\eqref{eq:finite} satisfying $|a|\not\approx 1$ are denoted by ``{\bf edge states}''. In particular, we define {\bf two-sided edge states} as eigenstates that are edge states but are neither left nor right edge states. Although our main focus in this work is on one-sided edge states, we also plot an example of two-sided localized state in panel (b) of Fig.~\ref{fig:edge state} and provide a short discussion of this type of edge states in Remark~\ref{rmk:two-sided}. 
\end{remark}

To specifically study the left edge states, we assume without loss of generality $c_1=1$ in \eqref{eq:bc_1} and \eqref{eq:bc_2} and investigate the dependence of $(a, c_2)$ on $(k_{3,1}, k_{3,2})$ through a perturbative approach. Our strategy starts with noticing that $(\tilde{a}, \tilde{k}_{3,2})$ can be solved from \eqref{eq:bc_1} and \eqref{eq:bc_2} for given $(k_{3,1}, c_2)=(k_{3,1}, 0)$, which is essentially finding ``genuine left edge states'' from \eqref{eq:5a}. Since $|c_2|$ for ``left edge states'' in long chains should be small, we view $(a, c_2)$ as perturbed values from $(\tilde{a}, 0)$ for generic $(k_{3,1}, k_{3,2})$ where $k_{3,2}$ is deviated from the special value $\tilde{k}_{3,2}$. According to the Implicit Function Theorem, $(a, c_2)$ for a left edge state in \eqref{eq:bc_1} and \eqref{eq:bc_2} are functions of $(k_{3,1}, k_{3,2})$ near $(a, c_2)(k_{3,1}, \tilde{k}_{3,2})=(\tilde{a}, 0)$. By Remark~\ref{remark:root_a}, equation \eqref{eq:5} becomes degenerate if $k_{3,1}=k_{2}$, while $|k_{3,1}-k_2|=k_2$ leads to another special case with $\tilde{a}=\pm 1$. In the generic case with $(k_{3,1}-k_2)[(k_{3,1}-k_2)^2-k_2^2]\neq 0$, equation \eqref{eq:5} always has a root $|\tilde{a}_1|<1$. With that being said, we can discuss the ``left edge states'' in three cases based on whether $k_{3,1}\approx k_{2}$ or $|k_{3,1}-k_2|\approx k_2$. Particularly we analyze the generic case $\{ k_{3,1}\not\approx k_2, |k_{3,1}-k_2|\not\approx k_2 \}$ in the following and briefly discuss the other two special cases (with more detail in the Appendices).
The results for all three cases are summarized in Theorem~\ref{theorem:edgestates} and Corollary~\ref{corollary:number_of_edgestates} towards the end of this section. 

\subsection{\texorpdfstring{Generic case: One-sided and two-sided edge states ($k_{3,1}\not\approx k_{2}$ and $|k_{3,1}-k_{2}|\not\approx k_2$)}{Generic case: One-sided and two-sided edge states (k3,1 not approx k2 and |k3,1-k2|not approx k2)}}

\label{subsec:generic}

\begin{figure}[!htp]
\centering
\subfigure[]{
\includegraphics[width=6.5cm, height=5.5cm]{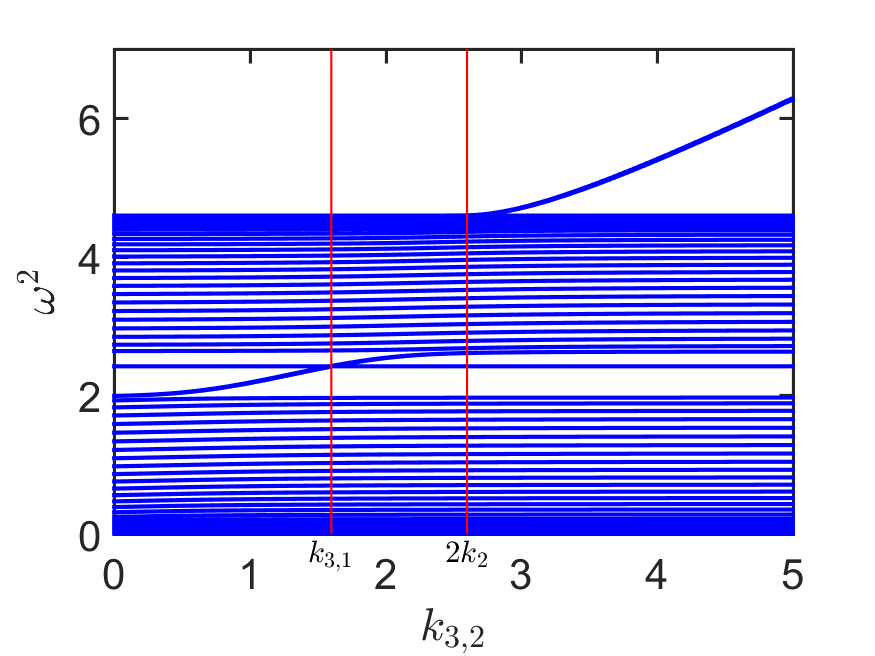}
}
\hspace{2mm}
\subfigure[]{
\includegraphics[height = 5.5cm, width = 6.5cm]{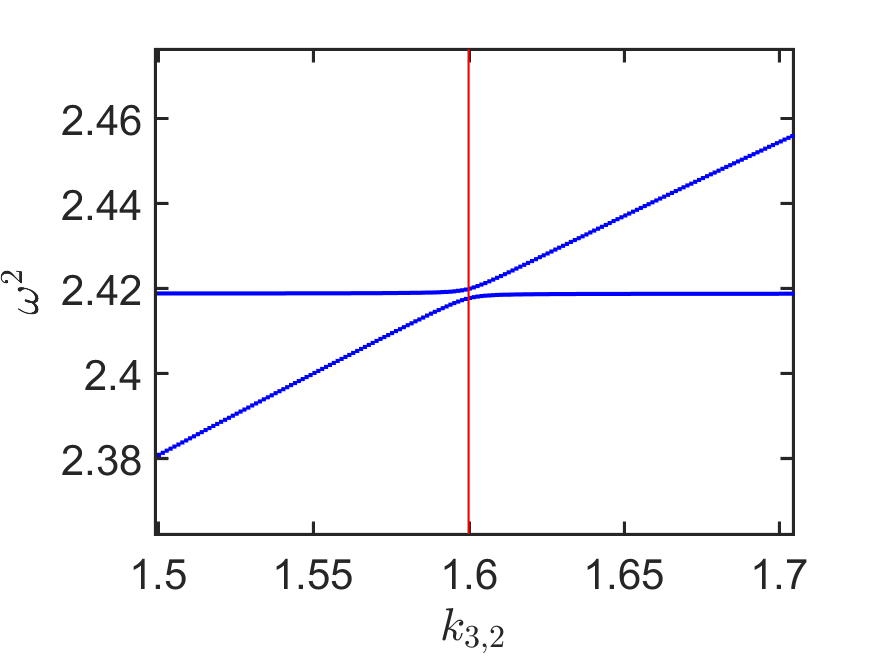}
}
\caption{
Here panel (a) shows the eigenfrequencies in the linear diatomic chain \eqref{eq:finite} with $\{ 2n=50, k_{1}=1, k_{2}=1.3, k_{3,1}=1.6 \}$ over varying stiffness constant $k_{3,2}$. Panel (b) zooms in on the region near $k_{3,1}=k_{3,2}$ in panel (a).
}
\label{fig:omega_k32}
\end{figure}

When $k_{3,1}$ is away from the special values $\{0, k_2, 2k_2 \}$, Remark~\ref{remark:root_a} states that equation \eqref{eq:5} always has a root $\tilde{a}$ which is away from $\pm 1$ and $-\frac{k_1}{k_2}$. 
When plotting the change of eigenfrequencies in the diatomic chain \eqref{eq:finite} with varying $k_{3,2}$ in Fig.~\ref{fig:omega_k32}, immediately we notice that two eigenfrequencies are usually outside the two bands. One of these two frequencies is always close to $\omega(\tilde{a})$ while the other frequency is highly dependent on $k_{3,2}$. It can be easily checked that these two eigenfrequencies correspond to a left edge state and a right edge state respectively, demonstrating the importance of $k_{3,1}$ and $k_{3,2}$ in determining one-sided edge states. 

Since the system \eqref{eq:bc_1} and \eqref{eq:bc_2} has a solution (genuine left edge state) $(a,c_2)=(\tilde{a}, 0)$ when $(k_{3,1}, k_{3,2})=(k_{3,1}, \tilde{k}_{3,2})$, we perturb $\tilde{a}$ as $a=\tilde{a}+\Delta a$ for generic $k_{3,2}\neq \tilde{k}_{3,2}$. According to \eqref{eq:v1v2_0}, we can choose $v_{1}$ and $v_{2}$ as
\begin{equation}
\begin{split}
  v_{1,1}(a) &= -{\rm sgn}(a)\sigma v_{2,2}(a)=\sqrt{k_{1}+k_{2}/{a}}=\sqrt{k_{1}+k_{2}/\tilde{a}}
  +\Theta(\Delta a), \\
  v_{2,1}(a) &= -{\rm sgn}(a) \sigma v_{1,2}(a)=\sqrt{k_{1}+k_{2}{a}}=\sqrt{k_{1}+k_{2}\tilde{a}}
  +\Theta(\Delta a).
  \label{eq:v1_v2_generic}
\end{split}
\end{equation}
By the Implicit Function Theorem (IFT), there exists a branch of solutions $(a, c_2)$ of \eqref{eq:bc_1} and \eqref{eq:bc_2} near $(k_{3,1}, k_{3,2})=(k_{3,1}, \tilde{k}_{3,2})$. In fact, it can be explicitly obtained from \eqref{eq:bc_1} that $c_2$ can be written as a function of $a$ and $k_{3,1}$.
\begin{equation}
\label{eq:c2}
    c_{2}(a) = \frac{ {\rm sgn}(a) (k_{3,1}-k_2)\sqrt{k_1+k_2/a}-\sigma k_2/a\sqrt{k_1+k_2 a}}{\sigma k_2 a\sqrt{k_1+k_2/a}-{\rm sgn}(a)(k_{3,1}-k_2)\sqrt{k_1+k_2 a}}
    \approx c_2'(\tilde{a})(\Delta a)=\Theta(\Delta a).
\end{equation}
Substituting this into \eqref{eq:bc_2}, we can express $a=a(k_{3,2})$ (with $k_{3,1}$ being fixed) or the other way around.
\begin{equation}
    k_{3,2}(a) =
    \frac{k_{1}(v_{1,1}+c_{2}a^{2-2n}v_{2,1})}
    { v_{1,2}+ c_{2}a^{2-2n}v_{2,2}}
    +\omega^{2}-k_{1}.
\label{eq:k32_of_a}    
\end{equation}

\begin{figure}[!htp]
\centering
\subfigure[]{
\includegraphics[height=5.5cm,width=6.5cm]{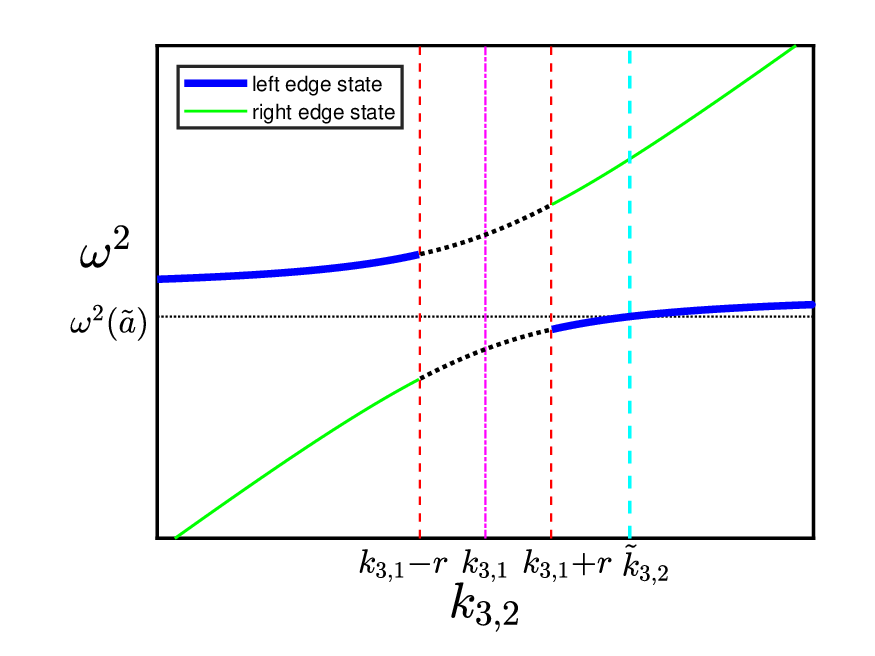}
}
\hspace{2mm}
\subfigure[]{
\includegraphics[height = 5.5cm, width = 6.5cm]{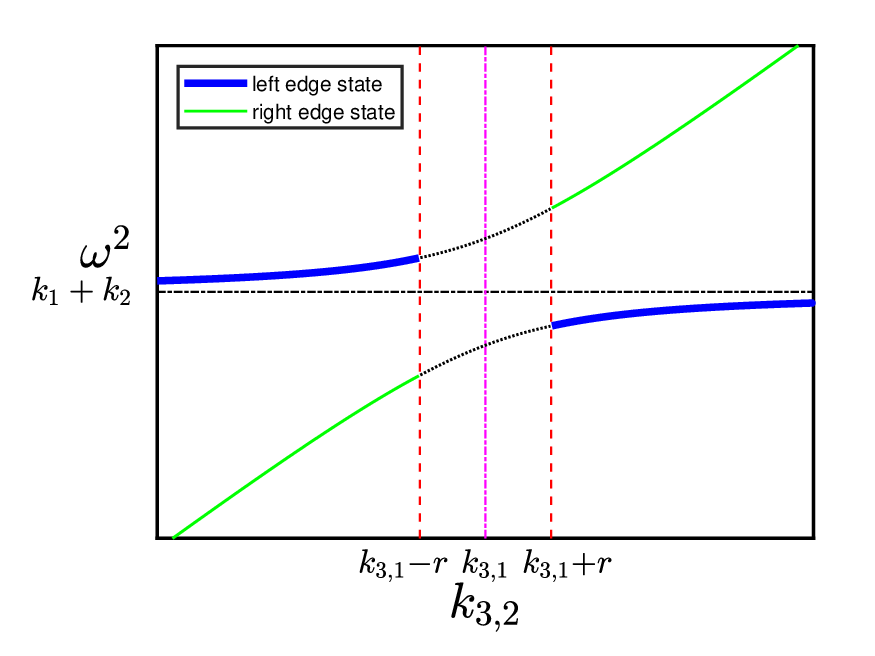}
}
\caption{Here we plot the schematic diagram to demonstrate the correspondence between $k_{3,2}$ and $\omega^{2}$ as $a$ approaches $\tilde{a}$ for \eqref{eq:finite}, where panel (a) is for generic case ($2k_{2}\not\approx k_{3,1}\not\approx k_{2}$) and panel (b) is for special case ($k_{3,1}=k_{2}$). The horizontal black dashed line represents $\omega^{2}(\tilde{a})$ corresponding to $a=\tilde{a}$, while the green solid curves and the blue bold solid curves illustrate $\omega^{2}$ corresponding to variations in $k_{3,2}$. The bold blue curves represent $\omega^{2}$ of left edge states, while the green curves correspond to right edge states. The two red vertical dashed lines indicate the regions where the edge states are more ``two-sided'' with $k_{3,2}\approx k_{3,1}$. The cyan vertical dashed line in panel (a) represents the special value $k_{3,2}=\tilde{k}_{3,2}$ such that $\omega^2=\omega^2(\tilde{a})$.
}
\label{schematic diagram:omega_k32}
\end{figure}

\begin{remark}
\label{rmk:IFT_generic}  ({\bf IFT}) 
\begin{itemize}
\item By directly checking the derivatives of  \eqref{eq:bc_1} and \eqref{eq:bc_2} in $(a, c_2)$, it can inferred that the Implicit Function Theorem holds for $(a, c_2)$ near $(k_{3,1},k_{3,2})=(k_{3,1}, \tilde{k}_{3,2})$. In addition, \eqref{eq:k32_of_a} implies that one can also easily write $k_{3,2}=k_{3,2}(a)$. 
\item To understand the range for the IFT to hold, we can calculate the radius of convergence $r$ for the series expansion in the form of $a(k_{3,2})=a(\tilde{k}_{3,2})+\sum_{i=1}^{+\infty}a_i (\Delta k_{3,2})^i$ or $k_{3,2}(a)=k_{3,2}(\tilde{a})+\sum_{i=1}^{+\infty}k_{3,2, i}(\Delta a)^i$. Although this is nontrivial in general, calculations of the first several derivatives $\frac{d^j k_{3,2}}{d a^j}|_{a=\tilde{a}}$ lead to the expectation that the radius of convergence for $\Delta a$ is at the order of $r=\mathcal{O}(|\tilde{a}|^{2n})$. Since $c_2=\Theta(\Delta a)$ by \eqref{eq:c2} and $|a|^{2n}=|\tilde{a}+\Delta a|^{2n} \approx |\tilde{a}|^{2n}$ for $0\not\approx a\not\approx 1$ and $|\Delta a|=\mathcal{O}(|\tilde{a}|^{2n})\ll \frac{1}{n}$, this implies the reasonableness of the condition $|c_2 a^{-2n}|=\mathcal{O}(1)$ in our definition of left edge states in Remark~\ref{rmk:edge states}.
\end{itemize}
\end{remark}
Since the order of $k_{3,2}$ in \eqref{eq:k32_of_a} depends on the relation between $a^{2n}$ and $\Delta a$, we consider the following three cases.
\begin{itemize}
    \item Case $|a^{2n}|\gg|\Delta a|$:
    
    In this case $|\Delta a|$ is so small that $|c_2 a^{2-2n}|\approx |c_2 \tilde{a}^{2-2n}|\ll 1$. Therefore \eqref{eq:k32_of_a} now implies
    \begin{equation}
        k_{3,2}\approx k_2+\frac{k_2 \tilde{a}\sqrt{k_1+k_2/\tilde{a}}}{\sigma {\rm sgn}(a) \sqrt{k_1+k_2 \tilde{a}}}=\frac{k_{3,1}k_2}{k_{3,1}-k_2}=\tilde{k}_{3,2}
    \end{equation}
    where equation~\eqref{eq:5a} gives the last step. In other words, an eigenstate 
 ``very localized'' at the left edge exists when $k_{3,2}$ is near the special value $\tilde{k}_{3,2}=\frac{k_{3,1}k_2}{k_{3,1}-k_2}$.

 \item Case $|a^{2n}|=\Theta(|\Delta a|)$: 
    
    When $|c_2 a^{2-2n}|=\Theta(1)$ or $\Delta a=\Theta(a^{2n})=\Theta(\tilde{a}^{2n})$, the eigenstate $u$ is still ``localized enough'' at the left edge and $k_{3,2}$ in \eqref{eq:k32_of_a} can fortunately take most of the values ($k_{3,1}\not\approx k_{3,2}\not\approx \tilde{k}_{3,2}$). This result effectively explains the commonness of edge states in finite chains with generic boundary coefficients (see Fig.~\ref{fig:edge state}(a) for an example). 
    
    \item Case $|a^{2n}|\ll|\Delta a|$:
    
    When $|\Delta a|$ is relatively large, $k_{3,2}$ becomes
    \begin{equation}
        k_{3,2}\approx k_2+{\rm sgn}(a)\frac{k_2 \sqrt{k_1+k_2\tilde{a}}}{\sigma\tilde{a}\sqrt{k_1+k_2/ \tilde{a}}}=k_{3,1}.
    \end{equation}
    That is to say, $k_{3,2}\approx k_{3,1}$ leads to $|c_2 a^{2n}|\gg 1$ hence the right end of the eigenstate becomes more nonnegligible. In fact, as $k_{3,2}$ gradually approaches $k_{3,1}$, the eigenstate grows to a two-sided localized state. When particularly $k_{3,2}=k_{3,1}$, the eigenstate is symmetric by Lemma~\ref{lm:symmetry} such that $c_2^2=a^{2n-2}$ and $\Delta a=\Theta(a^{n})$ (see Fig.~\ref{fig:edge state}(b) as an example). Similar to the definition of left edge states in Remark~\ref{rmk:edge states}, two-sided edge states are those with again $|a|\not\approx 1$ and larger $|c_2 a^{-2n}|$.

\begin{remark} ({\bf Mystery about two-sided edge states})
\label{rmk:two-sided}
Although our focus is on one-sided edge states, there are some misleading phenomena in the numerical computation of two-sided states worthy of explanation. For example, when $k_{3,2}\approx k_{3,1}$ the two eigenfrequencies in the bandgap in panel (a) of Fig.~\ref{fig:omega_k32} are very close and even look the same ``at the crossing''. More intriguingly, the numerically-obtained eigenstates for these two eigenfrequencies are possibly one-sided edge states at $k_{3,2}=k_{3,1}$ for large $n$, which does not follow our expectation of seeing two-sided edge states from the discussion above. This is because the difference between two eigenfrequencies numerically ``vanishs'' at $k_{3,1}=k_{3,2}$ in a long chain (the difference is approximately at the order $|\Delta a|\sim |a|^n$) and the numerical one-sided edge states are formed as linear combinations of actual two-sided edge states. In fact, Lemma~\ref{lm:L} guarantees that all the eigenfrequencies are distinct so that the crossing of two eigenfrequencies can never happen (see panel (b) of Fig.~\ref{fig:omega_k32} for the actual change of eigenfrequencies near $k_{3,2}=k_{3,1}$).  
\end{remark}
\end{itemize}
In the panel (a) of Fig.~\ref{schematic diagram:omega_k32}, we present a schematic diagram of the eigenfrequencies for the edge states in long diatomic chains \eqref{eq:finite} over the change of $k_{3,2}$. When $0\not\approx k_{3,1}\not\approx 2k_2$ (the case $k_{3,1}\approx k_2$ will be discussed in the next subsection), there is usually a left edge state with eigenfrequency close to $\omega(\tilde{a})$ where $\tilde{a}$ is determined by $k_{3,1}$ in \eqref{eq:5}. It should be noticed that the observed left edge state actually comes from two branches and its shape gradually becomes two-sided localized as $k_{3,2}$ approaches $k_{3,1}$. We summarize the conclusions in this subsection as the following Lemma.
\begin{lemma}
\label{lm:edge_states_generic}
In a long ($n\gg 1$) diatomic chain \eqref{eq:finite} with $0<k_1<k_2$, $k_{3,1}\not\approx k_2$ and $|k_{3,1}-k_2|\not\approx k_2$, 
\begin{itemize}
\item if $k_{3,2}\not\approx k_{3,1}$, then there exists a left edge state with $a\approx \tilde{a}(k_{3,1})$, where $\tilde{a}$ is from \eqref{eq:5}. To be more specific, the left edge state satisfies $|\Delta a|=|a-\tilde{a}|=\mathcal{O}(|\tilde{a}|^{2n})=\mathcal{O}(|a|^{2n})$ and $|c_2 a^{-2n}|=\mathcal{O}(1)$.
\item if $k_{3,2}\approx k_{3,1}$, then there exist two eigenstates with $|\Delta a|=|a-\tilde{a}|\gg |\tilde{a}|^{2n}$ and $|c_2 a^{-2n}|\gg 1$ that are more localized on two sides. In particular, when $k_{3,1}=k_{3,2}$, the eigenstates have $|\Delta a|=\Theta(|\tilde{a}|^{n})$.
\end{itemize} 
\end{lemma}

\subsection{\texorpdfstring{Special case I: Edge states with $k_{3,1}\approx k_{2}$ and $a\approx-\frac{k_1}{k_2}$}{Edge states with k{3,1}approx k{2} and a approx-frac{k1}{k2}}}
\label{subsec:sp1}


The situation with $k_{3,1}\approx k_2$ is similar to the generic case in Sec.~\ref{subsec:generic} since it also has one-sided or two-sided edge states (see the panel (c) of Fig.~\ref{fig:edge state} for an example of left edge state). However, this case is more special in the following aspects.
\begin{itemize}
\item When $k_{3,1}=k_2$, the left edge state in the semi-infinite chain has $\tilde{a}=-\frac{k_1}{k_2}$ and $\omega(\tilde{a})=\sqrt{k_1+k_2}$, which corresponds to the zero-energy state in the SSH model. However, direct calculation of \eqref{eq:tilde_k32} yields $\tilde{k}_{3,2}=\infty$, which implies that no genuine left edge states can exist for $k_{3,1}=k_2$. 

In order to study edge states in finite chains as perturbations of genuine edge states as in Sec.~\ref{subsec:generic}, we can define $\bar{k}_{3,2}=\frac{1}{k_{3,2}}$ or $\bar{k}_{3,2}=\frac{1}{k_{3,2}-k_2}$. Therefore IFT again can be applied to \eqref{eq:bc_1} and \eqref{eq:bc_2} to obtain $a(\bar{k}_{3,2})$ and $\bar{k}_{3,2}(a)$ near $\bar{k}_{3,2}=0$. For series expansion, we define $a_s=-\frac{k_1}{k_2}-\sqrt{-\frac{k_1}{k_2}-a}=-\frac{k_1}{k_2}+\Delta a_s$ and consider $\bar{k}_{3,2}(a_s)=\bar{k}_{3,2}(-\frac{k_1}{k_2})+\sum_{i=1}^{+\infty}\bar{k}_{3,2,i}(\Delta a_s)^i$.
After examining the first several terms in the series expansion, our guess is that the radius of convergence for $|\Delta a|=|\Delta a_s|^2$ is at the order of $r=\mathcal{O}(|\tilde{a}|^{4n})$. Since $c_2(a)=\Theta(\sqrt{|\Delta a|})$ in this case, the expected radius of convergence again coincides with the condition $|c_2 a^{-2n}|=\mathcal{O}(1)$ for left edge states in Remark~\ref{rmk:edge states}.

To better illustrate the dependence of edge states on $k_{3,2}$ in this case, we again show a schematic diagram (Fig.~\ref{schematic diagram:omega_k32} (b)) of the eigenfrequencies for the edge states over the change of $k_{3,2}$. The branch of eigenfrequency with $\sigma=1$ is always above $\omega=\sqrt{k_1+k_2}$ while the other branch with $\sigma=-1$ is always below that line. Compared with panel (a) of Fig.~\ref{schematic diagram:omega_k32}, the eigenfrequencies from neither branch in this diagram can reach $\omega(\tilde{a})=\sqrt{k_1+k_2}$ since $\tilde{k}_{3,2}$ is not applicable in this case! 

\item When $k_{3,1}\approx k_2$, the edge states can be similarly studied as for $k_{3,1}=k_2$ although the analysis becomes more complicated. At first we define $k_{3,1}=k_2+\delta k_{3,1}$ and $\tilde{a}=-\frac{k_1}{k_2}+\delta a$ for the left genuine edge state. Then we can write $a=\tilde{a}+\Delta a=-\frac{k_1}{k_2}+\delta a+\Delta a$ to include the effects of $k_{3,2}$ in finite chains. By comparing the order of $\delta a$ and $\Delta a$, we can obtain the results about one-sided and two-sided edge states as concluded in the following lemma (the proof is provided in the Appendix \ref{proof_edge_state_k31_appro_k2}).
\end{itemize}

\begin{lemma}
\label{lm:edge_states_sp1}
In a long ($n\gg 1$) diatomic chain \eqref{eq:finite} with $0<k_1<k_2$ and $k_{3,1}\approx k_2$, 
\begin{itemize}
\item if $k_{3,2}\not\approx k_{3,1}$, then there exists a left edge state with $a\approx \tilde{a}(k_{3,1})\approx -\frac{k_1}{k_2}$. To be more specific, the left edge state satisfies $|\Delta a|=|a-\tilde{a}|=\mathcal{O}(|\tilde{a}|^{2n})=\mathcal{O}(|a|^{2n})$ and $|c_2 a^{-2n}|=\mathcal{O}(1)$.
\begin{itemize}
    \item In particular, if $|k_{3,1}-k_2|=\mathcal{O}(|a|^{2n})=\mathcal{O}(|\frac{k_1}{k_2}|^{2n})$ and $k_{3,2}\not\approx k_2$, the left edge state has $|\Delta a|=\mathcal{O}(|\frac{k_1}{k_2}|^{4n})$. 
\end{itemize}
\item if $k_{3,2}\approx k_{3,1}$, then there exist two eigenstates with $|\Delta a|=|a-\tilde{a}|\gg |\tilde{a}|^{4n}$ and $|c_2 a^{-2n}|\gg 1$ that are more localized on two sides. In particular, when $k_{3,1}=k_{3,2}$, the eigenstates have $|\Delta a|=\Theta(|\tilde{a}|^{2n})$.
\end{itemize} 
\end{lemma}

\subsection{\texorpdfstring{Special case II: Eigenstates with $|k_{3,1}-k_2|\approx k_{2}$ and $a\approx \pm 1$}{Special case II: Eigenstates with |k{3,1}-k2|approx k{2} and a approx pm 1}}
\label{subsec:sp2}

It is clear from the expression \eqref{eq:omega} of $\omega$ that eigenfrequencies with $a=\pm 1$ are exactly at the band edges while those with $a\approx \pm 1$ are close to the band edges. According to Remark.~\ref{rmk:edge states}, eigenstates with $a\approx \pm 1$ are neither one-sided nor two-sided edge states in finite chains since they are either non-localized at all (panels (e) and (f) of Fig.~\ref{fig:edge state}) or not localized enough (panel (d) of Fig.~\ref{fig:edge state}). Since the number of eigenvectors for $\mathcal{T}$ particularly becomes one at $a=\pm 1$, we first consider the special case $a=\pm 1$ and then discuss the case $a\approx \pm 1$ in the following.

\begin{itemize}
    \item 
We first characterize the eigenstates with exactly $a=\pm 1$. When ${a}=\pm 1$, the matrix $\mathcal{T}$ from \eqref{eq:iteration} has an eigenvector $\vec{v}_{1}=(1,-a\sigma)^{\top}$ and a generalized eigenvector $\vec{v}_{2}=(1, a\sigma)^{\top}$. 
If we span 
$(u_{1},u_{2})^{\top}$ as
$(u_{1},u_{2})^{\top}
=c_{1}\vec{v}_{1}+c_{2}\vec{v}_{2}$, 
then $(u_{2n-1},u_{2n})^{\top}$ has the form
\begin{equation}
\label{relation:iteration2}
\begin{split}
  (u_{2n-1},u_{2n})^{\top}
&=c_{1}\vec{v}_{1}+c_{2}(n-1)(\frac{-2(k_{1}+k_{2})}{k_{2}})\vec{v}_{1}+c_{2}\vec{v}_{2},\quad a=1;\\
(u_{2n-1},u_{2n})^{\top}
&=(-1)^{n-1}[c_{1}\vec{v}_{1}-c_{2}(n-1)\frac{2(k_{2}-k_{1})}{k_{2}}\vec{v}_{1}+c_{2}\vec{v}_{2}],\quad a=-1.
\end{split}
\end{equation}
Substituting $\vec{v}_{1}$ and $\vec{v}_{2}$ into $-\omega^2 u=\mathcal{L}u$, we obtain the conditions for the existence of eigenstates with $a=\pm 1$ in the following Lemma~\ref{lm:band_edge_states} (please see the Appendix \ref{proof:a=pm1} for a more detailed derivation).

\begin{lemma}
\label{lm:band_edge_states}
Suppose the long chain \eqref{eq:finite} bears an eigenstate with $a=\pm 1$ and $\omega^2=k_1+k_2+a\sigma(k_1+k_2 a)$. For every given $k_{3,1}$, the corresponding $k_{3,2}$ (if exists) for such an eigenstate is uniquely determined $k_{3,2}=k_{3,2}(a,\sigma, k_{3,1})$. Some further properties about the admissible pair $\{ k_{3,1}, k_{3,2} \}$ are as follows.
\begin{itemize}

        \item If $k_{3,1}$ is not close enough to $0$ or $2k_2$, say $|k_{3,1}-k_2(1+\sigma )|\gg \frac{1}{n}$, then $k_{3,2}\approx k_2(1+\sigma )+a \sigma \frac{k_1k_2}{n(k_2+a k_1)}$.
        \item If $k_{3,1}$ is very close to $0$ or $2k_2$, say $|k_{3,1}-k_2(1+\sigma )|\ll \frac{1}{n}$, then $k_{3,2}-k_2(1+\sigma )\approx -(k_2(1+\sigma )-k_{3,1})$.
        \item If $|k_{3,1}-k_2(1+\sigma )|=\Theta(\frac{1}{n})$ and $k_{3,1}-k_2(1+\sigma )\not\approx a\sigma\frac{k_1 k_2}{n(k_2+a k_1)}$, then $|k_{3,2}-k_2(1+\sigma )|=\Theta(\frac{1}{n})$ and $k_{3,2}-k_2(1+\sigma )\not\approx a\sigma\frac{k_1 k_2}{n(k_2+a k_1)}$.
        \begin{itemize}
    \item If $a(k_{3,1}-k_{2}(1+\sigma))>0$, then $\zeta(k_{3,2}-k_2(1+\sigma))<0$ where $\zeta=a\sigma(1-c_{2}-c_{2}(n-1)\frac{2k_{2}+2ak_{1}}{k_{2}})$ and $c_{2}=\frac{-k_{3,1}+k_{2}(1+\sigma)}{k_{3,1}-k_{2}(1+\sigma)-2a\sigma k_{1}}$.

    \item If $a(k_{3,1}-k_{2}(1+\sigma))<0$, then $a(k_{3,2}-k_{2}(1+\sigma))>0$;

\end{itemize}
   \end{itemize}
\end{lemma} 
Combining the results from Lemma~\ref{lm:edge_states_generic}, Lemma~\ref{lm:edge_states_sp1} and Lemma~\ref{lm:band_edge_states}, we 
can identify the number of eigenfrequencies at band edges or away from the bands as in Prop.~\ref{prop:number_band_edge}.
\begin{proposition}
    \label{prop:number_band_edge}
    In the long diatomic chain \eqref{eq:finite}, the following statements hold:
    \begin{itemize}
        \item If $|k_{3,1}-k_2|\not\approx k_2$ and $|k_{3,2}-k_2|\not\approx k_2$, no eigenfrequencies are at band edges. At the same time, there are two eigenfrequencies outside the bands, which correspond to one-sided or two-sided edge states.
        \item  If either $|k_{3,1}-k_2|\approx k_2$ or $|k_{3,2}-k_2|\approx k_2$, there is at most one eigenfrequency at some band edge. At the same time, there exists another eigenfrequency outside the bands, which corresponds to a left or right edge state.
        \item  If $|k_{3,1}-k_2|\approx k_2$ and $|k_{3,2}-k_2|\approx k_2$, there are at most two eigenfrequencies at band edges and no eigenfrequencies for edge states. 
    \end{itemize}
\end{proposition}

\item 
Since $|k_{3,1}-k_2|\not\approx k_2$ and $|k_{3,2}-k_2|\not\approx k_2$ in the long chain \eqref{eq:finite} correspond to a left edge state and a right edge state, respectively, the existence of eigenstates with $a\approx\pm 1$ requires either $|k_{3,1}-k_2|\approx k_2$ or $|k_{3,2}-k_2|\approx k_2$, which is similar to the situation for $a=\pm 1$. Moreover, the eigenfrequencies with $a\approx\pm 1$ can usually be viewed as eigenfrequencies perturbed from the band edges (with $a=\pm 1$) by varying $k_{3,1}$ or $k_{3,2}$. In particular, we can derive the following results from Prop.~\ref{prop:number_band_edge}, Lemma~\ref{lm:L} and Lemma.~\ref{lm:increasing}.
\begin{lemma}
\label{lemma:a_approx_pm1}
(1). A necessary condition for the existence of an eigenstate $u$ in \eqref{eq:finite} with frequency outside the bands and near some band edge ($\pm 1\approx a\in\mathbb{R}$ and $\omega^2\approx k_1+k_2-\sigma(k_1\pm k_2)$) is $k_{3,1}\approx k_2(1\mp \sigma)$ or $k_{3,2}\approx k_2(1\mp \sigma)$.    

(2). If $|k_{3,1}-k_2|\not\approx k_2$, then there exists at most one eigenstate $u$ with $a\approx \pm 1$.

(3). If $k_{3,1}\approx 2k_2 \approx k_{3,2}$ ($k_{3,1}\approx 0 \approx k_{3,2}$), then there exists at most two eigenstates with frequencies outside the bands and near the edges of the optical (acoustic) band.

(4). If $k_{3,1}\approx 2k_2$ and $k_{3,2}\approx 0$, then there exists at most two eigenfrequencies outside the bands, with one near some edge of the optical band and the other near some edge of the acoustic band.
\end{lemma}
Please note that the discussion above can be made more precise by applying the asymptotic analysis as in Sec.~\ref{subsec:generic}. For instance, suppose the chain \eqref{eq:finite} with $|k_{3,1}-k_2|\not\approx k_2$ bears an eigenstate with eigenfrequency near some edge of the optical band ($\omega^2\approx 2k_2$ or $\omega^2\approx 2k_1+2k_2$), then we can estimate the value of $k_{3,2}$ with accuracy at $\mathcal{O}(\frac{1}{n})$ (proof is in the Appendix \ref{proof:k31_2k2}).
\begin{lemma}
\label{lemma:a_approx_pm1_2k2}
In the long diatomic chain \eqref{eq:finite} with $k_{2}>k_{1}$, if $0\not\approx k_{3,1}\not\approx 2k_{2}$, then the following statements hold:
\begin{itemize}
    \item If there exists any eigenstate $u$ with $\omega^2\approx 2k_2$ and $-1<a\approx -1$, then $\Delta k_{3,2}=k_{3,2}-2k_2<0$ and $\Omega(\frac{1}{n})= |\Delta k_{3,2}|\ll 1$.

    \item If there exists any eigenstate $u$ with $\omega^2\approx 2k_1+2k_2$ and $1>a=1+\Delta a\approx 1$, then $\Delta k_{3,2}=k_{3,2}-2k_2>0$ and $\Omega(\frac{1}{n})= \Delta k_{3,2}\ll 1$.
    
\end{itemize}
\end{lemma}
The estimates on eigenfrequencies near other band edges can be derived following similar strategy. However, further results are beyond the scope of this current work hence will not be included here.
\end{itemize}

\subsection{\texorpdfstring{Section summary ($k_1<k_2$)}{Section summary (k1 < k2)}}
\label{subsec:summary_k1k2}
We now summarize the main results in Sec.~\ref{subsec:generic}, Sec.~\ref{subsec:sp1} and Sec.~\ref{subsec:sp2} intuitively as the theorem below:

\begin{theorem}
\label{theorem:edgestates}
In a long ($n\gg 1$) diatomic chain~\eqref{eq:finite} with $k_1<k_2$, 
\begin{itemize}
\item if $|k_{3,1}-k_2|\not\approx k_2$,
\begin{itemize}
\item if $k_{3,2}\approx k_{3,1}$, there exist two eigenstates more localized at both edges ($|c_2 a^{-2n}|\gg 1$).
\item if $k_{3,2}\not\approx k_{3,1}$, there exists a left edge state.
\begin{itemize}
\item if $k_{3,1}=k_2$, then $a\approx -\frac{k_1}{k_2}$ and $|a+\frac{k_1}{k_2}|= \mathcal{O}(|\frac{k_1}{k_2}|^{4n})$
\item if $k_{3,1}\approx k_2$, then $a\approx -\frac{k_1}{k_2}$ and $|a+\frac{k_1}{k_2}|= \mathcal{O}(|\tilde{a}|^{2n})$.
\item if $k_{3,1}\not\approx k_2$, then $a\approx \tilde{a}\not\approx -\frac{k_1}{k_2}$ and $|a-\tilde{a}|= \mathcal{O}(|\tilde{a}|^{2n})$.
\end{itemize}
\item if $|k_{3,2}-k_2|\not\approx k_2$, there are no ``slow-decaying'' or ``non-localized'' eigenstates with $a\approx\pm 1$.
\item if $|k_{3,2}-k_2|\approx k_2$, then there exists at most one eigenstate with $a\approx \pm 1$. 
\begin{itemize}
    \item if $k_{3,2}-k_2\approx \mp\sigma k_2$, then the eigenstate with $a\approx \pm 1$ (if exists) satisfies $\omega^2\approx k_1+k_2-\sigma(k_1\pm k_2)$.
\end{itemize}
\end{itemize}
\item if $|k_{3,i}-k_2|\approx k_2$ for $i=1,2$, then there exist at most two eigenstates with $a\approx \pm 1$. 
\begin{itemize}
    \item if $k_{3,i}-k_2\approx \mp\sigma k_2$, then each eigenstate with $a\approx \pm 1$ (if exists) satisfies $\omega^2\approx k_1+k_2-\sigma(k_1\pm k_2)$. 
\end{itemize}
\end{itemize}
\end{theorem}

\begin{corollary}
\label{corollary:number_of_edgestates}
In a long ($n\gg 1$) diatomic chain~\eqref{eq:finite} with $k_1<k_2$, there are at most two eigenfrequencies outside the bands or at the band edges.
\end{corollary}
Therefore, observations (L1) and (L2) in Sec.~\ref{subsec:def_edgestates} are theoretically verified and explained. It can be perceived that there are different categories of ``eigenstates with outside-band eigenfrequencies'' in long chains.
\begin{itemize}
\item {\bf One-sided edge states}: When $k_{3,1}$ and $k_{3,2}$ are away from special values $\{0, 2k_2\}$ and $k_{3,1}\not\approx k_{3,2}$, there exist left and right edge states that are similar to those in semi-infinite chains.
\item {\bf Two-sided edge states}: When $k_{3,1}$ and $k_{3,2}$ are away from special values $\{0, 2k_2\}$ but $k_{3,1}\approx k_{3,2}$, there exist edge states more localized at both ends that do not appear in semi-infinite chains.
\item {\bf Slow-decaying eigenstates}: When $k_{3,1}$ or $k_{3,2}$ is close to some special value $0$ or $2k_2$, there may exist eigenstates with $a\approx\pm 1$. Although any eigenstate with $|a|<1$ in semi-infinite chains is an edge state, it is usually not localized enough for the same recognition in finite chains.
\end{itemize}

\begin{remark}
\label{rem:min_size}
\textbf{(Minimum lattice size and localization length)} The asymptotic conditions derived in Section~\ref{subsec:generic} and \ref{subsec:sp1} can be physically interpreted as defining the minimum lattice size required for observing robust edge states. In a finite system of $n$ unit cells, the ``purity'' of a left-localized edge state (measured by the amplitude ratio $|c_2/c_1|$) is limited by the finite-size hybridization with the right boundary, which is exponentially suppressed by the system size.

Generally, for a robust edge state (e.g., $|c_2/c_1| < \epsilon$), the condition for the chain length $2n$ can be derived from \eqref{eq:bc_2} as 
\beq
|\frac{c_2}{c_1}|=| a^{2n-2}\frac{ k_1 v_{1,1} (k_1+k_{3,2}-\omega^2)v_{1,2} }{ (k_1+k_{3,2}-\omega^2)v_{2,2}-k_1 v_{2,1} } |<\epsilon
\eeq
where $\omega$ and $v_{i,j}$ $(1\leq i,j\leq 2)$ are functions of $n$. This inequality explicitly quantifies the competition between the system size and boundary conditions. In order to obtain more specific results, it is important to note the distinction between the general and special case:
\begin{itemize}
    \item \textbf{General case (Section \ref{subsec:generic}):} When $k_{3,1}$ is far from $k_2$ (and other special values), the perfect matching stiffness $\tilde{k}_{3,2}$ is finite. In this regime, we have the relation between the frequency shift and the stiffness deviation:
    \beq
    \Delta{a}\sim a^{2n} \Delta k_{3,2}
    \eeq
    where $\Delta a=a-\tilde{a}$ and $\Delta k_{3,2}=k_{3,2}-\tilde{k}_{3,2}$. Since $c_2\sim \Delta a$ (assuming $c_1=1$) by \eqref{eq:c2} and $a^{2n}\sim \tilde{a}^{2n}$ for small $\Delta a$, it yields 
    \beq
    c_2\sim \tilde{a}^{2n}\Delta k_{3,2}
    \eeq
    which can approximately determine the minimal $n$ for $|c_2|<\epsilon$. In particular, if we adopt the condition $c_2=\mathcal{O}(a^{2n})$ in Remark~\ref{rmk:edge states} for left edge states, then $n$ can be any large number ($n\gg 1$ or $n\gg \xi=-\frac{1}{\ln{|a|}}$ with $\xi$ being the localization length) that makes asymptotic analysis possible.
    \item \textbf{Special case (Section \ref{subsec:sp1}):} When $k_{3,1}$ approaches $k_2$, the required boundary stiffness $\tilde{k}_{3,2}$ diverges ($\tilde{k}_{3,2} \to \infty$). Near this singularity, the leading-order term in the asymptotic expansion may vanish or become ill-conditioned. In particular, when $k_{3,1}=k_2$, the expansion yields
    \beq
    \frac{1}{k_{3,2}}\sim a^{-2n} \sqrt{\Delta a}.
    \eeq 
    Since $c_2\sim \sqrt{\Delta a}$ in this case, then 
    $c_2\sim \frac{a^{2n}}{k_{3,2}}$ can be used to approximately find $n$ such that $|c_2|<\epsilon$. It is worth noting that this implies a higher-order scaling for the eigenfrequency shift, $\Delta a \sim \mathcal{O}(a^{4n})$, indicating a stronger suppression of finite-size splitting in this regime. 

\end{itemize}
\end{remark}

\subsection{Discussion on band inversion}

When we consider the chain \eqref{eq:finite} with $k_{1}>k_{2}$, the Zak phase in \eqref{Zak} indicates an altered bulk topology compared to case $k_{2}>k_{1}$. In the same spirit of previous discussion, we can employ our analytical framework to specifically reveal the connection between boundary conditions and edge states (or ``slow-decaying'' eigenstates) for $k_1>k_2$.

\begin{theorem}
\label{thm:edgestates_2}

In a long ($n\gg 1$) diatomic chain~\eqref{eq:finite} with $k_{1}>k_{2}$, 
\begin{itemize}
\item if $|k_{3,1}-k_2|\not\approx k_2$,
\begin{itemize}
\item if $k_{3,1}<2k_2$, there exist no left edge states or two-sided edge states.
\item if $k_{3,1}>2k_2$ and $k_{3,1}\not\approx k_{3,2}$, there exist two left edge states with $|a-\tilde{a}|=\mathcal{O}(|\tilde{a}|^{2n})$.
\item if $k_{3,1}>2k_2$ and $k_{3,1}\approx k_{3,2}$, there exist four eigenstates more localized at both edges ($|c_2 a^{-2n}|\gg 1$).
\item if $|k_{3,2}-k_2|\not\approx k_2$, then there are no ``slow-decaying'' or ``non-localized'' eigenstates with $a\approx \pm 1$.

\item if $|k_{3,1}-k_2|\approx k_2$, then there exists at most one eigenstate with $a\approx 1$ and at most one eigenstate with $a\approx -1$. 
\begin{itemize}
    \item if $k_{3,2}-k_2\approx \mp\sigma k_2$, then each eigenstate with $a\approx \pm 1$ (if exists) satisfies $\omega^2\approx k_1+k_2-\sigma(k_1\pm k_2)$.
\end{itemize}

\end{itemize}

\item if $|k_{3,i}-k_2|\approx k_2$ for $i=1,2$, then there exist at most two eigenstates with $a\approx 1$ and at most two eigenstates with $a\approx -1$. 
\begin{itemize}
    \item if $k_{3,i}-k_2\approx \mp\sigma k_2$, then each eigenstate with $a\approx \pm 1$ (if exists) satisfies $\omega^2\approx k_1+k_2-\sigma(k_1\pm k_2)$. 
\end{itemize}

\end{itemize}

\end{theorem}

By comparing the results from Theorem~\ref{theorem:edgestates} and Theorem~\ref{thm:edgestates_2}, the existence and count of edge states in finite-size systems explicitly manifest the topological contrast.
In the panels (b)(c)(d) of Fig.~\ref{fig:num_edge_states}, we showcase the count of (both one-sided and two-sided) edge states in long chains under different parametric settings. It is evident that the count depends on three quantities: $(k_{2}-k_1)$, $(|k_{3,1}-k_2|-k_2)$ and $(|k_{3,2}-k_2|-k_2)$.
\begin{itemize}
\item Some thick lines in panels (b)(c)(d) are drawn to represent the special regions (``tubes'') near them. 
\begin{itemize}
\item $\{ |k_{3,1}-k_2|\approx k_2\}$ corresponds to the situation where the left edge states do not exist ($a\approx\pm 1$).
\item $\{ k_2\approx k_1=1 \}$ represents the case where the bandgap nearly vanishes and eigenstates with inside-bandgap eigenfrequencies are not sufficiently localized ($a\approx \pm 1$). 
\end{itemize}
\item In panel (d), when $|k_{3,2}-k_2|\approx k_2$, there can only exist left edge states hence its count is similar to that in panel (a) for semi-infinite chains (based on Remark~\ref{remark:root_a} and Remark~\ref{remark:root_a_2}).
\item In panels (b) and (c), when $k_{3,1}$ and $k_{3,2}$ are away from the special values, the number of edge states can be counted as the sum of edge states in the left and right semi-infinite chains, respectively.
\end{itemize}
\begin{remark}
     We notice that a long diatomic chain has three different regimes according to the boundary conditions. 
     \begin{itemize}
         \item {\bf nearly infinite regime}: If $|k_{3,1}-k_2|\not\approx k_2$ and $|k_{3,2}-k_2|\not\approx k_2$, then the long chain can be almost viewed as two semi-infinite chains to identify edge states.
         \item {\bf nearly semi-infinite regime}: If $|k_{3,1}-k_2|\not\approx k_2$ and $|k_{3,2}-k_2|\approx k_2$ ($|k_{3,1}-k_2|\approx k_2$ and $|k_{3,2}-k_2|\not\approx k_2$), then the left (right) edge states in a long chain can be studied as those in a left (right) semi-infinite chain.
         \item {\bf finite regime}: If $|k_{3,1}-k_2|\approx k_2$ and $|k_{3,2}-k_2|\approx k_2$, then the finiteness of the chain dominates over other factors and there are no sufficiently localized edge states.
     \end{itemize} 
     Particularly, in ``nearly infinite regime'' and ``nearly semi-infinite regime'', the existence of edge states manifests the ideal ``bulk-boundary correspondence'' as in semi-infinite chains. The disappearance of edge states in some circumstances does not indicate the failure of the topological principle. Instead, it arises from two physical mechanisms quantified by our asymptotic analysis:
     \begin{itemize}
     \item \textbf{Finite-size Hybridization:} When $n$ is small compared to the edge state decay length, the two edge modes couple and split, potentially pushing them into the bulk bands.
     \item \textbf{Boundary Symmetry Breaking:} The arbitrary stiffness $k_{3,1}$ acts as an effective on-site potential at the boundary. In the SSH context, this breaks the chiral symmetry that protects the zero-energy mode. Our analysis precisely determines the critical stiffness required to `push' the topological mode out of the bandgap.
     \end{itemize}
     
\end{remark}

\begin{remark}
\label{rmk:odd-number}
\textbf{(Odd-numbered lattices and sublattice imbalance)}
While our primary analysis focuses on chains with an even number of particles ($N=2n$), the framework can be directly applied to odd-numbered chains ($N=2n+1$) and their results exhibit distinct topological features due to sublattice imbalance.

In the even case, we showed that a mid-gap edge state with $\omega^2=k_1+k_2$ and $c_1c_2=0$ (analogous to the zero-energy mode in SSH model) cannot exist when the boundary stiffness matches the bulk coupling (i.e., $k_{3,1}=k_2$) because the characteristic equation degenerates. However, for an odd-numbered chain, the lattice possesses a mismatch between the two sublattices (e.g., $n+1$ sites on sub-lattice A and $n$ sites on sub-lattice B).

This imbalance forces the existence of a mid-gap state when the boundary conditions are symmetric with respect to the bulk alternation (specifically, $k_{3,1}=k_2$ and $k_{3,2}=k_1$). Under these conditions, the displacements on the even sub-lattice vanish identically ($u_{2m} \equiv 0$), allowing a non-trivial solution localized on the odd sub-lattice.

Furthermore, unlike the even chain where left and right boundaries are symmetric regarding the bond sequence, the odd chain is asymmetric (starting with $k_1$ and ending with $k_2$, assuming alternating bonds). Consequently, swapping the bulk stiffness values $k_1$ and $k_2$ in an odd chain switches the localization of the mid-gap mode from the left boundary to the right boundary, a phenomenon not present in the even-numbered counterpart.
\end{remark}

\section{Extended states with eigenfrequencies inside the bands}
\label{sec:extended_states}

\begin{figure}[!htp]
\centering
\subfigure[]{
\includegraphics[height=5.5 cm, width=6.5 cm ]{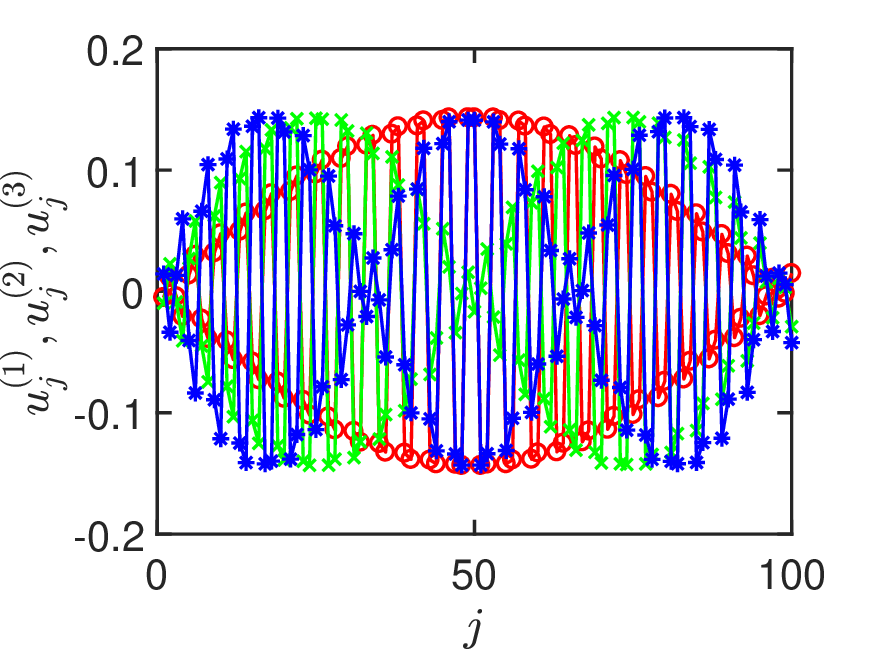}
}
\hspace{2mm}
\subfigure[]{
\includegraphics[height = 5.5 cm, width = 6.5 cm]{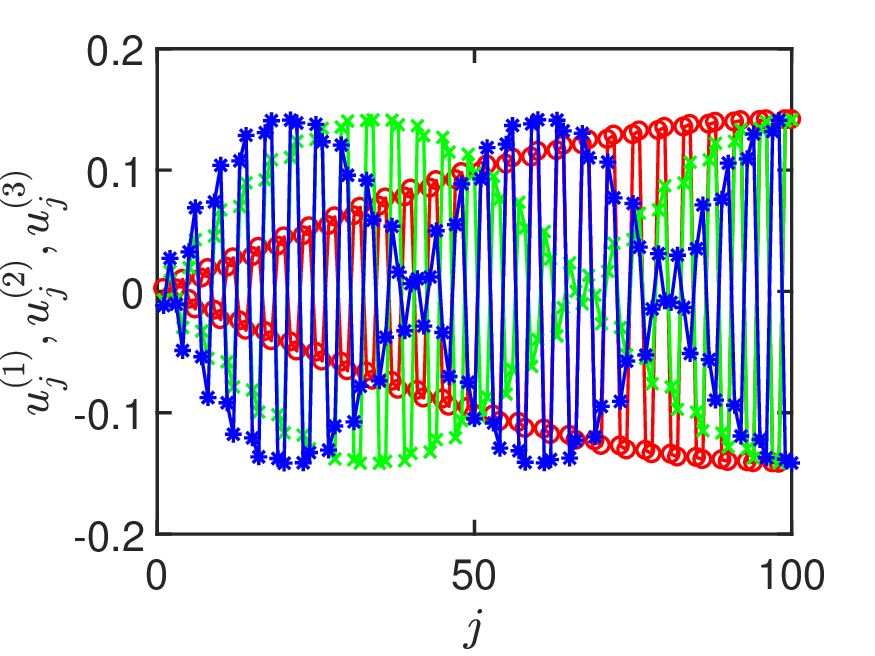}
}
\caption{In panel (a) (panel (b)), the eigenstates $u^{(k)}$ with in-band frequencies closest to the lower edge of the optical band are shown for the chain~\eqref{eq:finite} with $n=50$, $k_1=1$,
$k_{2}=2.3$, $k_{3,1}=1.3$ and $k_{3,2}=3.5$ ($k_{3,2}=2k_{2}$). Here $u^{(1)}$ is denoted by red $``\circ"$, $u^{(2)}$ is denoted by green $``\times"$ and $u^{(3)}$ is denoted by blue $``\ast"$.
}
\label{fig:examples_eigenstates_in_band}
\end{figure}

In Sec.~\ref{sec:finite_edgestates}, we have analyzed the eigenstates with frequencies outside the spectral bands in long diatomic chains \eqref{eq:finite}. Now we consider eigenfrequencies inside the bands and especially provide estimates for those near the band edges. 
Interestingly, the eigenstates for such frequencies usually feature shapes of sinusoidal waves that cover multiples of the half period or quarter period.
For example, the eigenstates $u^{(k)}$ with frequencies $\omega^{(k)}$ near the lower edge of the optical band are shown in Fig.~\ref{fig:examples_eigenstates_in_band}, where panels (a) and (b) demonstrate two different patterns, respectively.
We notice that these two patterns for near-band-edge eigenfrequencies partially resemble the situation for diatomic chains with periodic boundary conditions and free boundary conditions. However, it is worth mentioning that our findings hold for more generic boundary conditions in addition to oversimplified free ends and periodic boundaries.

\subsection{Mathematical formulation}
In the long chain~\eqref{eq:finite} with $k_1<k_2$, suppose that there are $n_1$ eigenfrequencies in the optical band ($(\omega^{(k)})^2\in (2k_2, 2k_1+2k_2)$) and $n_2$ eigenfrequencies in the acoustic band ($(\omega^{(k)})^2\in (0, 2k_1)$), then $2n-2\leq n_{1}+n_{2} \leq 2n$ by Corollary~\ref{corollary:number_of_edgestates}.
\begin{remark}
\label{remark:order_of_eigenfrequencies_optical}
({\bf The ordering of eigenfrequencies}) Since we use the eigenfrequencies near the lower edge of the optical band as our main examples in this section, the eigenfrequencies in two bands are ordered as follows for the convenience of discussion.
\begin{equation}
\label{eq:omega_order_1}
2k_2<(\omega^{(1)})^2<(\omega^{(2)})^2<\dots<(\omega^{(n_1)})^2<2k_1+2k_2,
\end{equation}
\begin{equation}
0<(\omega^{(2n)})^2<(\omega^{(2n-1)})^2<\cdots<(\omega^{(2n-n_2+1)})^2<2k_1.
\end{equation}
\end{remark}
Since the eigenfrequencies in the bands have $|a|=1$, we write $a=e^{i\theta}$ hence
\beq
\label{eq:omega_inband}
(\omega^{(k)})^2 = k_1+k_2+\sigma\sqrt{k_1^2+k_2^2+2k_1k_2\cos\theta^{(k)}}.
\eeq
If we assume $\theta\in (0, \pi)$ for eigenfrequencies in the bands, then
\begin{equation}
\begin{split}
\pi>\theta^{(1)}>\theta^{(2)}>&\dots>\theta^{(n_1)}>0, \quad \sigma=1; \\
\pi>\theta^{(2n-n_2+1)}>\theta^{(2n-n_2+2)}>&\cdots>\theta^{(2n)}>0, \quad \sigma=-1.
\end{split}
\end{equation}
Since the eigenstates with $\omega^{2}$ inside the bands have $|a|=1$ and $a\neq \pm1$, the eigenvectors \eqref{eq:v1v2_0} of $\mathcal{T}(\omega)$ can be chosen as
\beq
v_1= \left(\begin{array}{c}
 \sqrt{k_1+k_2 e^{-i\theta}} \\
 -\sigma\sqrt{k_1+k_2 e^{i\theta}}
 \end{array}\right), \quad 
v_2= \left(\begin{array}{c}
 \sqrt{k_1+k_2 e^{i\theta}} \\
 -\sigma\sqrt{k_1+k_2 e^{-i\theta}}
 \end{array}\right)
\eeq
When $\theta\in (0,\pi)$, we define $\alpha\in (-\frac{\pi}{2},0)$ such that 
\begin{equation}
\label{eq:alpha}
\sqrt{k_1+k_2 e^{-i\theta}}=\rho e^{i\alpha}, \quad \sqrt{k_1+k_2 e^{i\theta}}=\rho e^{-i\alpha}.
\end{equation}

Since the states $u^{(k)}$ are real, we suppose 
\begin{equation}
  c_{1}=re^{i\beta},\quad c_{2}=re^{-i\beta}
  \label{eq:c1c2}
\end{equation}
and substitute these into the \eqref{eq:bc_1} and \eqref{eq:bc_2} to obtain
\begin{eqnarray}
  \label{eq:bc_b1}
  \frac{\cos(\alpha+\beta)}{-\sigma\cos(\beta-\alpha)} &=&  \frac{k_{1}}{k_{1}+k_{3,1}-\omega^{2}},\\
  \frac{\cos(\alpha+\beta+(n-1)\theta)}{-\sigma\cos(\beta-\alpha+(n-1)\theta)} &=& \frac{k_{1}+k_{3,2}-\omega^{2}}{k_{1}}.
\label{eq:bc_b2}
\end{eqnarray}
Since \eqref{eq:bc_b1} and \eqref{eq:bc_b2} are invariant under the change $\beta\to\beta+\pi$, it suffices to consider $\beta\in [0,\pi)$. 
Similar to \eqref{eq:bc_1} and \eqref{eq:bc_2} in the previous section, here \eqref{eq:bc_b1} and \eqref{eq:bc_b2} have two unknowns $\theta$ ($\omega$ and $\alpha$ are functions of $\theta$) and $\beta$ depending on $k_{3,1}$ and $k_{3,2}$. 
In order to study eigenfrequencies near the lower edge of the optical band, we write $\theta=\pi-\Delta\theta$ with $0<\Delta\theta\ll 1$, hence \eqref{eq:alpha} and \eqref{eq:omega} yield
\begin{equation}
\label{Delta alpha}
\alpha=-\frac{\pi}{2}+\Delta\alpha\approx-\frac{\pi}{2}+\frac{k_2}{2(k_2-k_1)}\Delta \theta,  \quad \omega^2\approx 2 k_2+\frac{k_1 k_2 (\Delta\theta)^2}{2(k_2-k_1)}.
\end{equation}
At the same time, \eqref{eq:bc_b1} can be rewritten as
\begin{equation}
\label{eq:tan_beta}
\tan\beta=\frac{2k_1+k_{3,1}-\omega^2}{k_{3,1}-\omega^2}\cot\alpha=\frac{2k_1+k_{3,1}-\omega^2}{\omega^2-k_{3,1}}\tan\Delta\alpha. 
\end{equation}
Depending on the choice of $k_{3,1}$ and $k_{3,2}$, we can obtain different types of results from \eqref{eq:tan_beta} and \eqref{eq:bc_b2} as follows.

\subsection{Nearly infinite regime}

\begin{figure}[!htp]
\centering
\subfigure[]{
\includegraphics[height = 5.5cm, width = 6.5cm]{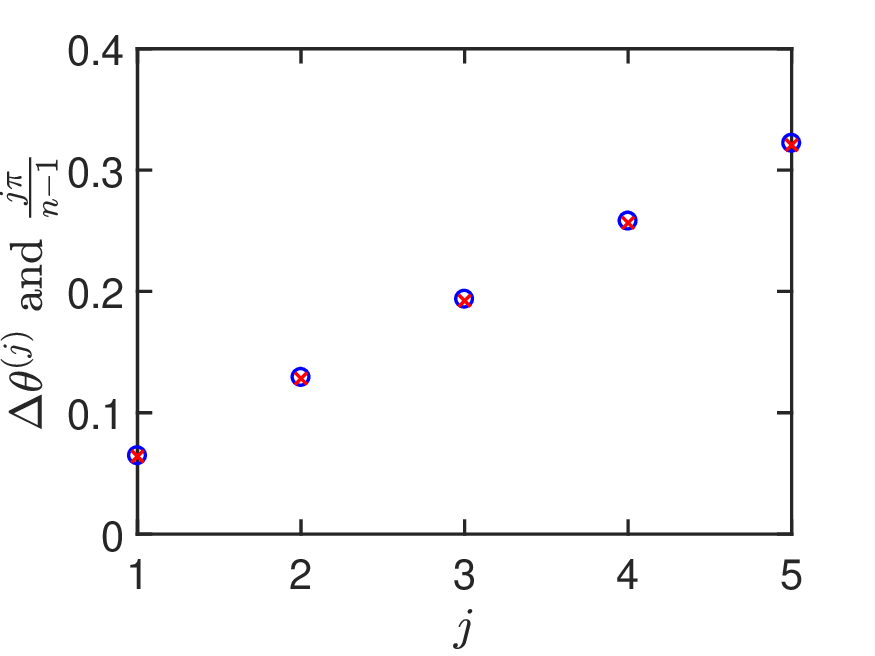}
}
\hspace{2mm}
\subfigure[]{
\includegraphics[height = 5.5cm, width = 6.5cm]{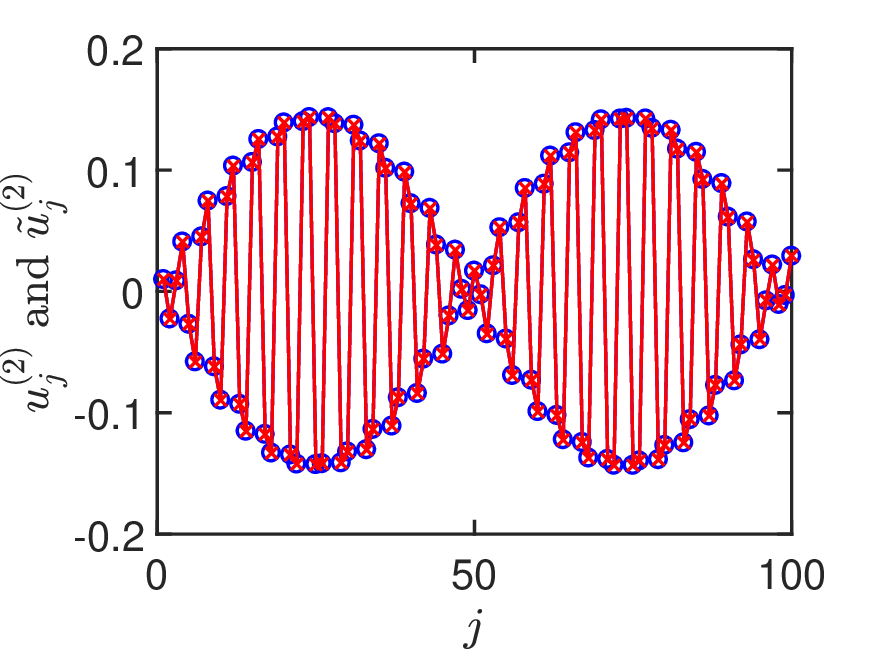}
}

\centering
\subfigure[]{
\includegraphics[height = 5.5cm, width = 6.5cm]{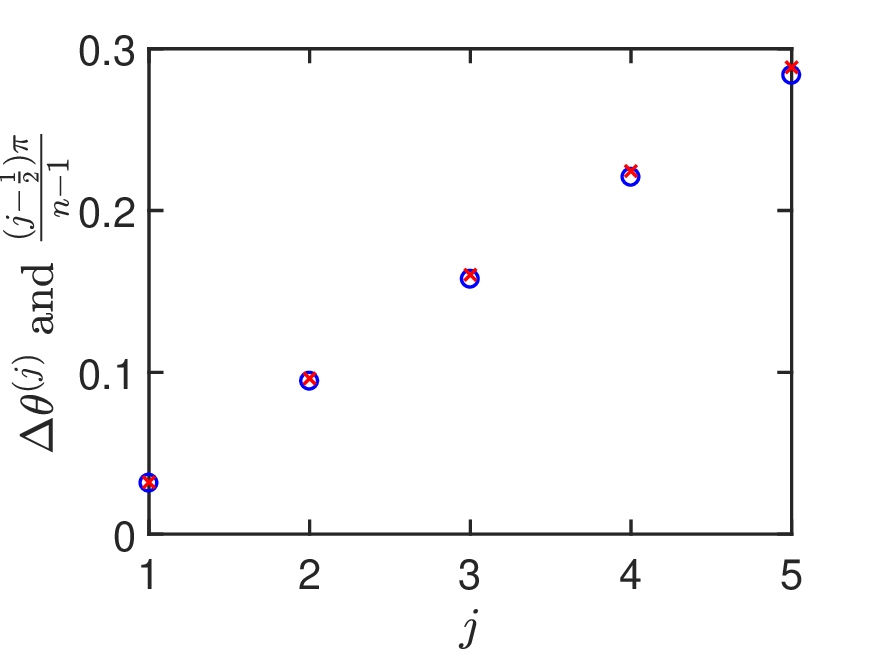}
}
\hspace{2mm}
\subfigure[]{
\includegraphics[height = 5.5cm, width = 6.5cm]{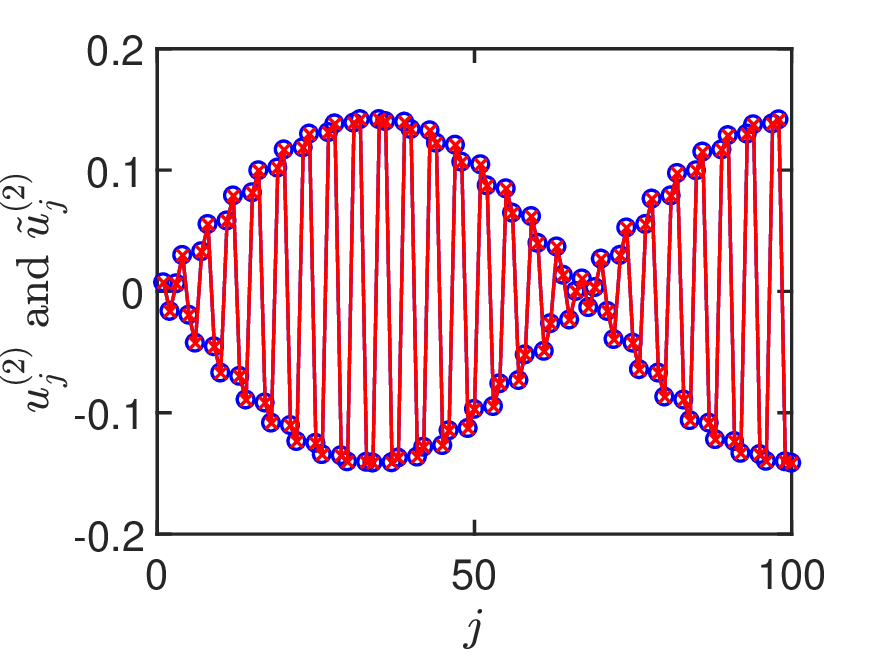}
}
\caption{
Here we show approximations of $\Delta\theta^{(j)}$ and eigenstates $u^{(j)}$ with frequencies near the lower edge of optical band ($\omega^2\approx 2k_2$) in the chain~\eqref{eq:finite} with $n=50$, $k_1=1$,
$k_{2}=2.3$ and $k_{3.1}=1.3$. In panel (a), we plot numerically obtained $\Delta\theta^{(j)}$ (denoted by blue ``$\circ$'') for $k_{3,2}=3.5\not\approx 2k_2$ and compare them with the approximations $\frac{j\pi}{n-1}$ (denoted by red ``$\times$''). In panel (b), the corresponding eigenstate $u^{(2)}$ (denoted by blue ``$\circ$'') and its approximation $\tilde{u}^{(2)}$ (denoted by red ``$\times$'') are illustrated respectively. The panels (c) and (d) follow the same structure as the top row. The difference is that the bottom row has $k_{3,2}=2k_2$ and $\Delta\theta^{(j)}$ is approximated by $\frac{(j-\frac{1}{2})\pi}{n-1}$.
}
\label{fig:eigenfrequencies_eigenvectors_comparison}
\end{figure}

With the generic boundary stiffness setting $0\not\approx k_{3,1}\not\approx 2k_2\not\approx k_{3,2}\not\approx 0$,  we know from Section~\ref{sec:finite_edgestates} that the existence of edge states are almost the same as that in two semi-infinite chains. 
For the in-band frequencies near the lower edge of the optical band, we have 
$|\tan\beta|=\mathcal{O}(|\Delta\alpha|)\ll 1$ from \eqref{eq:tan_beta} hence $\beta\approx 0$ or $\beta\approx \pi$. Suppose we define
\begin{equation}
\Delta\beta= 
\begin{cases}
\beta, & {\rm if}~~0\leq\beta<\frac{\pi}{2} \\
\beta-\pi, & {\rm if}~~\frac{\pi}{2}<\beta<\pi 
\end{cases}
\end{equation}
then
\begin{equation}
\label{Delta beta}
    \Delta\beta\approx \frac{2k_1+k_{3,1}-2k_2}{2k_2-k_{3,1}}\Delta\alpha.
\end{equation}
Next we notice that \eqref{eq:bc_b2} yields
\begin{equation}
\begin{split}
    \tan((n-1)\Delta\theta)=&\frac{(k_1+k_{3,2}-\omega^2)\sin(\Delta\beta-\Delta\alpha)-k_1\sin(\Delta\alpha+\Delta\beta)}{(k_1+k_{3,2}-\omega^2)\cos(\Delta\beta-\Delta\alpha)-k_1\cos(\Delta\alpha+\Delta\beta)} \\
    \approx &\frac{(k_{3,2}-2k_2)\Delta\beta-(2k_1+k_{3,2}-2k_2)\Delta\alpha}{k_{3,2}-2k_2}\\
    \approx & (\frac{k_{3,1}+2k_1-2k_2}{2k_2-k_{3,1}}+\frac{k_{3,2}+2k_1-2k_2}{2k_2-k_{3,2}})\Delta\alpha
\end{split}
\label{eq:n-1_theta}
\end{equation}
    hence
    \begin{equation}
    (n-1)\Delta\theta \approx m\pi+(\frac{2k_1+k_{3,1}-2k_2}{2k_2-k_{3,1}}+\frac{2k_1+k_{3,2}-2k_2}{2k_2-k_{3,2}})\Delta\alpha,~m\in\mathbb{Z}.
    \end{equation}
    Comparing the form of $\Delta\theta$ with the eigenfrequencies in the spectal band in order, it can be further inferred that
    \begin{equation}
    \label{eq:Delta theta}
    \Delta\theta^{(k)}=\frac{k\pi}{n-1}+\frac{\tilde{\theta}^{(k)}}{n-1} \approx \frac{k\pi}{n-1},~k\ll n
    \end{equation}
near the lower edge of the optical band (see the approximations of eigenfrequencies in panel (a) of Fig.~\ref{fig:eigenfrequencies_eigenvectors_comparison}). Moreover, in the generic case where $(\frac{2k_1+k_{3,1}-2k_2}{2k_2-k_{3,1}}+\frac{2k_1+k_{3,2}-2k_2}{2k_2-k_{3,2}})= \Theta(1)$, expanding \eqref{eq:n-1_theta} at higher orders enables us to obtain
\begin{equation}
\begin{split}
\label{formula:tilde theta k}
\tilde{\theta}^{(k)}
=&(\frac{2k_1+k_{3,1}-2k_2}{2k_2-k_{3,1}}+\frac{2k_1+k_{3,2}-2k_2}{2k_2-k_{3,2}})\frac{k_2}{2(k_2-k_1)}\frac{k\pi}{n-1}\\
&+[(\frac{2k_1+k_{3,1}-2k_2}{2k_2-k_{3,1}}+\frac{2k_1+k_{3,2}-2k_2}{2k_2-k_{3,2}})\frac{k_2}{2(k_2-k_1)}]^2\frac{k\pi}{(n-1)^2}+\Theta(|\frac{k}{n}|^3)
\end{split}
\end{equation}
and particularly
\begin{equation}
|\tilde{\theta}^{(k)}-k\tilde{\theta}^{(1)}|= \Theta(\frac{k^{3}}{n^{3}}), \quad k\ll n.
\label{theta_k}
\end{equation}

\begin{remark} 
\label{remark:eigenvectors_approximations_1}
({\bf Form and approximation of eigenvectors})
According to the periodic structure of diatomic chains and \eqref{eq:v1v2_0}, \eqref{eq:c1c2}, the eigenvector $u^{(k)}$ of $\mathcal{L}$ for eigenvalue $-(\omega^{(k)})^2$ can be explicitly expressed as
\begin{equation}
\label{matrix:uk op}
  \left(
  \begin{array}{c}
    u^{(k)}_{2j-1} \\
    u^{(k)}_{2j}
  \end{array}
  \right)
  =\left(
  \begin{array}{c}
    \cos(\alpha^{(k)}+\beta^{(k)}+(j-1)\theta^{(k)})\\
    -\cos(-\alpha^{(k)}+\beta^{(k)}+(j-1)\theta^{(k)})
  \end{array}
  \right), \quad 1\leq k\leq n_1
\end{equation}
which near the lower edge of the optical band ($k\ll n_1$) can be written as
\begin{equation}
\label{matrix:uk op lower}
\begin{aligned}
  \left(
  \begin{array}{c}
    u^{(k)}_{2j-1} \\
    u^{(k)}_{2j}
  \end{array}
  \right)
  &=(-1)^{j-1}\left(
  \begin{array}{c}
    \sin(\Delta\alpha^{(k)}+\Delta\beta^{(k)}-(j-1)\Delta\theta^{(k)})\\
    \sin(-\Delta\alpha^{(k)}+\Delta\beta^{(k)}-(j-1)\Delta\theta^{(k)})
  \end{array}
  \right), \\
\theta^{(k)}&=\pi-\Delta\theta^{(k)}, \quad \alpha^{(k)}=-\frac{\pi}{2}+\Delta\alpha^{(k)}, \quad \beta^{(k)}=\Delta\beta^{(k)}.
\end{aligned}
\end{equation}
Since $\Delta\theta^{(k)}\approx k\Delta\theta^{(1)}$ when $1\leq k\ll n$, we can approximate $u^{(k)}$ by $\tilde{u}^{(k)}$ (see panel (b) of Fig.~\ref{fig:eigenfrequencies_eigenvectors_comparison} for an example) where 
\begin{equation}
  \left(
  \begin{array}{c}
    \tilde{u}^{(k)}_{2j-1} \\
    \tilde{u}^{(k)}_{2j}
  \end{array}
  \right)
  =(-1)^{j-1}\left(
  \begin{array}{c}
    \sin(k(\Delta\alpha^{(1)}+\Delta\beta^{(1)}-(j-1)\Delta\theta^{(1)}))\\
    \sin(k(-\Delta\alpha^{(1)}+\Delta\beta^{(1)}-(j-1)\Delta\theta^{(1)}))
  \end{array}
  \right), \quad 1\leq k\ll n.
 \label{eq:approx eigenvector}
\end{equation}
In this situation we recall $|\Delta\alpha^{(k)}-k\Delta\alpha^{(1)}|= \Theta(|\Delta\theta^{(k)}|^3)$, $|\Delta\beta^{(k)}-k\Delta\beta^{(1)}|=\Theta(|\Delta\theta^{(k)}|^3)$ and $|\Delta\theta^{(k)}-k\Delta\theta^{(1)}|=\Theta(\frac{k^3}{n^4})$, then obtain 
\begin{equation}
\label{eq:approx eigenvector err}
|\tilde{u}^{(k)}_j-u^{(k)}_j|=\Theta(\frac{k^3}{n^3}).
\end{equation}

\end{remark}

\subsection{Nearly semi-infinite regime}

Assuming $0\not\approx k_{3,1}\not\approx 2k_2\approx k_{3,2}$ leads to another generic scenario discussed in Section~\ref{sec:finite_edgestates} where the existence of edge states resembles that in a semi-infinite chain with a left end. 
Suppose $k_{3,2}=2k_2+\delta k_{3,2}$ and $|\delta k_{3,2}|\ll 1$, then \eqref{eq:n-1_theta} reads
\begin{equation}
\tan((n-1)\Delta\theta)\approx\frac{-2 k_1\Delta\alpha }{ \delta k_{3,2}+2k_1\Delta\alpha\Delta\beta-\frac{k_1 k_2 (\Delta\theta)^2}{2(k_2-k_1)} }
\end{equation}
and
\begin{equation}
(n-1)\Delta\theta\approx m\pi+{\rm arctan}(\frac{-2 k_1\Delta\alpha }{ \delta k_{3,2}+2k_1\Delta\alpha\Delta\beta-\frac{k_1 k_2 (\Delta\theta)^2}{2(k_2-k_1)} }),~m\in\mathbb{Z}.
\end{equation}
This implies near the lower edge of the optical band we have $|\Delta\theta^{(k)}|= \mathcal{O}(\frac{k}{n})$ for $k\ll n$. To be more specific, the form of $\Delta\theta^{(k)}$ can be approximated in the following different situations.
\begin{itemize}
    \item If $\frac{1}{n}\ll |\delta k_{3,2}|\ll 1$, then $|\frac{-2 k_1\Delta\alpha }{ \delta k_{3,2}+2k_1\Delta\alpha\Delta\beta-\frac{k_1 k_2 (\Delta\theta)^2}{2(k_2-k_1)} }|\approx|\frac{-2k_1\Delta\alpha}{\delta k_{3,2}}|\ll 1$ thus $\Delta\theta^{(k)}\approx \frac{k\pi}{n-1}$.
    \item If $|\delta k_{3,2}|\ll \frac{1}{n}$, then $|\frac{-2 k_1\Delta\alpha }{ \delta k_{3,2}+2k_1\Delta\alpha\Delta\beta-\frac{k_1 k_2 (\Delta\theta)^2}{2(k_2-k_1)} }|\gg 1$ thus $\Delta\theta^{(k)}\approx \frac{(k-\frac{1}{2})\pi}{n-1}$.
    \item If $|\delta k_{3,2}|=\Theta (\frac{1}{n})$, then $|\frac{-2 k_1\Delta\alpha }{ \delta k_{3,2}+2k_1\Delta\alpha\Delta\beta-\frac{k_1 k_2 (\Delta\theta)^2}{2(k_2-k_1)} }|\approx|\frac{-2k_1\Delta\alpha}{\delta k_{3,2}}|= \Theta(1)$ thus $\Delta\theta^{(k)}\approx \frac{(k-1)\pi+\gamma}{n-1}$ where $\gamma\approx{\rm arctan}(\frac{-2 k_1\Delta\alpha }{ \delta k_{3,2} })$ is away from $0$ and $\frac{\pi}{2}$.
\end{itemize}
If $|\delta k_{3,2}|\ll \frac{1}{n}$, there is only one eigenfrequency outside the bands and the eigenfrequencies near the lower edge of the optical band satisfy 
\begin{equation}
\label{eq: Delta theta near lower opt for special k31}
\Delta\theta^{(k)}=\frac{(k-\frac{1}{2})\pi}{n-1}+\frac{\tilde{\theta}^{(k)}}{n-1}\approx\frac{(k-\frac{1}{2})\pi}{n-1}, \quad 1\leq k\ll n
\end{equation}
as illustrated in the panel (c) of Fig.~\ref{fig:eigenfrequencies_eigenvectors_comparison}. Moreover, when $|\delta k_{3,2}|$ is even smaller such as $|\delta k_{3,2}|\ll \frac{1}{n^4}$,  we have
\begin{equation}
\begin{split}
\tilde{\theta}^{(k)}&\approx
-(\frac{2k_{2}(2k_{1}+k_{3,1}-2k_{2})}{4k_{1}(k_{1}-k_{2})(2k_{2}-k_{3,1})}-\frac{1}{2})
\frac{(k-\frac{1}{2})\pi}{n-1}\\
 &+(\frac{2k_{2}(2k_{1}+k_{3,1}-2k_{2})}{4k_{1}(k_{1}-k_{2})(2k_{2}-k_{3,1})}-\frac{1}{2})^{2}
\frac{(k-\frac{1}{2})\pi}{(n-1)^{2}}+\Theta(|\frac{k}{n}|^{3})
\end{split}
\end{equation}
and
\begin{equation}
  |\tilde{\theta}^{(k)}-(2k-1)\tilde{\theta}^{(1)}|=\Theta(\frac{k^{3}}{n^{3}})
\end{equation}
Similar to Remark~\ref{remark:eigenvectors_approximations_1}, the eigenvectors $u^{(k)}$ can be written again as \eqref{matrix:uk op} or \eqref{matrix:uk op lower}. However, now $\tilde{u}^{(k)}$ will be defined in a slightly different way.

\begin{remark} 
\label{remark:eigenvectors_approximations_2}
({\bf Approximation of eigenvectors})
Under the setting $0\not\approx k_{3,1}\not\approx 2k_2$ and $|k_{3,2}-2k_2|\ll \frac{1}{n^4}$, we define $\tilde{u}^{(k)}$ as
\begin{equation}
  \left(
  \begin{array}{c}
    \tilde{u}^{(k)}_{2j-1} \\
    \tilde{u}^{(k)}_{2j}
  \end{array}
  \right)
  =(-1)^{j-1}\left(
  \begin{array}{c}
    \sin((2k-1)(\Delta\alpha^{(1)}+\Delta\beta^{(1)}-(j-1)\Delta\theta^{(1)}))\\
    \sin((2k-1)(-\Delta\alpha^{(1)}+\Delta\beta^{(1)}-(j-1)\Delta\theta^{(1)}))
  \end{array}
  \right), \quad 1\leq k\ll n.
 \label{eq:approx eigenvector speci}
\end{equation}
Since $|\Delta\alpha^{(k)}-(2k-1)\Delta\alpha^{(1)}|= \Theta(|\Delta\theta^{(k)}|^3)$, $|\Delta\beta^{(k)}-(2k-1)\Delta\beta^{(1)}|= \Theta(|\Delta\theta^{(k)}|^3)$ and $|\Delta\theta^{(k)}-(2k-1)\Delta\theta^{(1)}|=\Theta(\frac{k^3}{n^4})$, it can be derived that 
\begin{equation}
\label{eq:approx eigenvector err speci}
|\tilde{u}^{(k)}_j-u^{(k)}_j|=\Theta(\frac{k^3}{n^3}).
\end{equation}
A numerical example of the comparison between $u^{(k)}$ and $\tilde{u}^{(k)}$ is provided in panel (d) of Fig.~\ref{fig:eigenfrequencies_eigenvectors_comparison}.
\end{remark}

\subsection{Finite regime}

Suppose $k_{3,1}\approx 2k_2\approx k_{3,2}$, then the finiteness of the chain becomes more effective and the estimate of $\Delta\theta$ heavily depends on the choice of $\{ k_{3,1},k_{3,2}\}$. Here we only list a few examples instead of discussing all situations. Suppose $k_{3,1}=2k_2+\delta k_{3,1}$ and $k_{3,2}=2k_2+\delta k_{3,2}$, then \eqref{eq:tan_beta} now reads $\tan\beta\approx\frac{2k_1}{\frac{k_1 k_2(\Delta\theta)^2}{2(k_2-k_1)}-\delta k_{3,1}}\tan\Delta\alpha$ hence $|\beta|\gg |\Delta\alpha|= \Theta(|\Delta\theta|)$.
\begin{itemize}
    \item Example (I): $|\delta k_{3,1}|\ll \frac{1}{n^2} \gg |\delta k_{3,2}|$\\
    If we write $\beta=\frac{\pi}{2}+\delta\beta$, then $\delta\beta\approx\frac{\Delta\theta}{2}$ or $|\Delta\theta|\ll\frac{1}{n}$. For the former \eqref{eq:n-1_theta} becomes 
    \begin{equation}
    \tan((n-1)\Delta\theta)\approx\frac{ \frac{k_1 k_2 (\Delta\theta)^2}{k_2-k_1} }{ 2k_1\Delta\alpha  }\approx 0
    \end{equation}
    thus $\Delta\theta^{(k)}\approx \frac{k\pi}{n-1}$ or $\Delta\theta^{(k)}\approx \frac{(k-1)\pi}{n-1}$ for $1\leq k\ll n$.
    \item Example (II): $|\delta k_{3,1}|\gg \frac{1}{n} \ll |\delta k_{3,2}|$\\
    This means $|\Delta\alpha|\ll |\Delta\beta|\ll 1$. Then \eqref{eq:n-1_theta} becomes 
    \begin{equation}
    \tan((n-1)\Delta\theta)\approx-\frac{ 2k_1\Delta\alpha-\delta k_{3,2}\Delta\beta }{ \delta k_{3,2}  }\approx 0
    \end{equation}
    thus $\Delta\theta^{(k)}\approx \frac{k\pi}{n-1}$ or $\Delta\theta^{(k)}\approx \frac{(k-1)\pi}{n-1}$ for $1\leq k\ll n$.
    \item Example (III): $|\delta k_{3,1}|\gg \frac{1}{n}$ and $|\delta k_{3,2}|\ll\frac{1}{n^2}$\\
    This is similar to Example (II) but here
    \begin{equation}
    \tan((n-1)\Delta\theta)\approx-\frac{ 2k_1\Delta\alpha }{ \delta k_{3,2}+2k_1\Delta\alpha\Delta\beta }
    \end{equation}
    which leads to $\Delta\theta^{(k)}\approx \frac{(k-\frac{1}{2})\pi}{n-1}$ for $1\leq k\ll n$.
\end{itemize}

\subsection{\texorpdfstring{Section summary ($k_1<k_2$)}{Section summary (k1<k2)}}

In the following Lemma~\ref{lm:op_eigenfrequencies} for optical band and Lemma~\ref{lm:ac_eigenfrequencies} for acoustic band, we summarize the results about the eigenfrequencies near the band edges for representative boundary stiffness settings, which under nonlinear continuation yield nonlinear middle-localized states and nonlinear edge-localized states, respectively.

\begin{lemma}
\label{lm:op_eigenfrequencies}
Consider the eigenfrequencies \eqref{eq:omega_inband} in the optical band ($(\omega^{(k)})^2\in (2k_2, 2k_1+2k_2)$) for a long diatomic chain \eqref{eq:finite},
\begin{itemize}
\item Setting $k_{3,1}\not\approx 2k_2$ and $|k_{3,2}-2k_2|\gg \frac{1}{n}$:
\begin{itemize}
\item When $1\leq k\ll n$ and $\omega^2\approx 2k_2$ (near the lower band edge),
    \begin{equation}
    \label{eq:theta_op_lw}
        \theta^{(k)}\approx \pi-\frac{k}{n-1}\pi.
    \end{equation}
    Moreover, if $k_{3,2}\not\approx 2k_2$ and $(\frac{2k_1+k_{3,1}-2k_2}{2k_2-k_{3,1}}+\frac{2k_1+k_{3,2}-2k_2}{2k_2-k_{3,2}})=\Theta(1)$, 
    then 
    $$ |\tilde{\theta}^{(k)}-k\tilde{\theta}^{(1)}|= \Theta(\frac{k^{3}}{n^{3}}), \quad |\tilde{u}^{(k)}_j-u^{(k)}_j|= \Theta(\frac{k^3}{n^3}), \quad k\ll n. $$
\item When $1\leq k\ll n$ and $\omega^2\approx 2k_1+2k_2$ (near the upper band edge),
    \begin{equation}
    \label{eq:theta_op_up}
        \theta^{(n_1+1-k)}\approx \frac{k}{n-1}\pi.
    \end{equation}
\end{itemize}
\item Setting $k_{3,1}\not\approx 2k_2$ and $|k_{3,2}-2k_2|\ll \frac{1}{n}$:
\begin{itemize}
    \item When $1\leq k\ll n$ and $\omega^2\approx 2k_2$ (near the lower band edge),
    \begin{equation}
    \label{eq:theta_op_lw_2}
        \theta^{(k)}\approx \pi-\frac{k-\frac{1}{2}}{n-1}\pi.
    \end{equation}
    Moreover, if $|k_{3,2}-2k_2|\ll\frac{1}{n^4}$, then     
    $$ |\tilde{\theta}^{(k)}-(2k-1)\tilde{\theta}^{(1)}|= \Theta(\frac{k^{3}}{n^{3}}), \quad |\tilde{u}^{(k)}_j-u^{(k)}_j|= \Theta(\frac{k^3}{n^3}), \quad k\ll n. $$
    \item When $1\leq k\ll n$ and $\omega^2\approx 2k_1+2k_2$ (near the upper band edge),
    \begin{equation}
    \label{eq:theta_op_up_2}
    \theta^{(n_1+1-k)}\approx \frac{k-\frac{1}{2}}{n-1}\pi.
    \end{equation}
\end{itemize}

\end{itemize}
\end{lemma}

\begin{lemma}
\label{lm:ac_eigenfrequencies}
Consider the eigenfrequencies \eqref{eq:omega_inband} in the acoustic band ($(\omega^{((j))})^2\in (0, 2k_1)$) for a long diatomic chain \eqref{eq:finite},
\begin{itemize}
\item Setting $k_{3,1}\not\approx 0$ and $k_{3,2}\gg \frac{1}{n}$:
\begin{itemize}
\item When $1\leq k\ll n$ and $\omega^2\approx 0$ (near the lower band edge),
\begin{equation}
\label{eq:theta_ac_lw}
\theta^{(2n-k+1)}\approx \frac{k}{n-1}\pi.
\end{equation}
\item When $1\leq k\ll n$ and $\omega^2\approx 2k_1$ (near the upper band edge),
    \begin{equation}
    \label{eq:theta_ac_up}
        \theta^{(2n-n_2+k)}\approx \pi-\frac{k}{n-1}\pi.
    \end{equation}
\end{itemize}
\item Setting $k_{3,1}\not\approx 0$ and $k_{3,2}\ll \frac{1}{n}$:
\begin{itemize}
    \item When $1\leq k\ll n$ and $\omega^2\approx 0$ (near the lower band edge),
    \begin{equation}
    \label{eq:theta_ac_lw_2}
        \theta^{(2n-k+1)}\approx \frac{k-\frac{1}{2}}{n-1}\pi.
    \end{equation}
    \item When $1\leq k\ll n$ and $\omega^2\approx 2k_1$ (near the upper band edge),
    \begin{equation}
    \label{eq:theta_ac_up_2}
    \theta^{(2n-n_2+k)}\approx \pi-\frac{k-\frac{1}{2}}{n-1}\pi.
    \end{equation}
\end{itemize}

\end{itemize}
\end{lemma}
According to the results above, the eigenfrequencies inside the bands and near the band edges can be well approximated under different generic boundary stiffness settings. For example, the setting $k_{3,1}\approx k_2\approx k_{3,2}$ leads to $\Delta \theta\approx \frac{k\pi}{n-1}$ for eigenfrequencies near all band edges. On the other hand, keeping the same $k_{3,1}$ and changing the right end to a free end ($k_{3,2}=0$) makes $\Delta \theta\approx \frac{(k-\frac{1}{2})\pi}{n-1}$ near the band edges in the acoustic band.

\begin{figure}[!htp]
\centering
\subfigure[]{
\includegraphics[width=6.5cm, height=5.5cm]{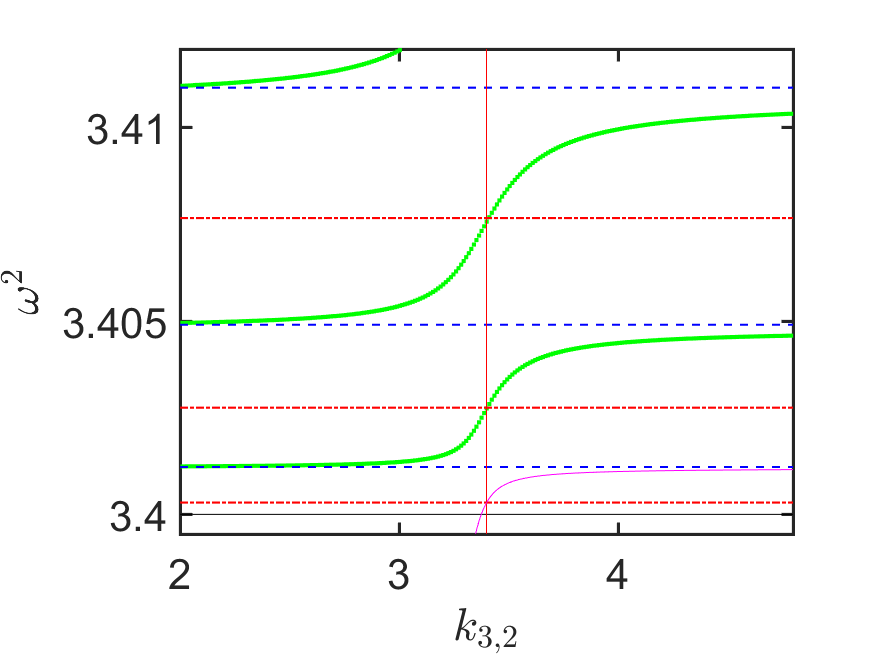}
}
\hspace{2mm}
\subfigure[]{
\includegraphics[height = 5.5cm, width = 6.5cm]{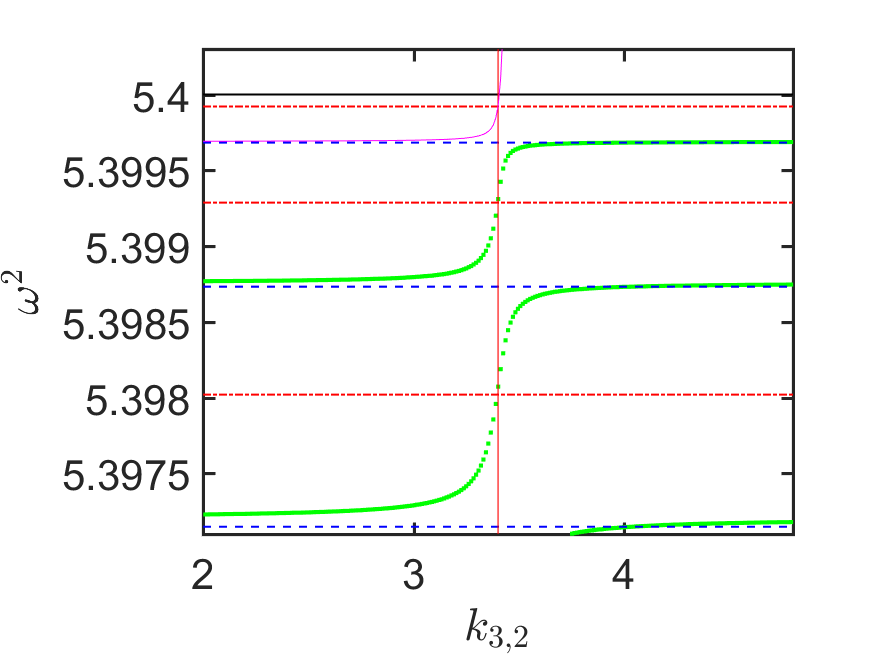}
}
\caption{Here we show the eigenfrequencies $\omega^{2}$ over the change of $k_{3,2}$ near the lower (upper) edge of the optical band in panel (a) (panel (b)) for the chain~\eqref{eq:finite} with $2n=200$, $k_{1}=1$, $k_{2}=1.7$ and $k_{3,1}=1.6$. 
In panel (a), the green dashed curves and magenta solid curve represent the in-band eigenfrequencies and crossing-band-edge eigenfrequency, respectively. At the same time, the red dash-dot lines and blue dashed lines represent $\omega^{2}=2k_{2}+\frac{k_{1}k_{2}}{2(k_{2}-k_{1})}\cdot(\frac{(2k-1)\pi}{2(n-1)})^{2}$ and $\omega^{2}=2k_{2}+\frac{k_{1}k_{2}}{2(k_{2}-k_{1})}\cdot(\frac{k\pi}{n-1})^{2}$, respectively. On the other hand, the red dash-dot lines and blue dashed lines in panel (b) represent $\omega^{2}=2k_{1}+2k_{2}-\frac{k_{1}k_{2}}{2(k_{1}+k_{2})}\cdot(\frac{(2k-1)\pi}{2(n-1)})^{2}$ and $\omega^{2}=2k_{1}+2k_{2}-\frac{k_{1}k_{2}}{2(k_{1}+k_{2})}\cdot(\frac{k\pi}{n-1})^{2}$.
In both panels, the red solid vertical lines represent $k_{3,2}=2k_{2}=3.4$.
}
\label{fig:freq_combine}
\end{figure}

Combining the results from Sec.~\ref{sec:finite_edgestates} and Sec.~\ref{sec:extended_states}, we now have a complete picture showing how an eigenfrequency enters (or exits) the bands and how the eigenfrequencies vary as $k_{3,1}$ (or $k_{3,2}$) changes. In Fig.~\ref{fig:freq_combine}, we show an example of eigenfrequencies near the edges of the optical band for growing $k_{3,2}$ (with estimates from Lemma~\ref{lm:band_edge_states} and Lemma~\ref{lm:op_eigenfrequencies}). 
\begin{itemize}
    \item It can be seen from panel (a) that an eigenfrequency outside the bands crosses the lower band edge ($\omega^2=2k_2$) at $k_{3,2}\approx 2k_2-\frac{k_1 k_2}{n(k_2-k_1)}$ and converges to $\omega^{2}\approx k_1+k_2+\sqrt{ k_1^2+k_2^2-2k_1 k_2\cos{\frac{\pi}{n-1}} }$ as $k_{3,2}$ increases. 
    \item Similarly in panel (b), as $k_{3,2}$ increases, an eigenfrequency near $\omega^{2}\approx k_1+k_2+\sqrt{ k_1^2+k_2^2+2k_1 k_2\cos{\frac{\pi}{n-1}} }$ crosses the upper band edge ($\omega^2=2k_1+2k_2$) at $k_{3,2}\approx 2k_2+\frac{k_1 k_2}{n(k_2+k_1)}$ to exit the bands.
    \item In both panels (a) and (b), the in-band eigenfrequencies near the lower edge change from $\omega^{2}\approx k_1+k_2+\sqrt{ k_1^2+k_2^2-2k_1 k_2\cos{\frac{k\pi}{n-1}} }$ to $\omega^{2}\approx k_1+k_2+\sqrt{ k_1^2+k_2^2-2k_1 k_2\cos{\frac{(k+1)\pi}{n-1}} }$ as $k_{3,2}$ grows.
\end{itemize}

\subsection{Discussion on band inversion}

Although all of the results above in this section are obtained for long chains with $k_1<k_2$, similar strategy can be applied to estimate the in-band eigenfrequencies for the case with $k_1>k_2$. 
For example, suppose the eigenfrequencies in the optical band are ordered as 
\begin{equation}
2k_1<(\omega^{(1)})^2<(\omega^{(2)})^2<\dots<(\omega^{(n_1)})^2<2k_1+2k_2,
\end{equation}
then the in-band eigenfrequencies near the lower edge again can be estimated as in the following Lemma.
\begin{lemma}
\label{lemma:lower_op}
For eigenfrequencies near the lower edge of the optical band ($(\omega^{(k)})^2\approx 2k_1$, $1\leq k\ll n$) for a long diatomic chain~\eqref{eq:finite}:
\begin{itemize}

\item When $k_{3,1}\not\approx 0$, 
\begin{itemize}
    \item If $k_{3,2}\not\approx0$, then $\theta^{(k)}\approx \pi-\frac{k}{n-1}\pi.$

    \item If $0<k_{3,2}\ll\frac{1}{n}$, then $ \theta^{(k)}\approx \pi-\frac{k-\frac{1}{2}}{n-1}\pi.$
    
\end{itemize}
    
\item When $0<k_{3,1}\ll\frac{1}{n}$,
\begin{itemize}
    \item If $0<k_{3,2}\ll\frac{1}{n}$, then $ \theta^{(k)}\approx \pi-\frac{k}{n-1}\pi.$

    \item If $k_{3,2}\not\approx 0$, then $\theta^{(k)}\approx \pi-\frac{k-\frac{1}{2}}{n-1}\pi.$

\end{itemize}
\end{itemize}

\end{lemma}

After band inversion, the in-band eigenfrequencies near the band edges still exhibit predominantly two patterns, a situation similar to the case of $k_{2}>k_{1}$. A detailed discussion is omitted here since the analytical strategy is essentially identical to that for $k_{2}>k_{1}$.

\section{Extended topics on finite-size lattices}
\label{sec:extended topics}

The rigorous asymptotic analysis presented in Sec.~\ref{sec:finite_edgestates} and \ref{sec:extended_states} focused on 1D linear lattices. In this section, we apply these foundational results to some more complex scenarios: nonlinear interactions, multi-layer chains and two-dimensional lattices. Our goal here is not to provide an exhaustive classification, but to illustrate the extensibility of our framework and to show how the linear 1D solutions serve as the 'seed' or 'building block' for understanding these more intricate systems.

\subsection{Nonlinear edge states in nonlinear diatomic chains}

Suppose $q^{(j)}(t)=u^{(j)}e^{i\omega^{(j)}t}$ is an eigenstate of the linear chain \eqref{eq:finite}, it can possibly bifurcate into nonlinear time-periodic solutions for the nonlinear chain satisfying \eqref{eq:finite_nonlinear}
\begin{equation}
  \frac{d^{2}}{dt^{2}}q(t)=\mathcal{L}q(t)+\mathcal{N}(q(t))
  \label{eq:finite_nonlinear}
\end{equation}
where $\mathcal{N}(q)$ represents some nonlinear function with $\mathcal{N}(0)=0$ and $\frac{\partial \mathcal{N}}{\partial q}|_{q=0}=0$. For example, if we consider the chain with an on-site potential such that $\mathcal{N}(q)=b q^3$ with $q^3$ representing the Hadamard product $q\circ q\circ q$, then the Hamiltonian (or energy) of the system \eqref{eq:finite_nonlinear} can be written as 
\begin{equation}
\label{eq:energy}
E=\sum_{j=1}^{2n}[\frac{1}{2}(\dot{q}_j)^2-\frac{b}{4}q_j^4] +\sum_{j=1}^{n}\frac{k_1}{2}(q_{2j-1}-q_{2j})^2+\sum_{j=1}^{n-1}\frac{k_2}{2}(q_{2j+1}-q_{2j})^2
+\frac{k_{3,1}}{2}q_{1}^2+\frac{k_{3,2}}{2}q_{2n}^2.
\end{equation}
Under \textbf{nonresonance condition}, an eigenstate $q^{(j)}$ of \eqref{eq:finite} can be nonlinearly continued into a family of nonlinear time-periodic solutions for \eqref{eq:finite_nonlinear} through Hopf bifurcation. 
\begin{itemize}
\item Typically, the edge states of the linear chain bifurcate into families of boundary-localized time-periodic solutions, or referred as ``nonlinear edge states''. 
\begin{itemize}
    \item In other words, eigenfrequencies away from the bands of the linear chain usually keep distance from the bands in the weakly nonlinear continuation. 
\end{itemize}
\item Intriguingly, the eigenstates of the linear chain with frequencies inside the bands and near the edges can also possibly be continued into nonlinear boundary-localized states. 
\begin{itemize}
    \item That is to say, eigenfrequencies of the linear chain inside the bands may exit the bands through the edges in the nonlinear continuation to generate new nonlinear edge states.
\end{itemize}
\end{itemize}
\begin{figure}[!htp]
\centering
\subfigure[]{
\includegraphics[height = 5.5cm, width = 6.5cm]{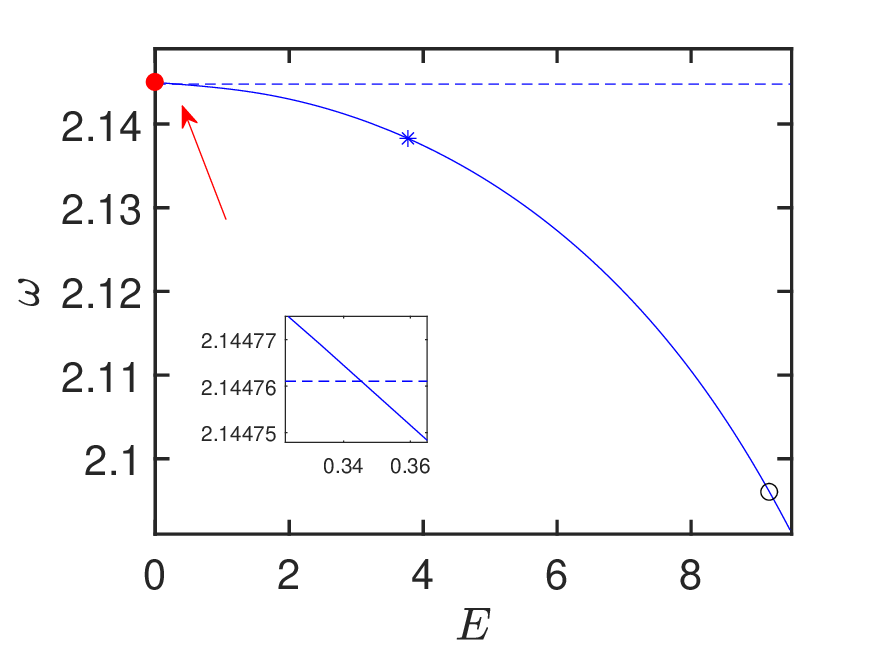}
}
\hspace{2mm}
\subfigure[]{
\includegraphics[height = 5.5cm, width = 6.5cm]{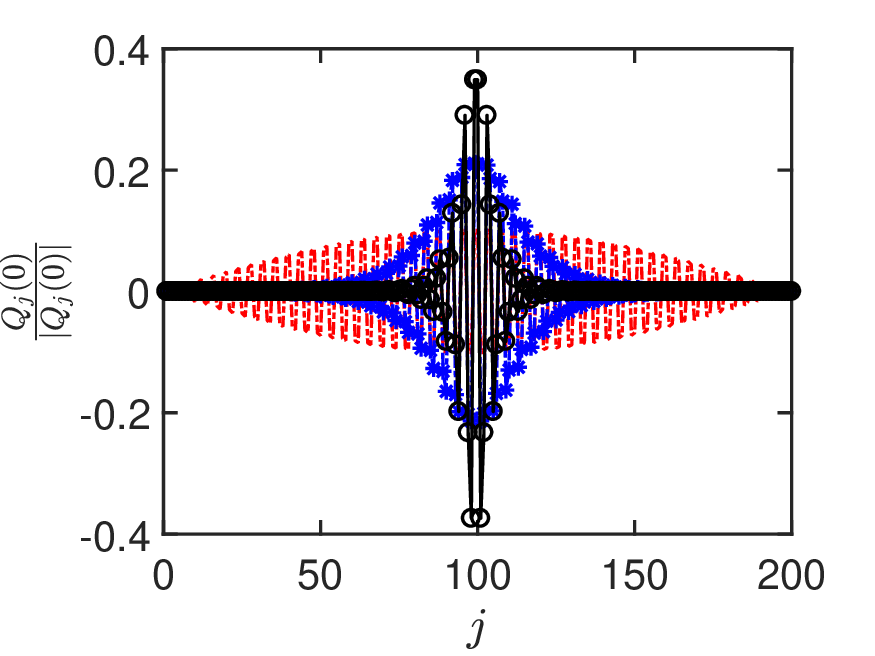}
}

\centering
\subfigure[]{
\includegraphics[height = 5.5cm, width = 6.5cm]{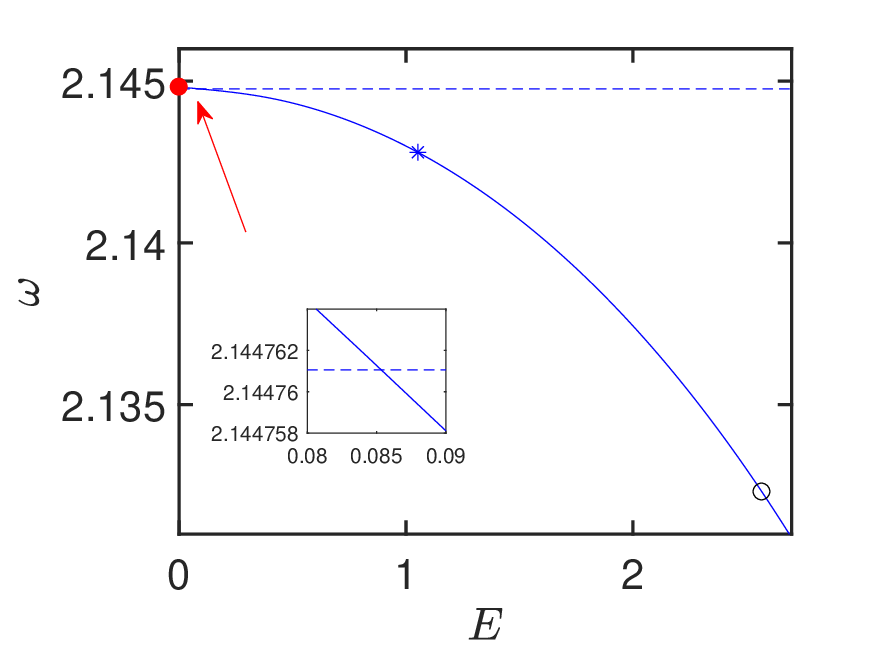}
}
\hspace{2mm}
\subfigure[]{
\includegraphics[height = 5.5cm, width = 6.5cm]{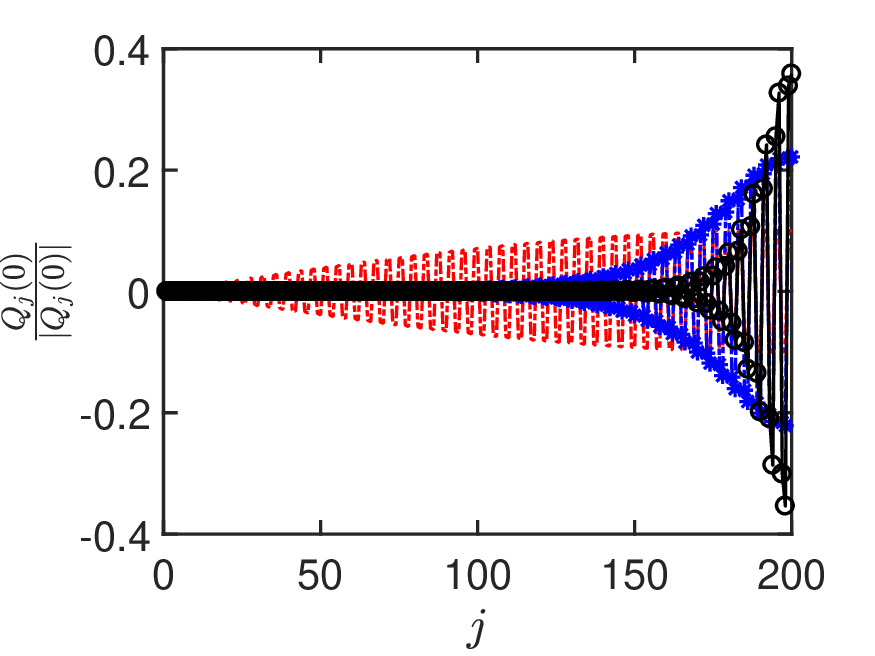}
}
\caption{ Here we show the change of frequency of $q$ in \eqref{eq:finite_nonlinear} continued from $q^{(1)}$ over Hamiltonian ``E'' in panel (a) (panel (c)) and its normalized vibrational modes in panel (b) (panel (d)) with $2n=200$, $k_1=1$,
$k_{2}=2.3$, $k_{3,1}=1.3$ and $k_{3,2}=3.5$ ($k_{3,2}=4.6=2k_{2}$). The normalized modes in panel (b) (panel (d)) are depicted with  red  ``$\cdot-$'', blue ``$\ast-$'', and black ``$\circ-$'', corresponding to three frequencies indicated in panel (a) (panel (c))  by  red solid circle ``$\bullet$'', blue ``$\ast$'', and black ``$\circ$'' respectively, where a gradual enhancement of locality is observed in the middle (boundary) area.}
\label{middle_edge_localized}
\end{figure} 
Although the nonlinear edge states continued from linear edge states are commonly observed with a relatively obvious explanation, the mechanism for the emergence of nonlinear edge states continued from linear extended states is still not fully decoded. Therefore, we place the emphasis on the latter in this subsection to show a powerful nonlinear application of our analytical framework.

Recent works~\cite{Chaunsali2023} provided an elegant description of nonlinear edge states in the small-bandgap limit~\cite{Chaunsali2019} using a continuum Dirac equation, interpreting them as trimmed bulk solitons. Our approach complements this by working directly with the discrete lattice for arbitrary bandgaps. Furthermore, our method highlights a distinct mechanism: nonlinear edge states can originate from linear extended modes whose frequencies are pushed into the bandgap by nonlinearity, a phenomenon naturally captured by continuing the discrete spectrum of the finite system.

Focusing on time-periodic solutions, we 
define $\tau=\omega t$ and $Q(\tau)=q(\frac{\tau}{\omega})=q(t)$.
Then the system \eqref{eq:finite_nonlinear} with on-site cubic nonlinearity becomes
\begin{equation}
  \omega^{2}\frac{d^{2}}{d\tau^{2}}Q(\tau)=\mathcal{L}Q(\tau)+b(Q(\tau))^{3}.
  \label{eq:finite_cubic_tau}
\end{equation}
\begin{remark}
Let $\omega^{(1)}$ denote the eigenfrequency in the optical band and near its lower edge ($\omega^{(1)}\approx\sqrt{2k_2}$) as in \eqref{eq:omega_order_1}, then the nonresonance condition \eqref{eq:nr_condition} holds
\beq
\label{eq:nr_condition}
\frac{\omega^{(j)}}{\omega^{(1)}}\not\in\mathbb{Z}, \quad 2\leq j\leq n
\eeq
for its nonlinear continuation since $2\omega^{(1)}\approx 2\sqrt{2k_2}>\sqrt{2k_1+2k_2}$. By the Implicit Function Theorem and Lyapunov-Schmidt Reduction, it can be proved that $q^{(1)}$ bifurcates into a family of time-periodic solutions in \eqref{eq:finite_cubic_tau} as the strength of nonlinearity (amplitude) increases. 
\end{remark}

Numerical observations reveal that as nonlinearity gradually increases, the solution's frequency can progressively detach from the spectral band and enter the bandgap (for $b>0$). Simultaneously, the corresponding nonlinear time-periodic solution exhibits enhanced localization, as illustrated in Fig.~\ref{middle_edge_localized}. In panels (a) and (b) of Fig.~\ref{middle_edge_localized} we employ the boundary coefficients $|k_{3,1}-k_2|\not\approx k_2 \not\approx |k_{3,2}-k_2|$ (belonging to the ``nearly infinite regime'') and witnessed the formation of nonlinear middle-localized states. On the other hand, the example in panels (c) and (d) in Fig.~\ref{middle_edge_localized} is equipped with $|k_{3,1}-k_2|\not\approx k_2 \approx k_{3,2}-k_2$ (belonging to the ``nearly semi-infinite regime'') and demonstrates the emergence of nonlinear boundary-localized states. These nonlinear middle-localized and boundary-localized states are exclusive and common for nonlinear finite chains. For instance, it can be directly checked that such nonlinear localized states also exist in finite chains with FPU-type nonlinearity instead of on-site cubic nonlinearity (results not shown here). 

Based on Lemma~\ref{lm:op_eigenfrequencies}, we can not only predict the emergence of these nonlinear localized states from the boundary setup, but also theoretically estimate the nonlinearly continued solutions $Q$ and frequency $\omega$ from the linear limit. However, since the analytical derivation for explaining the nonlinear continuation is nontrivial and beyond the scope of this work, we will leave the detailed discussion in a future report.

\subsection{Multi-layer diatomic chains}
\label{sec:multi_layer}

\begin{figure}[!htp]
\centering
\vspace{-0.2cm}  
\includegraphics[height = 2.5cm, width = 13.5cm]{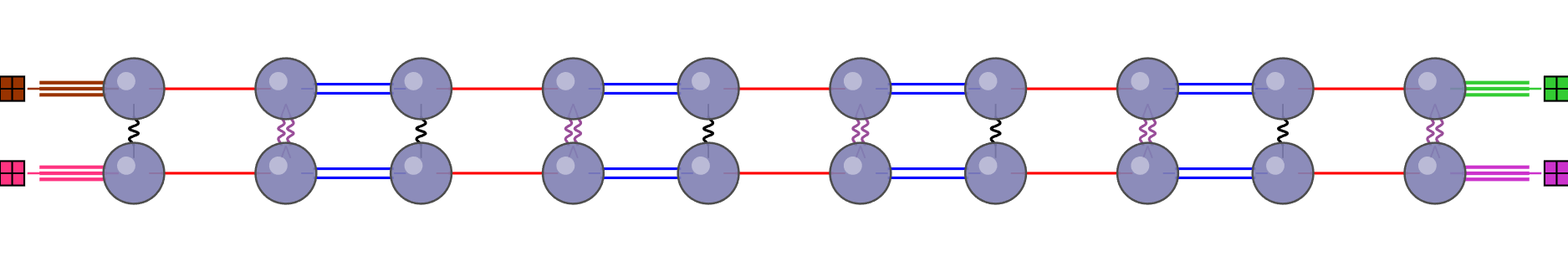}
\vspace{-0.5cm}  
\caption{Here we show a schematic of the two-layer diatomic chains \eqref{eq:two_layer_finite}. The interatomic couplings are represented as follows: $k_{1}$
  (red single lines, ``$-$"), $k_{2}$
  (blue double lines, ``="), $k_{5}$
  (black single curves, ``$\sim$"), and 
$k_{6}$ (purple double curves, ``$\approx$"). The boundary couplings at the left and right ends are shown as triple lines (``$\equiv$") and denoted by $k_{3,j}$ and $k_{4,j}$ ($j=1,2$), respectively.}
\vspace{-0.2cm}  
\label{fig:two_layer}
\end{figure}

\begin{figure}[!htp]
\centering
\vspace{-0.2cm}  
\includegraphics[height = 2.5cm, width = 13.5cm]{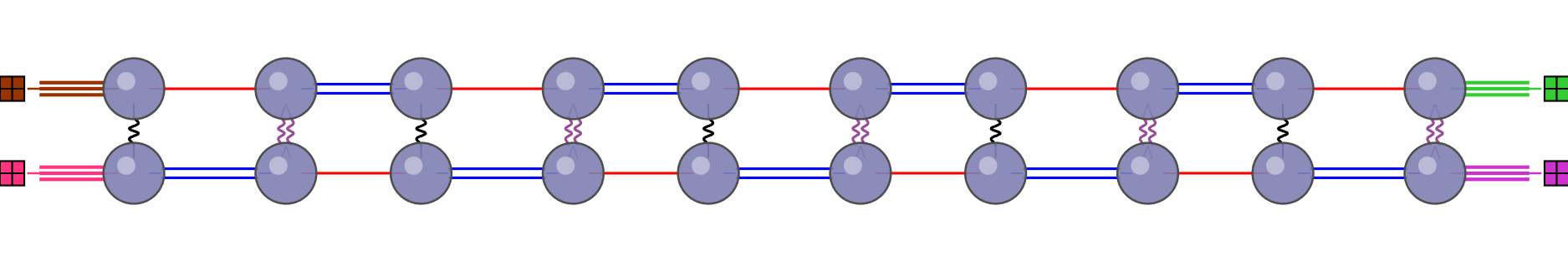}
\vspace{-0.5cm}  
\caption{This is the schematic of another setting of two-layer diatomic chains. The line styles (red single solid, blue double solid, black single curve, purple double curve, triple solid) are adopted from Fig.~\ref{fig:two_layer}.}
\vspace{-0.2cm}  
\label{fig:two_layer_finite_2}
\end{figure}

Before moving from one-dimensional chains to two-dimensional lattices, we first consider an intermediate circumstance with multiple parallel coupled chains. To be more specific, we can start from the following two-layer diatomic chain of length $2n$:
\begin{equation}
\begin{split}
   \ddot{q}_{1,1} &= k_{3,1}(0-q_{1,1})+k_{1}(q_{1,2}-q_{1,1})+k_5(q_{2,1}-q_{1,1}), \\
  \ddot{q}_{2,1} &= k_{3,2}(0-q_{2,1})+k_{1}(q_{2,2}-q_{2,1})+k_5(q_{1,1}-q_{2,1}), \\
  \ddot{q}_{1,2n} &= k_{1}(q_{1,2n-1}-q_{1,2n})+k_{4,1}(0-q_{1,2n})+k_6(q_{2,2n}-q_{1,2n}), \\
  \ddot{q}_{2,2n} &= k_{1}(q_{2,2n-1}-q_{2,2n})+k_{4,2}(0-q_{2,2n})+k_6(q_{1,2n}-q_{2,2n}), \\
  \ddot{q}_{1,2m} &= k_{1}(q_{1,2m-1}-q_{1,2m})+k_{2}(q_{1,2m+1}-q_{1,2m})+k_6(q_{2,2m}-q_{1,2m}), \\
  \ddot{q}_{2,2m} &= k_{1}(q_{2,2m-1}-q_{2,2m})+k_{2}(q_{2,2m+1}-q_{2,2m})+k_6(q_{1,2m}-q_{2,2m}), \\
  \ddot{q}_{1,2m+1} &= k_{2}(q_{1,2m}-q_{1,2m+1})+k_{1}(q_{1,2m+2}-q_{1,2m+1})+k_5(q_{2,2m+1}-q_{1,2m+1}), \\
  \ddot{q}_{2,2m+1} &= k_{2}(q_{2,2m}-q_{2,2m+1})+k_{1}(q_{2,2m+2}-q_{2,2m+1})+k_5(q_{1,2m+1}-q_{2,2m+1}), \quad 1\leq m\leq n-1
\end{split}
\label{eq:two_layer_finite}
\end{equation}
where $q_{j,k}$ represents the displacement of the atom located at the $j$-th row (layer) and $k$-th column, $\{ k_1, k_2 \}$ are the alternating linear coefficients for horizontal interactions, $\{ k_{5}, k_6 \}$ are the linear coefficients for vertical (inter-layer) interactions, $\{ k_{3,1}, k_{3,2}, k_{4,1}, k_{4,2} \}$ are the linear coefficients for boundary interactions. With a schematic shown in Fig.~\ref{fig:two_layer}, \eqref{eq:two_layer_finite} is just a simple example of multi-layer chains and other multi-layer models (e.g. the example in Fig.~\ref{fig:two_layer_finite_2}) can be similarly studied.

Similar to \eqref{eq:iteration}, we assume $q_{j, k}(t)=u_{j, k}e^{i\omega t}$ and obtain the iteration relation between adjacent unit cells.
\begin{equation}
   \left(
  \begin{array}{c}
    u_{1,2m+1} \\
    u_{2,2m+1} \\
    u_{1,2m+2} \\
    u_{2,2m+2}
  \end{array}
  \right)
  =
  \mathcal{T}(\omega)
   \left(
  \begin{array}{c}
    u_{1,2m-1} \\
    u_{2,2m-1} \\
    u_{1,2m}\\
    u_{2,2m}
  \end{array}
  \right)
  \label{eq:two_layer_iteration}
\end{equation}
where
\beq
\mathcal{T}(\omega)=
\left(\begin{array}{cccc}
    k_{2} & 0 & 0 & 0 \\
     0 &    k_2 & 0 & 0 \\
     \omega^2-k_1-k_2-k_5 & k_5 & k_1 & 0 \\
     k_5 & \omega^2-k_1-k_2-k_5 & 0 & k_1 
  \end{array}\right)^{-1}
  \left(\begin{array}{cccc}
    -k_{1} & 0 & k_1+k_2+k_6-\omega^2 & -k_6 \\
     0 &    -k_1 & -k_6 & k_1+k_2+k_6-\omega^2 \\
     0 & 0 & -k_2 & 0 \\
     0 & 0 & 0 & -k_2 
  \end{array}\right).
\eeq
Let the eigenvalues of $\mathcal{T}$ be denoted by $\{ a_1, \frac{1}{a_1}, a_2, \frac{1}{a_2} \}$ (where $|a_j|\leq 1$, $j=1,2$), then the corresponding eigenvectors are 
\beq
v_1=v(a_1)=\frac{1}{2}\left(\begin{array}{c}
    -\sigma_1 \\
    -\sigma_1  \\
     \frac{k_1+k_2 a_1}{\sqrt{ (k_1+k_2 a_1)(k_1+k_2/a_1) }} \\
     \frac{k_1+k_2 a_1}{\sqrt{ (k_1+k_2 a_1)(k_1+k_2/a_1) }} 
  \end{array}\right), 
\quad v_2=v(\frac{1}{a_1})=\frac{1}{2}\left(\begin{array}{c}
    -\sigma_1 \\
    -\sigma_1 \\
     \frac{k_1+k_2/ a_1}{\sqrt{ (k_1+k_2 a_1)(k_1+k_2/a_1) }} \\
     \frac{k_1+k_2/ a_1}{\sqrt{ (k_1+k_2 a_1)(k_1+k_2/a_1) }} 
  \end{array}\right)
\eeq
for
\beq
\label{eq:two_layer_omega_1}
\omega^2=k_1+k_2+\sigma_1\sqrt{(k_1+k_2 a_1)(k_1+ k_2/a_1)}
\eeq
and
\begin{eqnarray}
\begin{split}
v_3&=v(a_2)=\frac{1}{2d}
\left(\begin{array}{c}
    -[(k_5-k_6)+\sigma_2\sqrt{(k_5-k_6)^2+(k_1+k_2 a_2)(k_1+k_2/a_2)}]\\
     (k_5-k_6)+\sigma_2\sqrt{(k_5-k_6)^2+(k_1+k_2 a_2)(k_1+k_2/a_2)} \\
     k_1+k_2 a_2 \\
    -(k_1+k_2 a_2)
  \end{array}\right), \\
v_4&=v(\frac{1}{a_2})=\frac{1}{2d}
\left(\begin{array}{c}
    -[(k_5-k_6)+\sigma_2\sqrt{(k_5-k_6)^2+(k_1+k_2 a_2)(k_1+k_2/a_2)}]\\
     (k_5-k_6)+\sigma_2\sqrt{(k_5-k_6)^2+(k_1+k_2 a_2)(k_1+k_2/a_2)} \\
     k_1+k_2 /a_2 \\
    -(k_1+k_2 /a_2)
  \end{array}\right)
\end{split}
\end{eqnarray}
where  
$$d=\sqrt{ (k_5-k_6)^2 +(k_1+k_2 a_2)(k_1+k_2/a_2)+\sigma_2(k_5-k_6)\sqrt{(k_5-k_6)^2+(k_1+k_2 a_2)(k_1+k_2/a_2)}  }$$
for
\beq
\label{eq:two_layer_omega_2}
\omega^2=k_1+k_2+k_5+k_6+\sigma_2\sqrt{(k_5-k_6)^2+(k_1+k_2 a_2)(k_1+ k_2/a_2)}.
\eeq

\begin{figure}[!htp]
\centering
\subfigure[]{
\includegraphics[height = 5.5cm, width = 6.5cm]{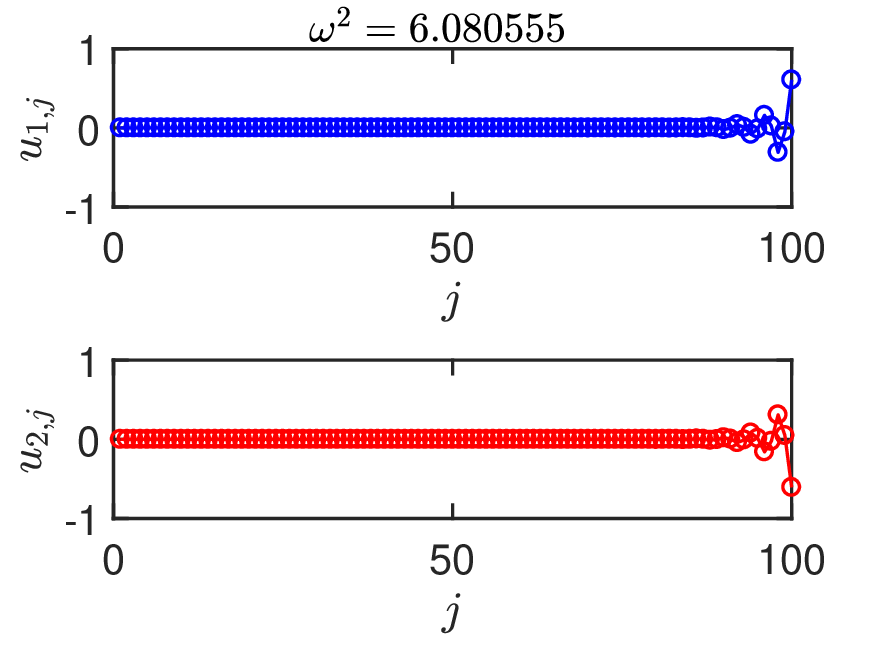}
}
\hspace{2mm}
\subfigure[]{
\includegraphics[height = 5.5cm, width = 6.5cm]{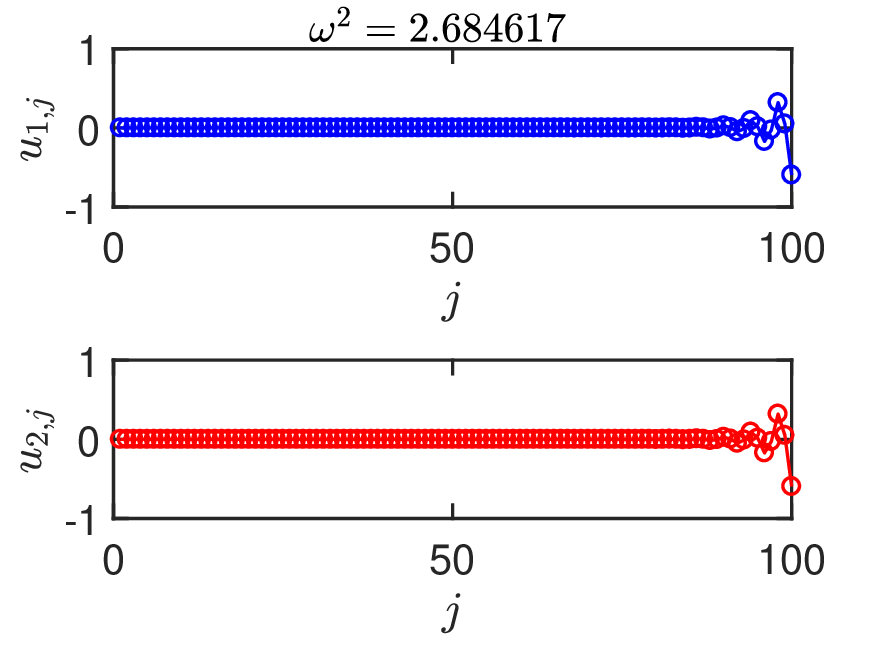}
}

\centering
\subfigure[]{
\includegraphics[height = 5.5cm, width = 6.5cm]{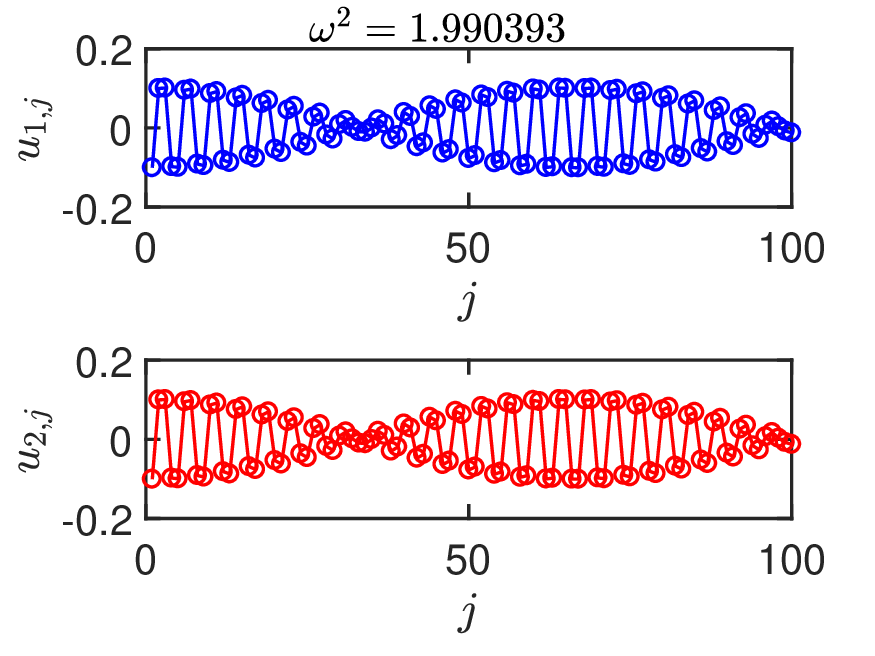}
}
\hspace{2mm}
\subfigure[]{
\includegraphics[height = 5.5cm, width = 6.5cm]{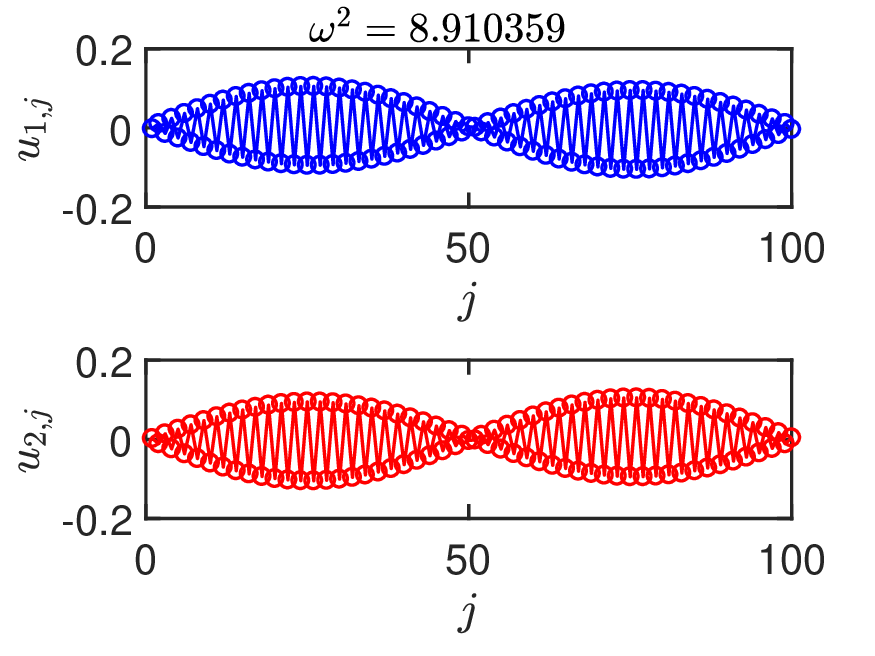}
}
\caption{For the two-layer model~\eqref{eq:two_layer_finite} (Fig.~\ref{fig:two_layer}) with 
$k_{1}=1, k_{2}=1.9, 
k_{5}=1.4, k_{6}=1.7, k_{3,1}=k_{3,2} =0$ and 
$k_{4,1}=k_{4,2}=1.6$, panels (a) and (b) depict two representative edge states. On the other hand, panels (c) and (d) show two representative extended states in the same model with frequencies near band edges.}
\label{2layer_states}
\end{figure}

In the same spirit of Sec.~\ref{sec:finite_edgestates} and \ref{sec:extended_states}, we can obtain the following observations of eigenstates in two-layer diatomic chains.
\begin{itemize}
\item According to \eqref{eq:two_layer_omega_1} and \eqref{eq:two_layer_omega_2}, this two-layer diatomic chain has at most four bands (two pairs): $[0,2k_1]\cup [2k_2,2k_1+2k_2]$ and $[k_1+k_2+k_5+k_6-\sqrt{(k_5-k_6)^2+(k_1+k_2)^2}, k_1+k_2+k_5+k_6-\sqrt{(k_5-k_6)^2+(k_1-k_2)^2}]\cup [k_1+k_2+k_5+k_6+\sqrt{(k_5-k_6)^2+(k_1-k_2)^2}, k_1+k_2+k_5+k_6+\sqrt{(k_5-k_6)^2+(k_1+k_2)^2}]$, which may overlap each other. An eigenstate is localized (edge state) if its frequency is outside at least one pair of bands. 
\item In semi-infinite two-layer diatomic chains, 
left edge states exist if $  \left(\begin{array}{c}
    u_{1,1} \\
    u_{2,1} \\
    u_{1,2}\\
    u_{2,2}
  \end{array}\right)=c_1 v_1+c_3 v_3$ (namely $c_2=c_4=0$). It can be seen that the number of left edge states mainly depends on $\{ k_{3,1}, k_{3,2}\}$ and ${\rm sgn}(k_1-k_2)$ but not ${\rm sgn}(k_5-k_6)$.
\item Zak phase for bands $[0,2k_1]\cup [2k_2, 2k_1+2k_2]$ changes as ${\rm sgn}(k_1-k_2)$ changes. If $k_5=k_6$, then the Zak phase for the other two bands also depends on ${\rm sgn}(k_1-k_2)$. However, ${\rm sgn}(k_5-k_6)$ does not affect the Zak phase for any band.
\item In finite (but long) two-layer diatomic chains, edge states commonly exist and their number depends on boundary coefficients and ${\rm sgn}(k_1-k_2)$ (Fig.~\ref{2layer_states} (a) and (b)). 
\item In finite (but long) two-layer diatomic chains, eigenfrequencies in the bands and near the band edges follow special patterns (Fig.~\ref{2layer_states} (c) and (d)).
\end{itemize}

\subsection{Two-dimensional diatomic lattices}

\begin{figure}[!htp]
\centering
\subfigure[]{
\includegraphics[height=4cm, width=5cm]{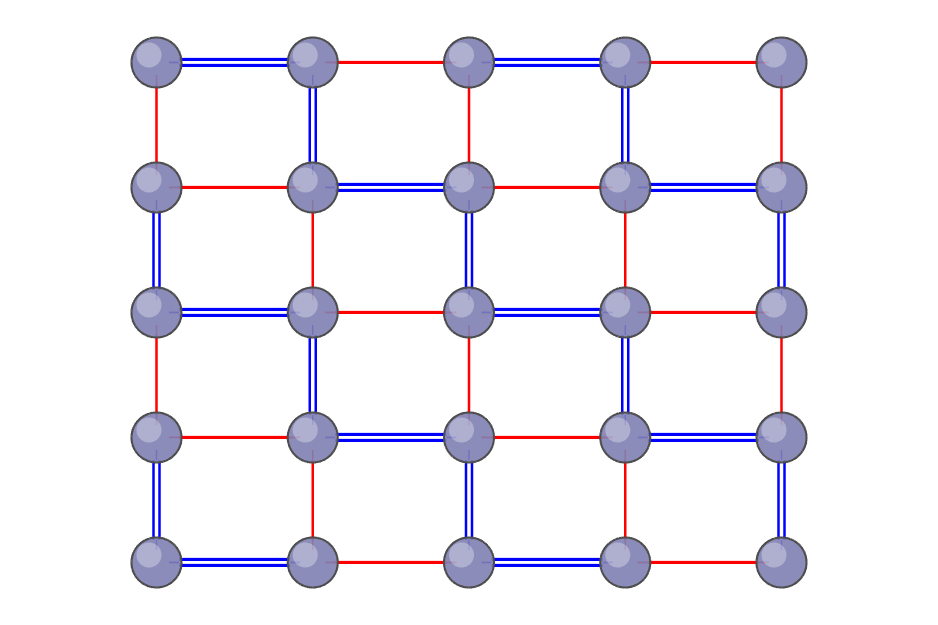}
}
\subfigure[]{
\includegraphics[height = 4cm, width = 5cm]{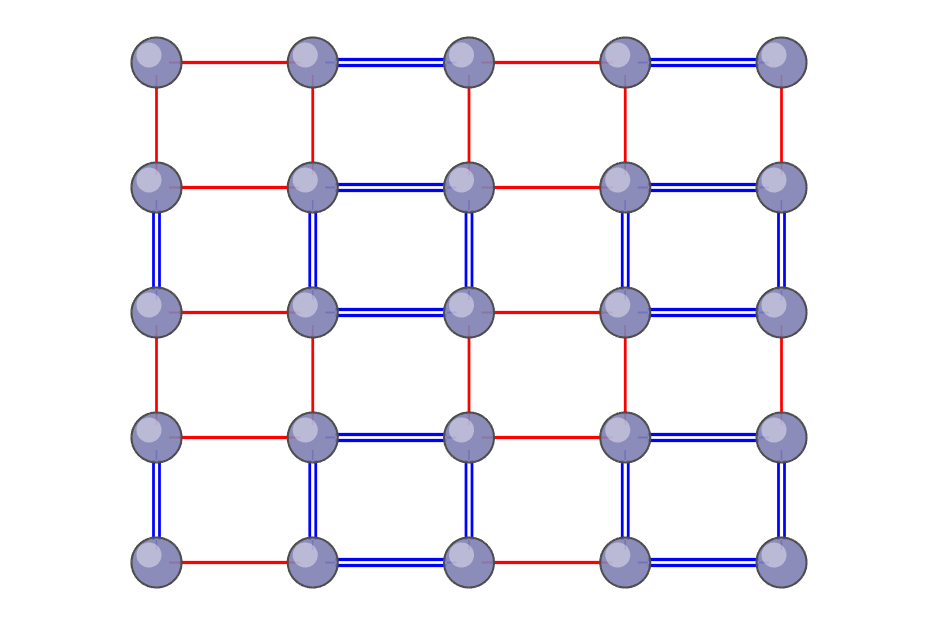}
}
\subfigure[]{
\includegraphics[height = 4cm, width = 5cm]{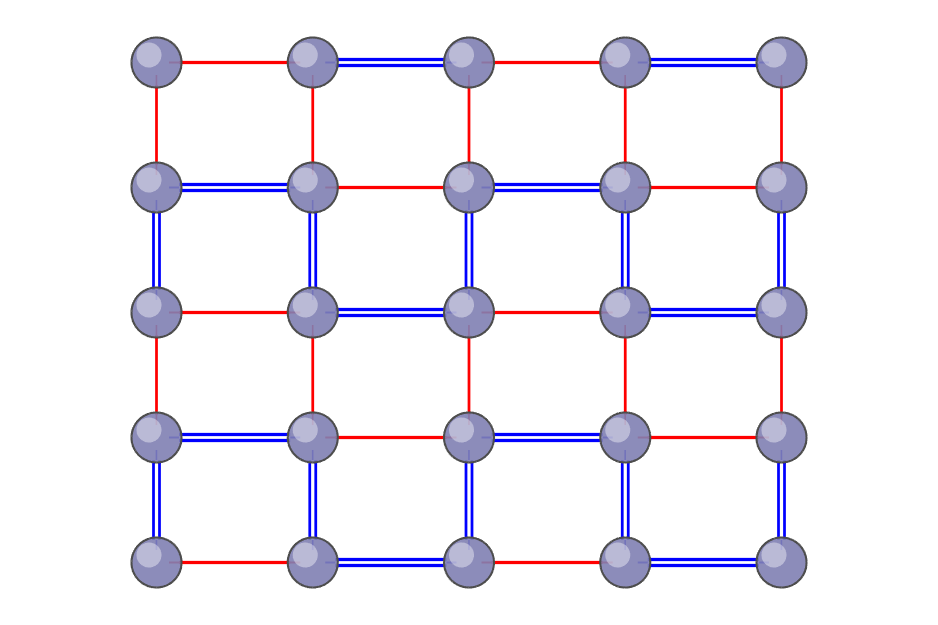}
}
\caption{
We show three different configurations of 2-D diatomic lattices. The two interatomic interactions are depicted with red single straight lines (``$-$") and blue double straight lines (``="), corresponding to parameters 
$k_{1}$ and $k_{2}$, respectively.
}
\label{fig:2d}
\end{figure}

After extending single-layer chains to multi-layer chains in Sec.~\ref{sec:multi_layer}, we now shift gear to investigate two-dimensional diatomic lattices and demonstrate the universality of our framework. Assuming each row and column is a diatomic chain, we can obtain 2-D diatomic lattices with different structures (with examples shown in Fig.~\ref{fig:2d}). Among them, the setting in Fig.~\ref{fig:2d} (a) is specifically considered in this section and it satisfies the following equations~\eqref{eq:2d}.
\begin{equation}
\begin{split}
  \ddot{q}_{2l-1,2m-1} = & k_{2}(q_{2l-1,2m-2}+q_{2l,2m-1})+k_{1}(q_{2l-1,2m}+q_{2l-2,2m-1})-(2k_1+2k_2)q_{2l-1,2m-1}, \\
  \ddot{q}_{2l-1,2m} = & k_{2}(q_{2l-1,2m+1}+q_{2l-2,2m})+k_{1}(q_{2l-1,2m-1}+q_{2l,2m})-(2k_1+2k_2)q_{2l-1,2m}, \\
  \ddot{q}_{2l,2m-1} = & k_2(q_{2l,2m}+q_{2l-1,2m-1})+k_1(q_{2l,2m-2}+q_{2l+1,2m-1})-(2k_1+2k_2)q_{2l,2m-1}, \\
  \ddot{q}_{2l,2m} = & k_{2}(q_{2l,2m-1}+q_{2l+1,2m})+k_{1}(q_{2l,2m+1}+q_{2l-1,2m})-(2k_1+2k_2)q_{2l,2m}, 
\end{split}
\label{eq:2d}
\end{equation}

If we take $( q_{2l-1,2m-1}, q_{2l-1,2m} )$ as a unit cell, then the lattice can be generated by its translations in the direction of $( q_{2l-2,2m}, q_{2l-2,2m+1} )$ or $(q_{2l,2m}, q_{2l, 2m+1} )$. Suppose the adjacent unit cells satisfy
$$
( q_{2l-2,2m}, q_{2l-2,2m+1} )=a_1( q_{2l-1,2m-1}, q_{2l-1,2m} ), \quad (q_{2l,2m}, q_{2l, 2m+1} )=a_2( q_{2l-1,2m-1}, q_{2l-1,2m} )
$$
 with $|a_1|\leq 1 \geq |a_2|$ (while
$$( q_{2l-1,2m+1}, q_{2l-1,2m+2} )=a_1 a_2( q_{2l-1,2m-1}, q_{2l-1,2m} ), \quad ( q_{2l+1,2m-1}, q_{2l+1,2m} )=a_1^{-1} a_2( q_{2l-1,2m-1}, q_{2l-1,2m} )$$
under this setting) 
and $q_{j,k}(t)=e^{i\omega t}u_{j,k}$, then the eigenfrequency can be expressed as
$$
\omega^2=2k_1+2k_2+\sigma\sqrt{(k_1+k_2 a_1)(k_1+k_2/a_1)(1+a_2)(1+a_2^{-1})}
$$ 
with eigenvectors for $( u_{2l-1,2m-1}, u_{2l-1,2m} )$ being
$$
v_{j_1, j_2}=
\frac{1}{\sqrt{2}}(-\sigma, \frac{ (k_1+k_2 a_1^{(-1)^{j_1+1}})(1+a_2^{(-1)^{j_2+1}}) }{\sqrt{(k_1+k_2 a_1)(k_1+k_2/a_1)(1+a_2)(1+a_2^{-1})} }), \quad 1\leq j_1\leq 2, 1\leq j_2\leq 2
$$
\begin{figure}[!htp]
\centering
\subfigure[]{
\includegraphics[height = 4.cm, width = 5cm]{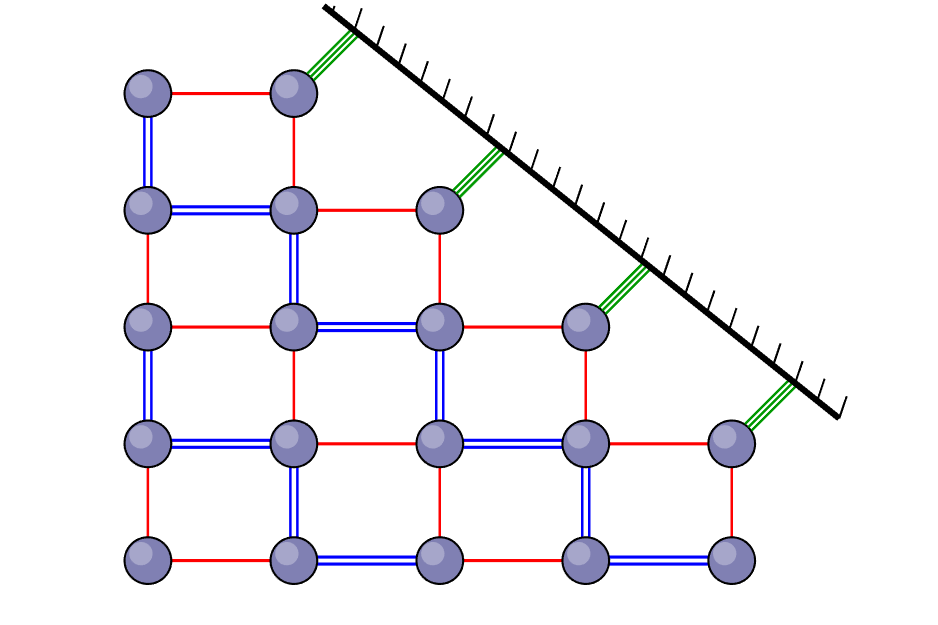}
}
\subfigure[]{
\includegraphics[height=4cm, width=5cm]{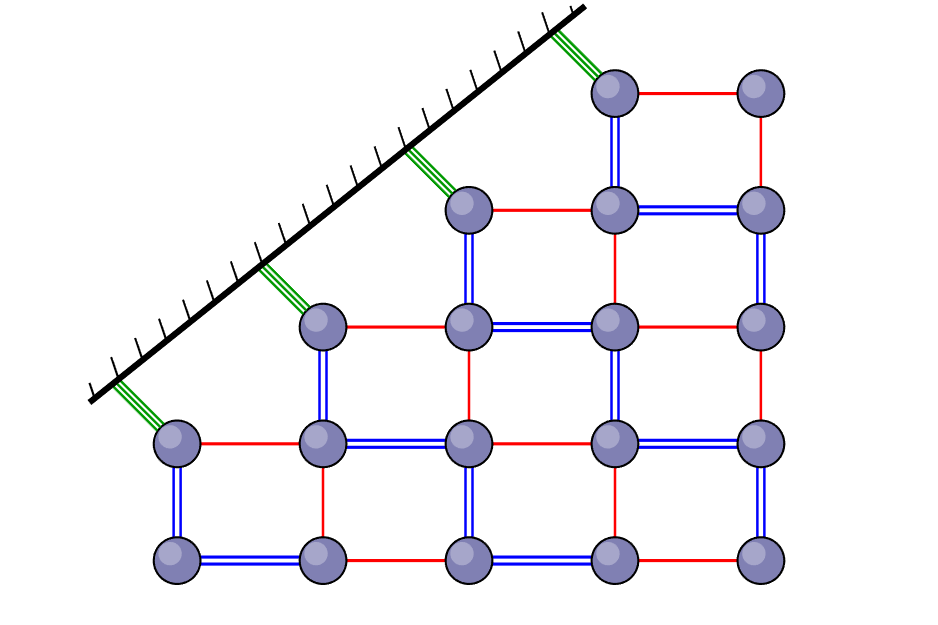}
}
\subfigure[]{
\includegraphics[height = 4cm, width = 5cm]{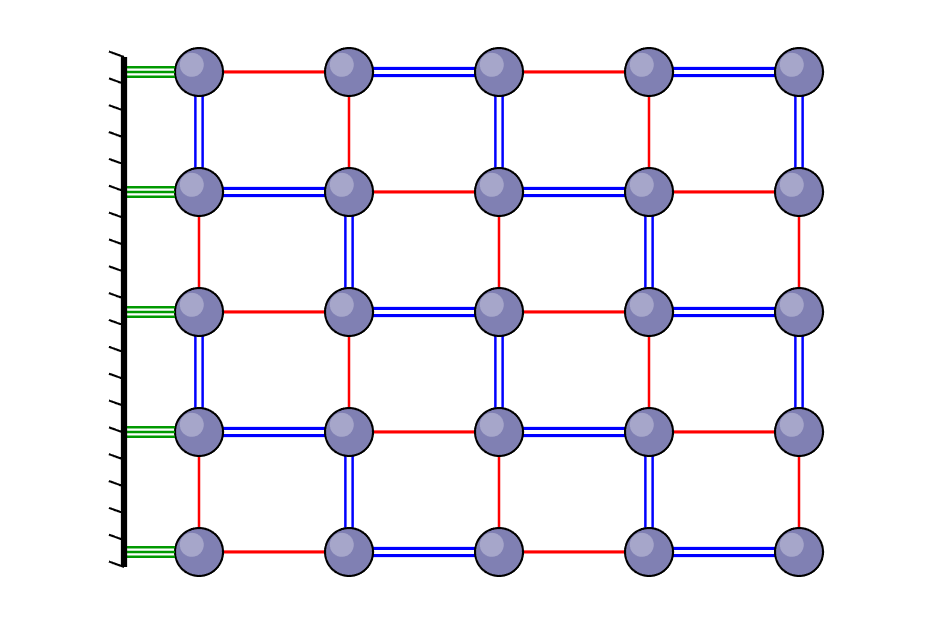}
}
\caption{Here are illustrations of three representative boundaries for the 2-D diatomic lattice~\eqref{eq:2d} (Fig.~\ref{fig:2d} (a)). 
}
\label{fig:2d_edge}
\end{figure}

\begin{figure}[!htp]
\centering
\subfigure[]{
\includegraphics[height = 5.5cm, width = 6.5cm]{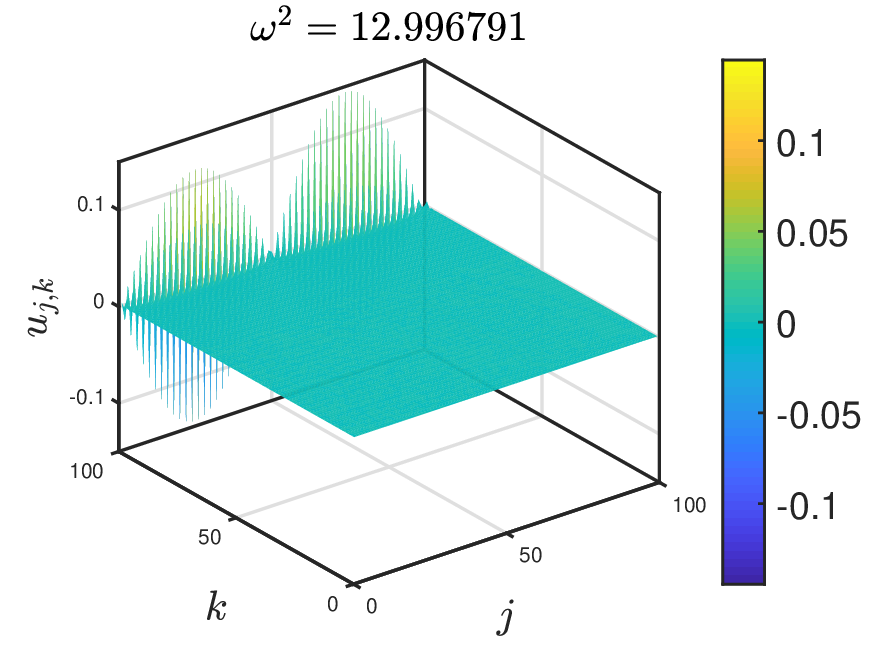}
}
\hspace{2mm}
\subfigure[]{
\includegraphics[height = 5.5cm, width = 6.5cm]{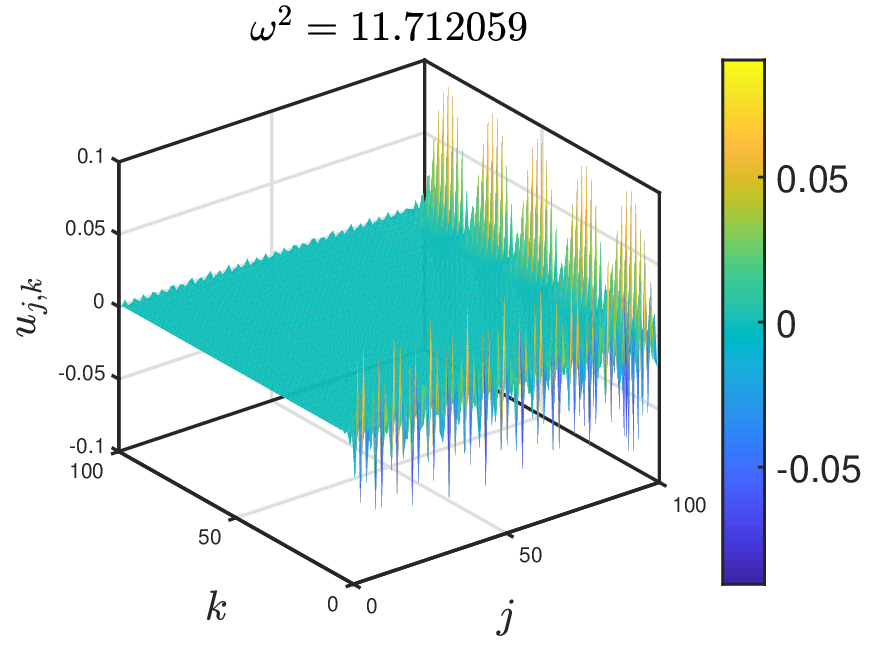}
}
\centering
\subfigure[]{
\includegraphics[height = 5.5cm, width = 6.5cm]{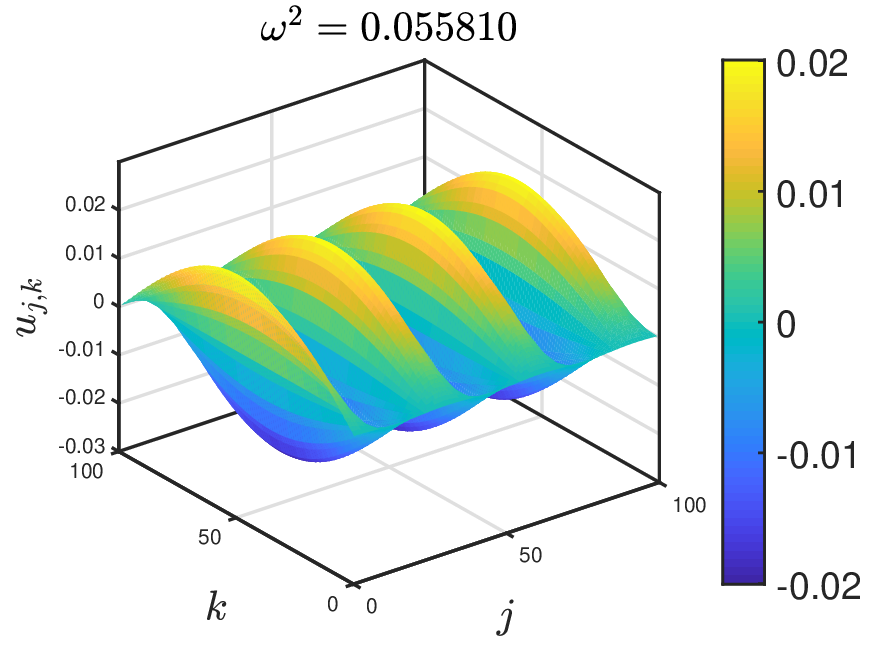}
}
\hspace{2mm}
\subfigure[]{
\includegraphics[height = 5.5cm, width = 6.5cm]{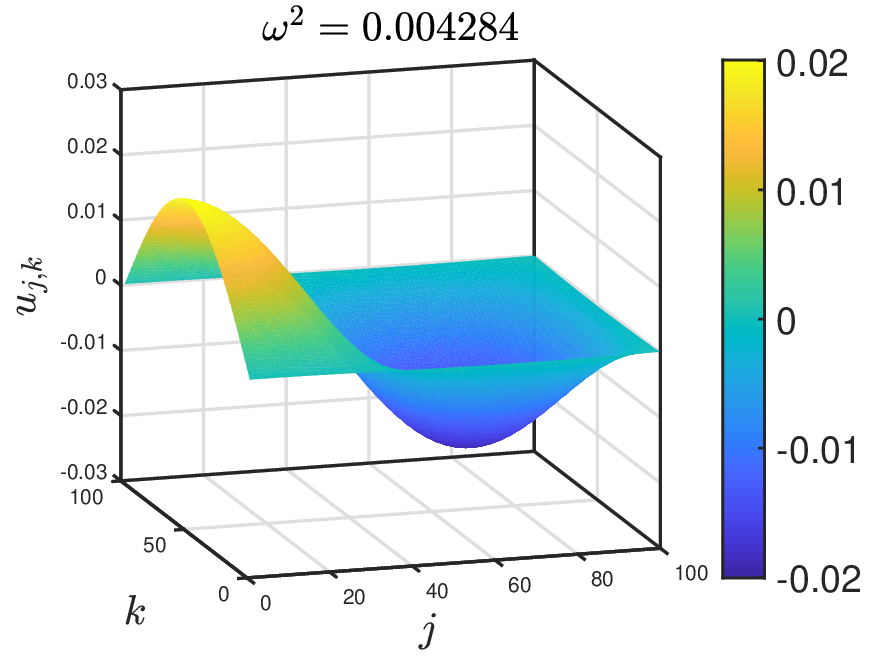}
}
\caption{With the parameters $k_{1}=1$, $k_{2}=1.9$, $k_{3}=0$, $k_{4}=4.3$, $k_{5}=3.9$ and $k_{6}=5.1$, panels (a) and (b) illustrate two representative edge states of the two-dimensional $100\times 100$ diatomic lattice~\eqref{eq:2d} with boundary \eqref{eq:2d_boundary} (``square boundary"). Meanwhile, panels (c) and (d) show two representative extended states with eigenfrequencies near the band edges in the same system.}
\label{fig:2d_states}
\end{figure}

The edge states in semi-infinite diatomic lattices \eqref{eq:2d} can be studied based on the boundary setting.  Some representative boundary choices for \eqref{eq:2d} are shown in Fig.~\ref{fig:2d_edge}.
\begin{itemize}
\item If one edge of the lattice is set as 
$$E_1=(\cdots, q_{2l-1, 2m-1}, q_{2l-1, 2m}, q_{2l, 2m}, q_{2l, 2m+1}, q_{2l+1, 2m+1}, q_{2l+1, 2m+2}, \cdots)$$
(setting $q_{j, k}=0$ for $j-k < 2l-2m-1$) with illustration in Fig.~\ref{fig:2d_edge} (a), then states localized at this edge should satisfy $(u_{2l-1,2m-1}, u_{2m-1,2m})=c_1 v_{1,1}+c_2 v_{1,2}$. 
\item If the lattice is equipped with the edge
$$E_2=(\cdots, q_{2l-1, 2m-1}, q_{2l-1, 2m}, q_{2l-2, 2m}, q_{2l-2, 2m+1}, q_{2l-3, 2m+1}, q_{2l-3, 2m+2}, \cdots)$$
(setting $q_{j, k}=0$ for $j+k < 2l+2m-2$) with illustration in Fig.~\ref{fig:2d_edge} (b), then states localized at this edge should satisfy $(u_{2l-1,2m-1}, u_{2m-1,2m})=c_1 v_{1,1}+c_2 v_{2,1}$. 
\item Suppose the lattice has the vertical edge
$$
E_3=(\cdots, q_{2l-1,2m}, q_{2l, 2m}, q_{2l+1, 2m}, q_{2l+2, 2m}, \cdots)
$$
(setting $q_{j,k}=0$ for $j<2m$) with illustration in Fig.~\ref{fig:2d_edge} (c), then the states localized at this edge should satisfy $(u_{2l-1,2m-1}, u_{2m-1,2m})=c_1 v_{1,1}+c_2 v_{1,2}$ with $|a_1 a_2^{-1}|<1$ or $(u_{2l-1,2m-1}, u_{2m-1,2m})=c_1 v_{1,1}+c_2 v_{2,1}$ with $|a_1^{-1} a_2|<1$.
\end{itemize}
From above examples, the 2D lattice with a straight boundary can usually be effectively decoupled into a set of effective 1D chains along the transverse direction. Consequently, the results of 1D edge states in previous sections can be naturally applied to study the existence of 2D edge modes along the specific boundary. However, since the number of boundary stiffness coefficients for 2D lattices is also large ($\Theta(n)$), in the following we demonstrate some representative numerical observations rather than providing a detailed analysis of the existence of 2D edge states.
\begin{itemize}
\item In large finite 2-D diatomic lattices, it can be examined that edge states exist for a wide range of boundary coefficients. 
\item Similarly to the situation for diatomic chains, the existence of edge states depends on boundary coefficients and ${\rm sgn}(k_1-k_2)$ (Fig.~\ref{fig:2d_states} (a) and (b)).  
\item At the same time, the eigenfrequencies in the bands and near the band edges for large diatomic lattices follow patterns  (Fig.~\ref{fig:2d_states} (c) and (d)) as in Lemma.~\ref{lm:op_eigenfrequencies} and Lemma.~\ref{lm:ac_eigenfrequencies}. 
\end{itemize}

\section{Conclusion}
\label{sec:conclusion}

In this work, a framework for analytically characterizing edge states and extended states is introduced in large-size lattices. We start with edge states in semi-infinite linear diatomic chains and then asymptotically extend the results to finite but long chains, demonstrating the dependence of edge states on lattice size and boundary conditions. Our analytical estimations bridge the gap between ideal topological invariants defined in infinite lattices and their spectral realizations in finite mechanical systems. We demonstrated that the bulk-boundary correspondence remains robust provided the boundary stiffness does not excessively break the underlying lattice symmetry and the system size exceeds the edge state localization length.

It is worth noting the analogy between our mechanical model and the quantum SSH model. The boundary stiffness $k_{3,1}$ in our system is mathematically equivalent to an on-site potential at the boundary in the tight-binding approximation. Consequently, the ``disappearance" of edge states due to boundary stiffness corresponds to the breaking of chiral symmetry in the quantum system, which removes the topological protection of the zero-energy mode.

In addition to that, we conduct a detailed theoretical and numerical investigation of eigenfrequencies inside the spectral bands but near the band edges for long chains. Interestingly, these near-band-edge eigenfrequencies are observed to follow special patterns under generic boundary settings. Moreover, the obtained knowledge of linear eigenfrequencies also paves the way for studying nonlinear edge states where the frequencies can exit the bands in nonlinear continuation. In fact, our numerical experiments reveal that linear extended states can possibly be continued to either new nonlinear edge states or middle-localized states, depending on their boundary settings. Although our discussion starts with the model of one-dimensional diatomic chains for simplicity, we also numerically show that similar results hold in lattices with more complicated structures and higher dimensions. Besides, our framework for large-size lattices has potential in more applications such as studying the stability of nonlinear edge states, which will be reported in future studies.

\vspace{5mm}

{\it Acknowledgements.} HX gratefully acknowledges that this work is partially supported by NSFC (Grant No. 11801191).

~\\~\\
\begin{appendices}

\section{Proof of Lemma.~\ref{lm:increasing}:}
\label{proof:lemma_increasing_freq}

\begin{proof}
Let $E={\rm diag}(0,0,\cdots,0,1)$. Suppose $k_{3,1}$ is fixed and $k_{3,2}$ varies, then we consider the eigenvalue problem 
\beq
\label{eq:perturbation_L}
\mathcal{L}(k_{3,2})u(k_{3,2})=\lambda(k_{3,2})u(k_{3,2})
\eeq
where $\| u(k_{3,2}) \|_2^2=1$. Differentiating \eqref{eq:perturbation_L} in $k_{3,2}$ yields 
\beq
\frac{\partial\lambda(k_{3,2})}{\partial k_{3,2}}=-u^T({k_{3,2}})E u(k_{3,2})=-|u_{2n}(k_{3,2})|^2\leq 0.
\eeq
It can be directly examined every eigenvector $u$ of $\mathcal{L}$ satisfies $u_{2n}\neq 0$. Therefore, we have proved $\frac{\partial\lambda(k_{3,2})}{\partial k_{3,2}}< 0$ hence $\frac{\partial\omega}{\partial k_{3,2}}> 0$. The proof for $\frac{\partial\omega}{\partial k_{3,1}}> 0$ can be obtained in a similar way.
\end{proof}

\section{Proof of Lemma.~\ref{lm:edge_states_sp1}:}
\label{proof_edge_state_k31_appro_k2}

\begin{proof}
When $k_{2}>k_{1}$ and $a\approx -\frac{k_{1}}{k_{2}}$,  $\omega^{2}=k_{1}+k_{2}-\sigma\sqrt{k_{1}+k_{2}a}\sqrt{k_{1}+k_{2}/a}$. Thus, 
\beq
\label{eq: v1v2_min}
v_1= \left(\begin{array}{c}
 \sqrt{k_1+k_2/{a}} \\
 \sigma\sqrt{k_1+k_2 a}
 \end{array}\right), \quad 
v_2= \left(\begin{array}{c}
 \sqrt{k_1+k_2 {a}} \\
 \sigma\sqrt{k_1+k_2/ a}
 \end{array}\right)
\eeq

We write $k_{3,1}=k_2+\delta k_{3,1}$ and $\tilde{a}=-\frac{k_1}{k_2}+\delta a$ (here $\delta a<0$). Based on equation~(\ref{eq:5}), it can be obtained that 
\begin{equation}
\delta k_{3,1}\approx \sigma k_2^2\sqrt{\frac{k_2\delta a}{k_1(k_1^2-k_2^2)}}= \Theta(|\delta a|^{1/2}).
\end{equation}
Different from the subcase $k_{3,1}=k_2$, now $c_2=0$ and $a=\tilde{a}$ in (\ref{eq:bc_2}) yields an edge state with $k_{3,2}=\Theta(\frac{1}{|\delta a|^{1/2}})=\Theta(\frac{1}{|\delta k_{3,1}|})$ that can be implemented in a finite chain.

In order to consider edge states with $c_2\neq 0$, we write $$a=\tilde{a}+\Delta a=-\frac{k_1}{k_2}+\delta a+\Delta a$$ and 

\begin{equation}
\label{eq:eigenvector_app}
v_{1,2}=\sigma\sqrt{k_2(\delta a+\Delta a)}, \quad v_{1,1}=\sqrt{\frac{k_1^2-k_2^2}{k_1}}+\Theta(\delta a+\Delta a)\approx i\sqrt{\frac{k_2^2-k_1^2}{k_1}}.
\end{equation}
Accordingly 
(\ref{eq:bc_1}) yields 
\begin{equation}
\label{eq:c_2}
c_2\approx \frac{\delta k_{3,1}\sqrt{\frac{k_1^2-k_2^2}{k_1}}-\sigma\frac{k_2^2}{k_1}\sqrt{k_2(\delta a+\Delta a)}}{\sigma\sqrt{k_1(k_1^2-k_2^2)}}\approx\frac{k_2^2}{k_1}\sqrt{\frac{k_2}{k_1(k_1^2-k_2^2)}}(\sqrt{\delta a}-\sqrt{\delta a+\Delta a}).
\end{equation}  

\begin{itemize}
\item If $|\delta a|\ll |\Delta a|\ll 1$ (here $\Delta a<\delta a<0$), then $c_2\approx -\frac{k_2^2}{k_1}\sqrt{\frac{k_2\Delta a}{k_1(k_1^2-k_2^2)}}$. This is the situation where $\delta a$ and $\delta k_{3,1}$ are small and the results are basically the same as those in the subcase $k_{3,1}=k_2$. 
\begin{itemize}
\item When $\sqrt{|\Delta a|}\ll a^{2n-2}$:\\
Then $1\ll k_{3,2}\approx \frac{k_1 a^{2n-2}}{\sigma c_2}= \Theta(\frac{a^{2n-2}}{\sqrt{|\Delta a|}})\ll \frac{a^{2n-2}}{\sqrt{|\delta a|}}$. Since $|c_2 a^{2-2n}|\ll 1$, the corresponding state $u$ is a left edge state.
\item When $\sqrt{|\Delta a|}=\Theta( a^{2n-2})$:\\
Then $k_{3,2}\approx \omega^2-k_1+ \frac{k_1 a^{2n-2}}{\sigma c_2}= \mathcal{O}(1)$ and $k_{3,2}\not\approx \omega^2-k_1\approx k_2$. Since $|c_2 a^{2-2n}|=\Theta (1)$, the eigenstate $u$ in this case can also be considered as left localized.
\item When $\sqrt{|\Delta a|}\gg a^{2n-2}$:\\
Then $k_{3,2}\approx \omega^2-k_1+ \frac{k_1( v_{1,1} a^{2n-2}+c_2 v_{2,1})}{ c_2 v_{2,2}}\approx \omega^2-k_1\approx k_2$ but now $|c_2 a^{2-2n}|\gg 1$.
\end{itemize}
Here we notice that $|\delta a+\Delta a|=\mathcal{O}(a^{2n-2})\ll\frac{1}{n}$ for left edge-state, namely $a^{2n-2}\approx (\frac{k_{1}}{k_{2}})^{2n-2}$.

\item If $|\delta a|=\Theta( |\Delta a|)$, then $\Theta (|\sqrt{\delta a}-\sqrt{\delta a+\Delta a}|)=|c_2|=\Theta (\sqrt{|\delta a|})$. At first we consider a special situation that $|\delta a+\Delta a|\ll |\delta a|$. In this case $ \sqrt{|\delta a+\Delta a|}\ll|c_2|\approx \frac{k_2^2}{k_1}\sqrt{|\frac{k_2\delta a}{k_1(k_1^2-k_2^2)}|}=\Theta(\sqrt{|\delta a|}) $. Another (and more generic) situation is $\Theta(|\delta a+\Delta a|)=|\delta a|=\Theta (|\Delta a|)$ and accordingly $\Theta(|c_2|)=\sqrt{|\delta a}|=\Theta(\sqrt{|\Delta a|})$.
\begin{itemize}
\item When $|\delta k_{3,1}|=\Theta( \sqrt{|\Delta a|})\ll a^{2n-2}$, $k_{3,2}\approx \omega^2-k_1+\frac{k_1 a^{2n-2}}{\sigma c_2}=\Theta(\frac{ a^{2n-2}}{|\delta k_{3,1}|})$.
\item When $|\delta k_{3,1}|=\Theta( \sqrt{|\Delta a|})=\Theta( a^{2n-2})$, $k_{3,2}\approx \omega^2-k_1+\frac{k_1 a^{2n-2}}{\sigma c_2}=\mathcal{O} (1)$ and $k_{3,2}\not\approx \omega^2-k_1\approx k_2$.
\item When $|\delta k_{3,1}|=\Theta( \sqrt{|\Delta a|})\gg a^{2n-2}$, $k_{3,2}\approx \omega^2-k_1+\frac{k_1( v_{1,1} a^{2n-2}+c_2 v_{2,1})}{ c_2 v_{2,2}}\approx \omega^2-k_1\approx k_2$ but here the eigenstate $u$ may not be left localized.
\end{itemize}
In addition, if there exists a edge state, then $|\delta a+\Delta a|=\mathcal{O}(a^{2n-2})\ll\frac{1}{n}$ and $a^{2n-2}\approx (\frac{k_{1}}{k_{2}})^{2n-2}$.

\item If $|\Delta a|\ll |\delta a|\ll 1$, then $|c_2|\approx |-\frac{k_2^2}{2k_1}\sqrt{\frac{k_2}{k_1(k_1^2-k_2^2)}}\frac{\Delta a}{\sqrt{\delta a}}|\ll \sqrt{|\delta a|}=\Theta(\sqrt{|\delta a+\Delta a}|)$. 
\begin{itemize}
\item When $|\frac{\Delta a}{\sqrt{\delta a}}|\ll a^{2n-2}$, $|c_2| a^{2-2n}\ll 1$\quad ($|\Delta a|\ll \sqrt{|\delta a|}a^{2n-2}\ll a^{2n-2}$) hence the corresponding state is a left-edge state. To characterize $k_{3,2}$, we consider the following cases:
\begin{itemize}
\item $|\frac{\Delta a}{\delta a}|\ll a^{2n-2}$, $k_{3,2}\approx \frac{k_1 v_{1,1}}{ v_{1,2}}=\Theta(\frac{1}{\sqrt{|\delta a|}})$.
\item $|\frac{\Delta a}{\delta a}|=\Theta( a^{2n-2})$, $k_{3,2}\approx \omega^2-k_1+\frac{k_1 v_{1,1} a^{2n-2}}{v_{1,2}a^{2n-2}+c_2 v_{2,2}} = \Omega(\frac{1}{\sqrt{|\delta a|}})$.
\item $|\frac{\Delta a}{\delta a}|\gg a^{2n-2}$, $k_{3,2}\approx \omega^2-k_1+\frac{k_1 a^{2n-2}}{\sigma c_2} =\Theta (\frac{a^{2n-2}\sqrt{|\delta a|}}{|\Delta a|})$ hence $|\frac{a^{2n-2}}{\sqrt{\delta a}}|\ll k_{3,2}\ll \frac{1}{\sqrt{|\delta a|}}$.
\end{itemize}
In above cases, if $|\delta a+\Delta a|=\mathcal{O}(a^{2n-2})$, then $a^{2n-2}\approx (\frac{k_{1}}{k_{2}})^{2n-2}$.
\item When $|\frac{\Delta a}{\sqrt{\delta a}}| \gg a^{2n-2}$ (hence $\Theta(|\sqrt{\delta a}|)= |\delta k_{3,1}| \gg  a^{2n-2}$), ${k}_{3,2}\approx \omega^2-k_1+ \frac{ k_1 a^{2n-2}}{\sigma c_2}+\frac{v_{2,1}}{v_{2,2}}\approx \omega^2-k_1\approx k_2$. Here we have $1\ll|c_2 a^{2-2n}|=\Theta( |\frac{\Delta a}{\sqrt{\delta a}}a^{2-2n}|)$.
\item When $|\frac{\Delta a}{\sqrt{\delta a}}| =\Theta( a^{2n-2})$ (hence $\sqrt{|\delta a}|=\Theta (|\delta k_{3,1}|) \gg  a^{2n-2}$), $k_{3,2}\approx \omega^2-k_1+\frac{k_1 a^{2n-2}}{\sigma c_2}= \mathcal{O}(1)$ and $k_{3,2}\not\approx \omega^2-k_1\approx k_2$. Here, we notice that $|c_{2}a^{2n-2}|=\Theta(1)$ and $|\Delta a|=\Theta(\sqrt{|\delta a|}a^{2n-2})\ll a^{2n-2}$. 

\end{itemize}

\end{itemize}
The discussion above has already proven \textbf{ Lemma~\ref{lm:edge_states_sp1}:}.
In order to better characterize $a$ (or $\Delta a$) of the eigenstate from given $k_{3,1}$ and $k_{3,2}$, the results for the left edge states with $k_{3,1}\approx k_2$ can be rephrased as follows:

\begin{itemize}
    \item If $k_{3,1}=k_2\not\approx k_{3,2}$:
    \begin{itemize}
        \item If $k_{3,2}\gg 1$, then  $\sqrt{|\Delta a|}=\Theta(\frac{(\frac{k_{1}}{k_{2}})^{2n-2}}{k_{3,2}})$;
        \item If $k_{3,2}= \mathcal{O}(1)$ and  $k_{3,2}\not\approx k_2$, then 
        $\sqrt{|\Delta a|}=\Theta((\frac{k_{1}}{k_{2}})^{2n-2})$; 
        
    \end{itemize}
    \item If $k_{3,1}=k_2+\delta k_{3,1}$ and $k_{3,2}\not\approx k_{2}$:
    \begin{itemize}  
\item If $|\delta k_{3,1}|\ll (\frac{k_{1}}{k_{2}})^{2n-2}$ and $\frac{(\frac{k_{1}}{k_{2}})^{2n-2}}{|\delta k_{3,1}|}\ll k_{3,2}\ll\frac{1}{|\delta k_{3,1}|}$, then $ |\delta k_{3,1}|(\frac{k_{1}}{k_{2}})^{n-1}\ll\sqrt{|\Delta a|}\ll |\delta k_{3,1}|$;

\item If $|\delta k_{3,1}|\ll (\frac{k_{1}}{k_{2}})^{2n-2}$ and $ k_{3,2}=\Omega(\frac{1}{|\delta k_{3,1}|})$, then $\sqrt{|\Delta a|}=\mathcal{O}( |\delta k_{3,1}|(\frac{k_{1}}{k_{2}})^{n-1})$;

        \item If $|\delta k_{3,1}|\ll (\frac{k_{1}}{k_{2}})^{2n-2}$ and        $k_{3,2}=\Theta(|\frac{(\frac{k_{1}}{k_{2}})^{2n-2}}{\delta k_{3,1}}|)$, then $\sqrt{|\Delta a|}=\Theta (|\delta k_{3,1}|)$;
        
        \item If $|\delta k_{3,1}|\ll (\frac{k_{1}}{k_{2}})^{2n-2}$ and
        $1\ll k_{3,2}\ll |\frac{(\frac{k_{1}}{k_{2}})^{2n-2}}{\delta k_{3,1}}|$, then  $|\delta k_{3,1}|\ll\sqrt{|\Delta a|}\ll (\frac{k_{1}}{k_{2}})^{2n-2} $;

   \item If $|\delta k_{3,1}|\ll (\frac{k_{1}}{k_{2}})^{2n-2}$, 
        $k_{3,2}= \mathcal{O}(1)$ and $k_{3,2}\not\approx k_2$, then  $|\delta k_{3,1}|\ll\sqrt{|\Delta a|}=\Theta ((\frac{k_{1}}{k_{2}})^{2n-2}) $;

\item If $|\delta k_{3,1}|= \Theta((\frac{k_{1}}{k_{2}})^{2n-2})$ and $\frac{(\frac{k_{1}}{k_{2}})^{2n-2}}{|\delta k_{3,1}|}\ll k_{3,2}\ll\frac{1}{|\delta k_{3,1}|}$, then $ |\delta k_{3,1}|(\frac{k_{1}}{k_{2}})^{n-1}\ll\sqrt{|\Delta a|}\ll |\delta k_{3,1}|=\Theta((\frac{k_{1}}{k_{2}})^{2n-2})$;

\item If $|\delta k_{3,1}|= \Theta((\frac{k_{1}}{k_{2}})^{2n-2})$ and $ k_{3,2}=\Omega(\frac{1}{|\delta k_{3,1}|})$, then $\sqrt{|\Delta a|}=\mathcal{O}( |\delta k_{3,1}|(\frac{k_{1}}{k_{2}})^{n-1})$;

  \item If $|\delta k_{3,1}|=\Theta((\frac{k_{1}}{k_{2}})^{2n-2})$, 
        $k_{3,2}=\mathcal{O}(1)$ and $k_{3,2}\not\approx k_2$, then  $ \sqrt{|\Delta a|}=\Theta(|\delta k_{3,1}|)$;

\item If $\tilde{a}^{2n-2}\ll|\delta k_{3,1}|\ll1 $ and $\frac{(\frac{k_{1}}{k_{2}})^{2n-2}}{|\delta k_{3,1}|}\ll k_{3,2}\ll\frac{1}{|\delta k_{3,1}|}$, then $ |\delta k_{3,1}||\tilde{a}|^{n-1}\ll\sqrt{|\Delta a|}\ll \sqrt{|\delta k_{3,1}|}|\tilde{a}|^{n-1}$;

\item If $\tilde{a}^{2n-2}\ll|\delta k_{3,1}|\ll1 $ and $ k_{3,2}=\Omega(\frac{1}{|\delta k_{3,1}|})$, then $\sqrt{|\Delta a|}=\mathcal{O}( |\delta k_{3,1}||\tilde{a}|^{n-1})$;

        \item If $\tilde{a}^{2n-2}\ll|\delta k_{3,1}|\ll1 $, 
        $k_{3,2}= \mathcal{O}(1)$ and $k_{3,2}\not\approx k_2$, then  $|\tilde{a}|^{n-1}\gg\sqrt{|\Delta a|}=\Theta(\sqrt{|\delta k_{3,1}|}|\tilde{a}|^{n-1})\gg \tilde{a}^{2n-2}$;

    \end{itemize}

\item  If $k_{3,1}\approx k_{2}\approx k_{3,2}$, then $|c_{2}a^{2n-2}|\gg1$ and corresponding state may be localized at two sides.

\end{itemize}

\end{proof}

\section{Proof of Lemma.~\ref{lm:band_edge_states}:}
\label{proof:a=pm1}

\begin{proof}
First we consider the case $a=1$, then $\omega^2=(k_1+k_2)(1+\sigma)$ and
\begin{eqnarray}
\frac{u_1}{u_2}&=&-\sigma\frac{c_1+c_2}{c_1-c_2}=\frac{k_1}{k_1+k_{3,1}-\omega^2}, \\
\frac{u_{2n-1}}{u_{2n}}&=&-\sigma\frac{c_1+c_2-c_2(n-1)\frac{2(k_1+k_2)}{k_2}}{c_1-c_2-c_2(n-1)\frac{2(k_1+k_2)}{k_2}}=\frac{k_1+k_{3,2}-\omega^2}{k_1}.
\end{eqnarray}
\begin{itemize}
\item If $c_1=0$, then 
\begin{itemize}
\item $\sigma=-1$, $\omega^{2}=0$: $k_{3,1}=-2k_{1}<0$ and $k_{3,2}=0-\frac{2k_{1}}{1+(n-1)\frac{2(k_{1}+k_{2})}{k_{2}}}<0$; 
\item $\sigma=1$, $\omega^{2}=2(k_{1}+k_{2})$: $k_{3,1}=2k_1+2k_2$ and $2k_{2}<k_{3,2}=2k_2+\frac{2k_1}{1+(n-1)\frac{2k_1+2k_2}{k_2}}\approx 2k_2+\frac{k_2k_2}{n(k_1+k_2)}\approx 2k_2$.
\end{itemize}
\item If $c_1=1$ and $\zeta_{1}=1-c_{2}-c_{2}(n-1)\frac{2k_{1}+2k_{2}}{k_{2}}$, then
\begin{itemize}
    \item $\sigma=-1$, $\omega^{2}=0$: $c_{2}=-\frac{k_{3,1}}{2k_{1}+k_{3,1}}$ and $k_{3,2}=\frac{2c_2k_1}{\zeta_{1}}$.
\begin{itemize}
    \item $c_{2}=0$: $k_{3,1}=k_{3,2}=0$.

\end{itemize}

    \item $\sigma=1$: $\omega^2=2k_1+2k_2$, $c_2=\frac{2k_2-k_{3,1}}{k_{3,1}-2k_1-2k_2}$, $k_{3,2}=2k_2-\frac{2c_2 k_1}{\zeta_{1}}$.
    \begin{itemize}
        \item $c_2=0$: $k_{3,1}=k_{3,2}=2k_2$.
        \item $c_2<0$: $k_{3,1}<2k_2$ or $k_{3,1}>2k_1+2k_2$, $k_{3,2}>2k_2$ and  $k_{3,2}-2k_2\approx \frac{-c_2 k_1}{\zeta_{1}+c_{2}}= \mathcal{O}(\frac{1}{n})$.
\begin{itemize}
    \item If $|c_{2}|\gg\frac{1}{n}$, ($2(k_{1}+k_{2})<k_{3,1}$ or $k_{3,1}<2k_{2}$ and $2k_{2}-k_{3,1}\gg\frac{1}{n}$), $k_{3,2}\approx 2k_{2}+\frac{k_{1}k_{2}}{n(k_{1}+k_{2})}\approx 2k_{2}$. Namely, $|2k_{2}-k_{3,1}|\gg1$ and $k_{3,2}=2k_{2}+\Theta(\frac{1}{n})$.

    \item If $|c_{2}|\ll\frac{1}{n}$, ($k_{3,1}<2k_{2}$ and $2k_{2}-k_{3,1}\ll\frac{1}{n}$) $k_{3,2}\approx 2k_{2}-2c_{2}k_{1}\approx 2k_{2}$ and $k_{3,2}>2k_{2}$. Meanwhile $2k_{2}-k_{3,1}=\Theta(k_{3,2}-2k_{2})$.

    \item If $|c_{2}|=\Theta(\frac{1}{n})$, ($k_{3,1}<2k_{2}$ and $2k_{2}-k_{3,1}=\Theta(\frac{1}{n})$): Since $0<\zeta_{1}=\Theta(1)$, then $k_{3,2}=2k_{2}+\Theta(\frac{1}{n})$, ($k_{3,2}>2k_{2}$). Meanwhile, $2k_{2}-k_{3,1}=\Theta(k_{3,2}-2k_{2})=\Theta(\frac{1}{n})$.

\end{itemize}

        \item $c_2>0$: $2k_2<k_{3,1}<2k_1+2k_2$,
        \begin{itemize}
            \item If $c_2\gg \frac{1}{n}$, 
            (namely $k_{3,1}-2k_{2}\gg\frac{1}{n}$) $2k_{2}<k_{3,2}\approx 2k_2+\frac{k_1 k_2}{n(k_1+k_2)}\approx 2k_2$.
            \item If $c_2\ll \frac{1}{n}$, $0<k_{3,1}-2k_2\ll \frac{1}{n}$, $2k_{2}>k_{3,2}\approx 2k_2-2c_2 k_1\approx 2k_2$ and $2k_{2}-k_{3,2}\ll\frac{1}{n}$.
            \item If $c_2=\Theta(\frac{1}{n})$: $0<k_{3,1}-2k_2=\Theta(\frac{1}{n})$. If $|\zeta_{1}|=\Theta(1)$ and $\zeta_{1}<0$, then $|k_{3,2}-2k_2|=\Theta(\frac{1}{n})$ and $k_{3,2}>2k_2$; If $|\zeta_{1}|=\Theta(1)$ and $\zeta_{1}>0$, then $|k_{3,2}-2k_2|=\Theta(\frac{1}{n})$ and $k_{3,2}<2k_2$;
            
            \item If $c_2=\Theta(\frac{1}{n})$: $0<k_{3,1}-2k_2=\Theta(\frac{1}{n})$. If $|\zeta_{1}|\ll 1$ and $\zeta_{1}>0$, then $|k_{3,2}-2k_2|\gg \frac{1}{n}$ and $k_{3,2}<2k_{2}$; If $|\zeta_{1}|\ll 1$ and $\zeta_{1}<0$, then $|k_{3,2}-2k_2|\gg\frac{1}{n}$ and $k_{3,2}>2k_{2}$.
        \end{itemize}
    \end{itemize}
\end{itemize}
\end{itemize}
Then it comes to the case $a=-1$ where $\omega^2=(1-\sigma)k_1+(1+\sigma)k_2$ and
\begin{eqnarray}
\frac{u_1}{u_2}&=&\sigma\frac{c_1+c_2}{c_1-c_2}=\frac{k_1}{k_1+k_{3,1}-\omega^2}, \\
\frac{u_{2n-1}}{u_{2n}}&=&\sigma\frac{c_1+c_2-c_2(n-1)\frac{2(k_2-k_1)}{k_2}}{c_1-c_2-c_2(n-1)\frac{2(k_2-k_1)}{k_2}}=\frac{k_1+k_{3,2}-\omega^2}{k_1}.
\end{eqnarray}
\begin{itemize}
\item If $c_1=0$, then 
\begin{itemize}
\item $\sigma=1$, $\omega^{2}=2k_{2}$: $k_{3,1}=2k_2-2k_1>0$, $k_{3,2}=2k_2-\frac{2k_1}{1+(n-1)\frac{2k_2-2k_1}{k_2}}\approx 2k_2-\frac{2k_1k_2}{n(k_2-k_1)}$.
\item $\sigma=-1$, $\omega^{2}=2k_{1}$: $k_{3,1}=2k_1$ and $0<k_{3,2}=\frac{2k_1}{1+(n-1)\frac{2k_2-2k_1}{k_2}}\approx \frac{k_2k_2}{n(k_2-k_1)}\approx 0$.
\end{itemize}
\item If $c_1=1$ and $\zeta_{2}=1-c_{2}-c_{2}(n-1)\frac{2k_{2}-2k_{1}}{k_{2}}$, then
\begin{itemize}
    \item $\sigma=1$ : $\omega^2=2k_2$, $c_2=\frac{2k_2-k_{3,1}}{k_{3,1}+2k_1-2k_2}$, $k_{3,2}=2k_2+\frac{2c_2k_1}{\zeta_{2}}$.
    \begin{itemize}
        \item $c_2=0$: $k_{3,1}=k_{3,2}=2k_2$.
        \item $c_2<0$: $k_{3,1}>2k_2$ or $k_{3,1}<2k_2-2k_1$, $0<2k_2-k_{3,2}= \mathcal{O}(\frac{1}{n})$.
\begin{itemize}
    \item If $|c_{2}|\gg\frac{1}{n}$, ( $k_{3,1}-2k_{2}\gg\frac{1}{n}$), $k_{3,2}\approx 2k_{2}-\frac{k_{1}k_{2}}{n(k_{2}-k_{1})}$; 

    \item If $|c_{2}|\ll\frac{1}{n}$, ($k_{3,1}>2k_{2}$ and $0<k_{3,1}-2k_{2}\ll\frac{1}{n}$), then $2k_{2}>k_{3,2}\approx 2k_{2}+2c_{2}k_{1}$ and $2k_{2}-k_{3,2}\ll\frac{1}{n}$, ($k_{3,1}-2k_{2}=\Theta(2k_{2}-k_{3,2})$);

    \item If $|c_{2}|=\Theta(\frac{1}{n})$, ($k_{3,1}>2k_{2}$ and $k_{3,1}-2k_{2}=\Theta(\frac{1}{n})$), then $2k_{2}-k_{3,2}= \Theta(\frac{1}{n})$ and $k_{3,2}<2k_{2}$;

\end{itemize}

        \item $c_2>0$: $2k_2-2k_1<k_{3,1}<2k_2$, 
        \begin{itemize}
            \item If $c_2\gg \frac{1}{n}$,($2k_{2}-k_{3,1}\gg\frac{1}{n}$), then $2k_{2}>k_{3,2}\approx 2k_2-\frac{k_1 k_2}{n(k_2-k_1)}\approx 2k_2$.
            \item If $c_2\ll \frac{1}{n}$, $2k_{2}<k_{3,2}\approx 2k_2+2c_2k_1\approx 2k_2$.
            \item If $c_2=\Theta(\frac{1}{n})$: $0<2k_2-k_{3,1}=\Theta(\frac{1}{n})$. If $|\zeta_{2}|=\Theta(1)$ and $\zeta_{2}>0$, then $|k_{3,2}-2k_2|=\Theta(\frac{1}{n})$ and $k_{3,2}>2k_{2}$; If $|\zeta_{2}|=\Theta(1)$ and $\zeta_{2}<0$, then $|k_{3,2}-2k_2|=\Theta(\frac{1}{n})$ and $k_{3,2}<2k_{2}$.
            \item If $c_2=\Theta(\frac{1}{n})$: $0<2k_2-k_{3,1}=\Theta(\frac{1}{n})$. If $|\zeta_{2}|\ll1$ and $\zeta_{2}>0$, then $|k_{3,2}-2k_2|\gg\frac{1}{n}$ and $k_{3,2}>2k_{2}$; If $|\zeta_{2}|\ll1$ and $\zeta_{2}<0$, then $|k_{3,2}-2k_2|\gg \frac{1}{n}$ and $k_{3,2}<2k_{2}$.
        \end{itemize}
    \end{itemize}
    \item $\sigma=-1$: $\omega^2=2k_1$, $c_2=\frac{-k_{3,1}}{k_{3,1}-2k_1}$, $k_{3,2}=-\frac{2c_2 k_1}{\zeta_{2}}$.
    \begin{itemize}
        \item $c_2=0$: $k_{3,1}=k_{3,2}=0$.
        \item $c_2<0$: $k_{3,1}>2k_1$ or $k_{3,1}<0$, $0<k_{3,2}=\mathcal{O}( \frac{1}{n})$.
\begin{itemize}
    \item If $|c_{2}|\gg\frac{1}{n}$, ($- k_{3,1}\gg\frac{1}{n}$), then $0<k_{3,2}\approx \frac{k_{1}k_{2}}{n(k_{2}-k_{1})}$;

    \item If $|c_{2}|\ll\frac{1}{n}$ (or  $|c_{2}|=\Theta(\frac{1}{n}))$, then $k_{3,1}\ll\frac{1}{n}$(or $k_{3,1}=\Theta(\frac{1}{n})$);

\end{itemize}
        
        \item  $c_2>0$: $0<k_{3,1}<2k_1$, $\zeta_{2}<0$ hence $c_2>\frac{1}{1+(n-1)\frac{2k_2-2k_1}{k_2}}\approx \frac{k_2}{2n(k_2-k_1)}$. This means $k_{3,1}=\Omega(\frac{1}{n})$.
        \begin{itemize}
            \item If $c_2\gg \frac{1}{n}$, ($k_{3,1}\gg\frac{1}{n}$), then $0<k_{3,2}\approx \frac{k_1 k_2}{n(k_2-k_1)}\approx 0$.

\item If $c_{2}\ll\frac{1}{n}$, ($0<k_{3,1}\ll\frac{1}{n}$), then $k_{3,2}\approx -2c_{2}k_{1}<0$ and $-k_{3,2}=\Theta(|c_{2}|)=\Theta(k_{3,1})$;

            \item If $c_2=\Theta(\frac{1}{n})$: $k_{3,1}=\Theta(\frac{1}{n})$. If $|\zeta_{2}|=\Theta(1)$ and $\zeta_{2}>0$, then $k_{3,2}=\Theta (\frac{1}{n})$ and $k_{3,2}<0$; If $|\zeta_{2}|=\Theta(1)$ and $\zeta_{2}<0$, then $k_{3,2}=\Theta (\frac{1}{n})$ and $k_{3,2}>0$; 
            \item If $c_2=\Theta(\frac{1}{n})$: $k_{3,1}=\Theta(\frac{1}{n})$. If $|\zeta_{2}|\ll 1$ and $\zeta_{2}<0$, then $|k_{3,2}|\gg \frac{1}{n}$ and $k_{3,2}>0$; If $|\zeta_{2}|\ll 1$ and $\zeta_{2}>0$, then $|k_{3,2}|\gg \frac{1}{n}$ and $k_{3,2}<0$.
        \end{itemize}
    \end{itemize}
\end{itemize}
\end{itemize}
Comparing $|k_{3,1}-k_2(1+\sigma )|$ and $|k_{3,2}-k_2(1+\sigma)|$ in each case above concludes the proof.
    
\end{proof}

\section{Proof of Lemma.~\ref{lemma:a_approx_pm1_2k2}:}
\label{proof:k31_2k2}

\begin{proof}
 Here we just show the case
$\omega^2\approx 2k_2$ and the conclusion on the case $\omega^{2}\approx2(k_{1}+k_{2})$ can be obtained similarly.   
Since 
\begin{equation}
\label{eq:c2_k31_a+}
c_2\approx\frac{\frac{k_1 k_2 \Delta a}{\sqrt{k_1-k_2}}-\delta k_{3,1}\sqrt{k_1-k_2}+k_1 k_2 (\Delta a)^2(\frac{1}{2\sqrt{k_1-k_2}})+\frac{k_2\delta k_{3,1}\Delta a}{2\sqrt{k_1-k_2}}}{\frac{k_1 k_2 \Delta a}{\sqrt{k_1-k_2}}+\delta k_{3,1}\sqrt{k_1-k_2}}
\end{equation}
depends on the relation between $\delta k_{3,1}$ and $\Delta a$, the following five cases will be investigated and $\Delta k_{3,2}$ 
\begin{equation}
    \Delta k_{3,2}\approx-k_1+k_1\frac{v_{1,1}+v_{2,1}c_{2}a^{2-2n}}{v_{1,2}+v_{2,2}c_{2}a^{2-2n}}
\end{equation}
will be obtained accordingly.
For $|\delta k_{3,1}|\gg |\Delta a|$ ($c_2\approx -1$): 
\begin{itemize}\item $|\Delta a|\ll \frac{1}{n}$ ($a^{-2n}\approx 1+2n\Delta a$ and $c_2+1= \Theta(|\frac{\Delta a}{\delta k_{3,1}}|)$):
\begin{itemize}
\item $|\delta k_{3,1}|\gg \frac{1}{n}$: Then $|1+c_2 a^{2-2n}|=\Theta(|n\Delta a|)$ and $|\Delta k_{3,2}|=\Theta(\frac{1}{n})$.
\item $|\delta k_{3,1}|=\Theta(\frac{1}{n})$: Then $|v_{1,2}+v_{2,2}c_2 a^{2-2n}|=\mathcal{O}(|n\Delta a|)$ and $|\Delta k_{3,2}|=\Omega(\frac{1}{n})$.
\item $|\delta k_{3,1}|\ll \frac{1}{n}$: Then $|1+c_2 a^{2-2n}|=\Theta(|\frac{\Delta a}{\delta k_{3,1}}|)$ and $|\Delta k_{3,2}|=\Theta(|\delta k_{3,1}|)$.
\end{itemize}
\item $|\Delta a|=\Theta (\frac{1}{n})$ ($1\not\approx a^{-2n}=\Theta(1)$): Then $|\Delta k_{3,2}|=\Theta(|\Delta a|)$.
 \item $|\Delta a|\gg \frac{1}{n}$ ($a^{-2n}\gg 1$): Then $|\Delta k_{3,2}|=\Theta(|\Delta a|)$.
\end{itemize}

As a result, this leads to $\Omega(\frac{1}{n})=|\Delta k_{3,2}|\ll 1$ and $ \Delta k_{3,2}>0$. 

\end{proof}

\section{Equations of motion for boundary atoms in Fig.~\ref{fig:2d_states}}
\label{appendix:2d_boundary}

\begin{equation}
\begin{split}
  \ddot{q}_{1,1} = & k_{2}q_{2,1}+k_{1}q_{1,2}-(k_1+k_2+k_5)q_{1,1}, \\
  \ddot{q}_{1,2j} = & k_{2}q_{1,2j+1}+k_{1}(q_{1,2j-1}+q_{2,2j})-(2k_1+k_2+k_5)q_{1,2j}, \\
  \ddot{q}_{1,2j+1} = & k_2(q_{1,2j}+q_{2,2j+1})+k_1 q_{1,2j+2}-(k_1+2k_2+k_5)q_{1,2j+1}, \\
  \ddot{q}_{1,100} = & k_{1}(q_{1,99}+q_{2,100})-(2k_1+k_4+k_5)q_{1,100}, \\
  \ddot{q}_{100,1} = & k_{2}q_{99,1}+q_{100,2})-(2k_2+k_6)q_{1,1}, \\
  \ddot{q}_{100,100} = & k_{2}q_{100,99}+k_{1}q_{99,100}-(k_1+k_2+k_5+k_6)q_{100,100}, \\
  \ddot{q}_{100,2j} = & k_2q_{100,2j-1}+k_1(q_{99,2j}+ q_{100,2j+1})-(2k_1+k_2+k_6)q_{100,2j}, \\
  \ddot{q}_{100,2j+1} = & k_1 q_{100,2j}+ k_{2}(q_{99,2j+1}+q_{100,2j+2})-(k_1+2k_2+k_6)q_{100,2j+1}, \\
  \ddot{q}_{2j+1,1} = & k_{2}q_{2j+2,1}+k_{1}(q_{2j,1}+q_{2j+1,2})-(2k_1+k_2)q_{2j+1,1}, \\
  \ddot{q}_{2j,1} = & k_{2}(q_{1,2j-1}+q_{2j-1,2})+k_{1}q_{2j+1,1}-(2k_2+k_1)q_{2j,1}, \\
  \ddot{q}_{2j+1,100} = & k_2q_{2j,100}+k_1 (q_{2j+1,99}+q_{2j+2,100})-(2k_1+k_2+k_4)q_{2j+1,100}, \\
  \ddot{q}_{2j,100} = & k_{1}q_{2j-1,100}+k_{2}(q_{2j,99}+q_{2j+1,100})-(k_1+2k_2+k_4)q_{2j,100}, ~(1\leq j\leq 49).
\end{split}
\label{eq:2d_boundary}
\end{equation}

\end{appendices}

\end{document}